\PassOptionsToPackage{normalem}{ulem} 
\documentclass[]{TEAI}
\usepackage{helvet}

\usepackage{amsmath} 
\usepackage{natbib}
\usepackage{graphicx}
\usepackage{subcaption} 

\usepackage[toc,page,header]{appendix}
\usepackage[utf8]{inputenc} 
\usepackage[T1]{fontenc}    
\usepackage{hyperref}       
\usepackage{url}            
\usepackage{booktabs}       
\usepackage{lmodern}        
\usepackage{amsfonts}       
\usepackage{nicefrac}       
\usepackage{microtype}      
\usepackage{wrapfig}

\usepackage{amssymb}  
\usepackage{fontawesome}  
\usepackage{url}  

\usepackage{titletoc}

\usepackage{tikz}  
\usepackage{comment}  
\usepackage{tabularx}  
\usepackage{booktabs}  

\usepackage{minitoc}

\usepackage{booktabs}
\usepackage{array}
\usepackage{etoolbox}

\newcommand{\numpapers}{500}

\definecolor{lightblue}{RGB}{200, 230, 255}  
\definecolor{headerblue}{RGB}{150, 200, 255} 

\usepackage{pgfplots}
\usepackage[utf8]{inputenc} 
\usepackage[T1]{fontenc}    
\usepackage{hyperref}       
\usepackage{xurl}           
\usepackage{url}            
\usepackage{booktabs}       
\usepackage{amsfonts}       
\usepackage{nicefrac}       
\usepackage{microtype}      
\usepackage{xcolor}         
\usepackage{graphicx}
\usepackage{float}
\usepackage{comment}
\usepackage{forest}
\useforestlibrary{edges}
\usepackage{multirow} 
\usepackage{amsmath} 
\usepackage{makecell} 
\usepackage{siunitx}  
\usepackage{tikz}
\usepackage{pgf-pie} 
\usepackage{subcaption}
\usepackage{wrapfig}
\usepackage[export]{adjustbox}
\usepackage{tikz}
\usepackage{times}
\usepackage{epsfig}
\usepackage{enumitem}
\usepackage{graphicx}

\usepackage{ragged2e}      
\usepackage{tabularx}       
\usepackage{array}          
\usepackage{caption}        
\usepackage{enumitem}
\usepackage{pifont}
\usepackage[hang,flushmargin]{footmisc} 

\usepackage{tcolorbox}

\usepackage{tcolorbox}    
\tcbuselibrary{breakable}  
\tcbuselibrary{skins}      

\usepackage{tabularx}
\usepackage{listings}

\definecolor{Gray}{gray}{0.92}
\definecolor{gaoyifeng-pink}{RGB}{255,235,245}
\usepackage{arydshln}
\usepackage{hhline}
\usepackage{makecell}
\usepackage{verbatim}
\usepackage{titlesec}
\usepackage{tabularx}
\usepackage{ltablex}

\newcommand*{\belowrulesepcolor}[1]{%
  \noalign{%
    \kern-\belowrulesep 
    \begingroup 
      \color{#1}%
      \hrule height\belowrulesep 
    \endgroup 
    \vspace{-0.03mm}
  }%
} 
\newcommand*{\aboverulesepcolor}[1]{%
  \noalign{%
  \vspace{-0.03mm}
    \begingroup 
      \color{#1}%
    \endgroup 
    \kern-\aboverulesep 
  }%
}

\definecolor{figorange}{RGB}{228,130,47}
\definecolor{figred}{RGB}{255,0,0}
\definecolor{figgreen}{RGB}{0,176,80}

\definecolor{perception-blue}{HTML}{527CDF}
\definecolor{perception-blue-light}{HTML}{EEF3FF}

\definecolor{cognition-green}{HTML}{86D4CD}
\definecolor{cognition-green-light}{HTML}{F2FBFA}

\definecolor{planning-orange}{HTML}{E9A859}
\definecolor{planning-orange-light}{HTML}{FFF7EB}

\definecolor{interaction-red}{HTML}{E2707C}
\definecolor{interaction-red-light}{HTML}{FFECEF}

\definecolor{agentic-purple}{HTML}{9B72CF}
\definecolor{agentic-purple-light}{HTML}{F0E6FF}

\definecolor{tablegray}{gray}{0.95}

\definecolor{perception-table-blue}{HTML}{E6EEFF} 

\definecolor{cognition-table-green}{HTML}{E2F7F5}

\definecolor{planning-table-orange}{HTML}{FFF5E3}

\definecolor{interaction-table-pink}{HTML}{FFE9EC}

\definecolor{agentic-table-purple}{HTML}{F0E6FF}

\title{Safety in Embodied AI: A Survey of Risks, Attacks, and Defenses}

\author{
Xiao Li\textsuperscript{1,*},
Xiang Zheng\textsuperscript{3,*},
Yifeng Gao\textsuperscript{1},
Xinyu Xia\textsuperscript{4},
Yixu Wang\textsuperscript{1},
Xin Wang\textsuperscript{1},
Ye Sun\textsuperscript{1},
Yunhan Zhao\textsuperscript{1},
Ming Wen\textsuperscript{1,2},
Jiayu Li\textsuperscript{1},
Zixing Chen\textsuperscript{1},
Xun Gong\textsuperscript{4},
Yi Liu\textsuperscript{3},
Yige Li\textsuperscript{5},
Yutao Wu\textsuperscript{6},
Cong Wang\textsuperscript{3},
Jun Sun\textsuperscript{5},
Yixin Cao\textsuperscript{1,2},
Zhineng Chen\textsuperscript{1},
Jingjing Chen\textsuperscript{1},
Tao Gui\textsuperscript{1,2},
Qi Zhang\textsuperscript{1},
Zuxuan Wu\textsuperscript{1,2},
Xipeng Qiu\textsuperscript{1,2},
Xuanjing Huang\textsuperscript{1},
Tiehua Zhang\textsuperscript{7},
Zhipeng Wei\textsuperscript{9},
Kun Wang\textsuperscript{10},
Xinfeng Li\textsuperscript{10},
Hanxun Huang\textsuperscript{12},
Sarah Erfani\textsuperscript{12},
James Bailey\textsuperscript{12},
Jianping Wang\textsuperscript{3},
Chaowei Xiao\textsuperscript{13},
Ran He\textsuperscript{11},
Bo Li\textsuperscript{8},
Xingjun Ma\textsuperscript{1,2,$\dagger$},
Yu-Gang Jiang\textsuperscript{1,$\dagger$}
}

\affiliation[1]{\mbox{Fudan University}}
\affiliation[2]{\mbox{Shanghai Innovation Institute}}
\affiliation[3]{\mbox{City University of Hong Kong}}
\affiliation[4]{\mbox{Jilin University}}
\affiliation[5]{\mbox{Singapore Management University}}
\affiliation[6]{\mbox{Deakin University}}
\affiliation[7]{\mbox{Tongji University}}
\affiliation[8]{\mbox{UIUC}}
\affiliation[9]{\mbox{UC Berkeley}}
\affiliation[10]{\mbox{Nanyang Technological University}}
\affiliation[11]{\mbox{Chinese Academy of Sciences}}
\affiliation[12]{\mbox{The University of Melbourne}}
\affiliation[13]{\mbox{Johns Hopkins University}}

\contribution[*]{Equal Contribution}
\contribution[\dagger]{Corresponding authors}

\abstract{
\begin{abstract}

Embodied Artificial Intelligence (Embodied AI) integrates perception, cognition, planning, and interaction into agents that operate in open-world, safety-critical environments. As these systems gain autonomy and enter domains such as transportation, healthcare, and industrial or assistive robotics, ensuring their safety becomes both technically challenging and socially indispensable. Unlike digital AI systems, embodied agents must act under uncertain sensing, incomplete knowledge, and dynamic human–robot interactions, where failures can directly lead to physical harm.
This survey provides a comprehensive and structured review of safety research in embodied AI, examining attacks and defenses across the full embodied pipeline, from perception and cognition to planning, action \& interaction, and agentic system. We introduce a multi-level taxonomy that unifies fragmented lines of work and connects embodied-specific safety findings with broader advances in vision, language, and multimodal foundation models. Our review synthesizes insights from over \numpapers{} papers spanning adversarial, backdoor, jailbreak, and hardware-level attacks; attack detection, safe training and robust inference; and risk-aware human–agent interaction.
This analysis reveals several overlooked challenges, including the fragility of multimodal perception fusion, the instability of planning under jailbreak attacks, and the trustworthiness of human–agent interaction in open-ended scenarios. By organizing the field into a coherent framework and identifying critical research gaps, this survey provides a roadmap for building embodied agents that are not only capable and autonomous but also safe, robust, and reliable in real-world deployment.
\end{abstract}
}

\correspondence{\email{xingjunma@fudan.edu.cn; ygj@fudan.edu.cn}}
\checkdata[Website]{\url{https://github.com/x-zheng16/Awesome-Embodied-AI-Safety}}
\checkdata[Key Words]{Embodied AI Safety; Trustworthy Embodied AI; Multimodal Safety; Attacks and Defenses}

\extrafloats{100}

\begin{document}
\maketitle


\vspace{-1.5em}

\section{Introduction}\label{sec:introduction}

Embodied Artificial Intelligence (Embodied AI) seeks to endow autonomous agents with the ability to perceive, reason, plan, and interact with the physical world \cite{turing2007computing}. Unlike purely digital AI systems, embodied agents operate in dynamic, uncertain, and safety-critical environments such as autonomous driving \cite{zhou2025opendrivevla,tian2024drivevlm,ma2024dolphins}, collaborative robotics \cite{black2410pi0,team2025gemini}, smart healthcare \cite{kaiser2020healthcare,holland2021service,fiske2019your}, and assistive robotics \cite{anderson2018vision,gurari2018vizwiz}. In these settings, unsafe perception, flawed reasoning, erroneous planning, or unsafe interaction can lead not only to degraded task performance but also to real-world accidents, physical harm, and loss of human trust.

\begin{figure*}[t]
    \centering
    \includegraphics[width=0.95\textwidth]{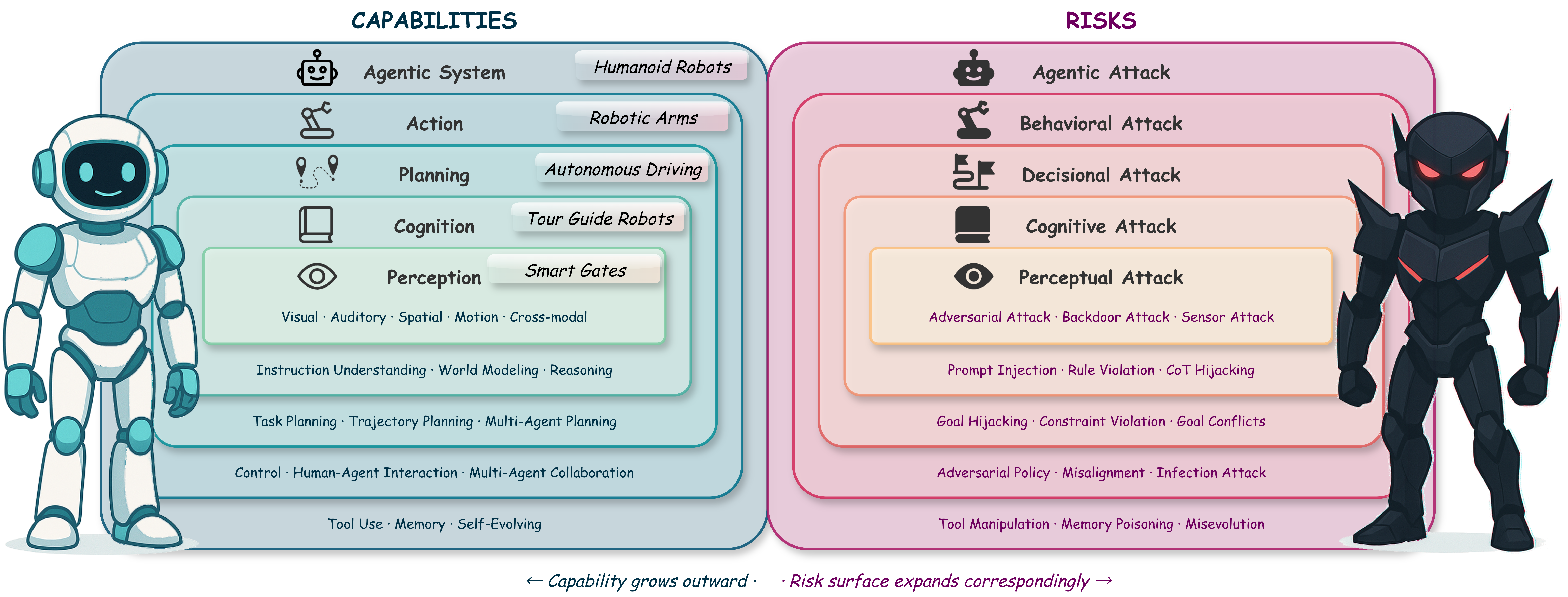}
    \caption{Capability vs.\ risk duality in embodied AI systems. \textbf{Left:} Nested capability layers from perception (innermost) to agentic systems (outermost), with representative embodiments at each level: sensor-only devices (e.g., face-recognition access controls), dialogue robots (e.g., museum guides), autonomous vehicles, robotic arms and humanoids, and future agentic robots with memory and tool use. \textbf{Right:} Corresponding safety risks at each layer. As capabilities expand outward, the attack surface grows correspondingly---vulnerabilities at inner layers cascade to outer layers, amplifying risks in more autonomous systems.}
    \label{fig:capability_risk}
\end{figure*}

\begin{figure*}[t]
    \centering
    \includegraphics[width=0.9\textwidth]{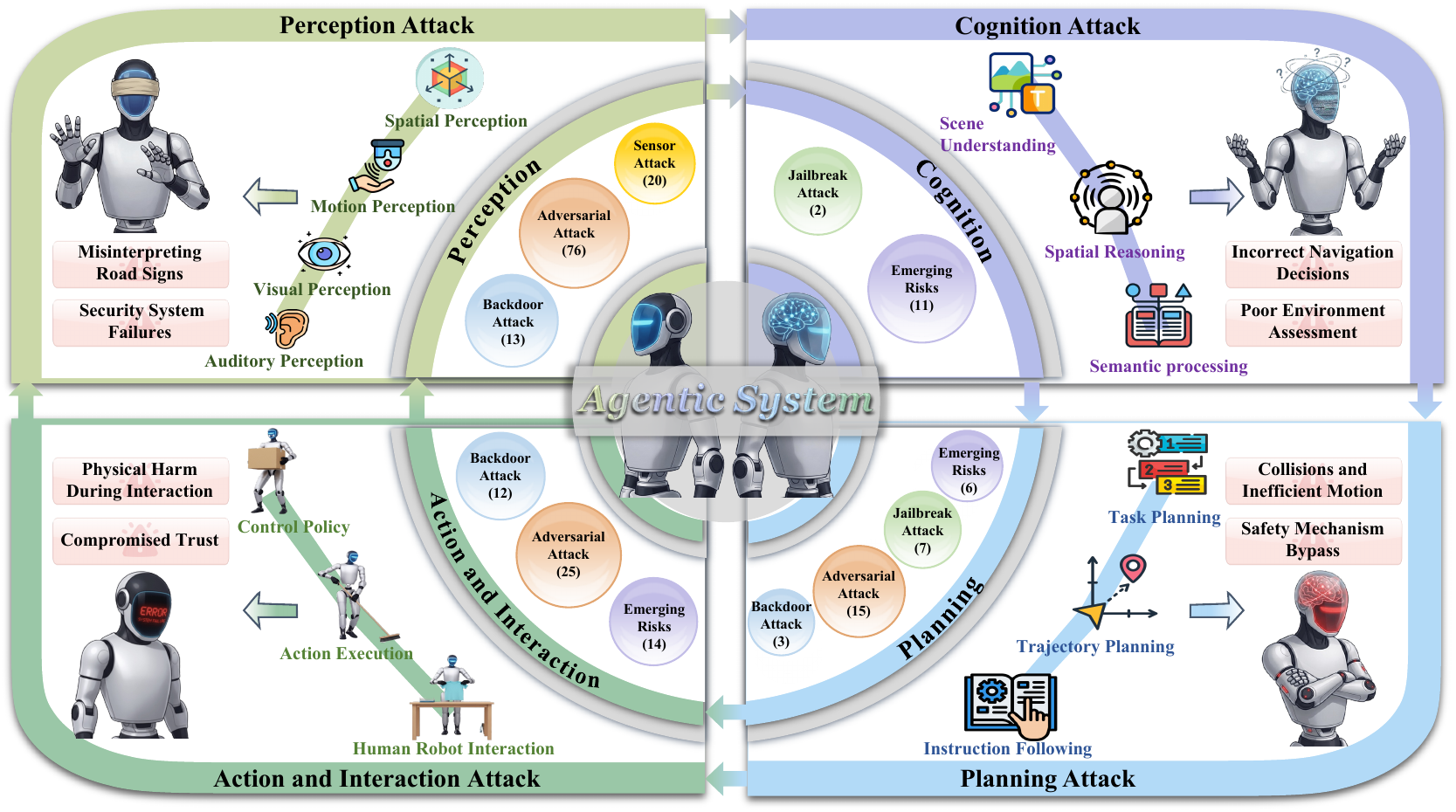}
    \caption{Illustration of safety threats and attack surfaces across capability layers of embodied AI systems.}
    \label{fig:illustration}
\end{figure*}

\begin{figure*}[t]
    \centering
    \includegraphics[width=0.95\textwidth]{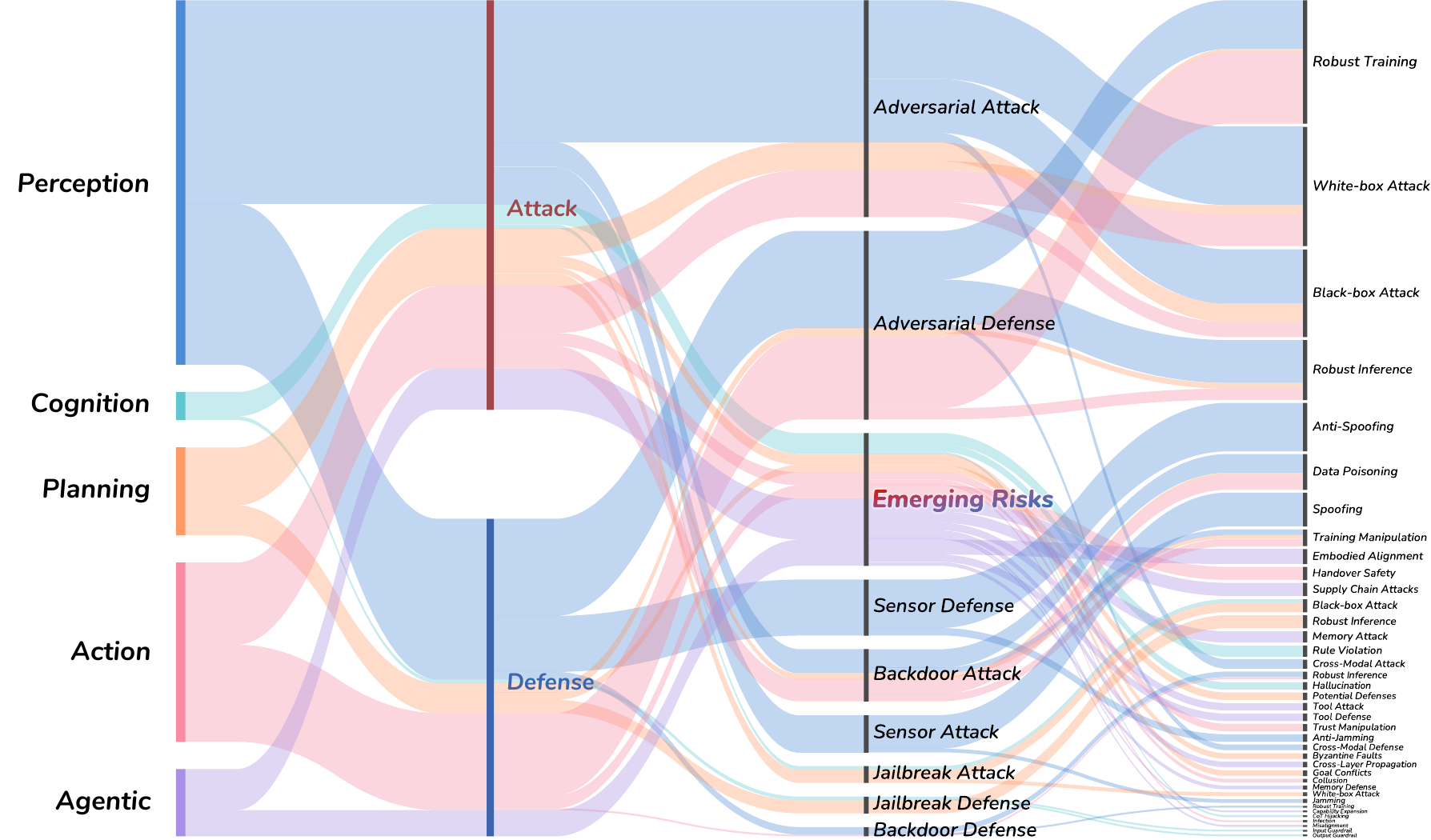}
    \caption{Overview of representative attack and defense methods across perception, cognition, planning, action, and agentic system layers. The width of the strips is proportional to the number of research works.}
    \label{fig:overview}
\end{figure*}

\newcommand{\boxfont}{\fontsize{5.7pt}{6.5pt}\selectfont} 

\newcommand{\EAS}{Embodied AI Safety}
\newlength{\secwidth}
\setlength{\secwidth}{5.5em}       
\newlength{\subsecwidth}
\setlength{\subsecwidth}{8.5em}    
\newlength{\subsubsecwidth}
\setlength{\subsubsecwidth}{8em}  
\newlength{\leafwidth}
\setlength{\leafwidth}{30.5em}       
\tikzstyle{node-cfg}=[
    rectangle,
    draw=brown!60!black,
    sharp corners,
    text opacity=1,
    inner sep=3pt,
    text=black,
    fill=yellow!5,
    fill opacity=.4,
    line width=1pt,
    font=\tiny,
    align=left
]

\tikzstyle{perception-bg-node}=[
    rectangle,
    rounded corners=2pt,
    fill=perception-blue-light,
    draw=perception-blue,
    inner sep=3pt,
    line width=1pt,
    text opacity=1,
    text=black,
    font=\boxfont,
    align=left
]

\tikzstyle{perception-line-node}=[
    rectangle,
    rounded corners=2pt,
    fill=none,
    fill opacity=.2,
    draw=perception-blue,
    inner sep=3pt,
    line width=0.5pt,
    text opacity=1,
    text=black,
    font=\tiny,
    align=left
]

\tikzstyle{cognition-bg-node}=[
    rectangle,
    rounded corners=2pt,
    fill=cognition-green-light,
    draw=cognition-green,
    inner sep=3pt,
    line width=1pt,
    text opacity=1,
    text=black,
    font=\boxfont,
    align=left
]

\tikzstyle{cognition-line-node}=[
    rectangle,
    rounded corners=2pt,
    fill=none,
    fill opacity=.2,
    draw=cognition-green,
    inner sep=3pt,
    line width=0.5pt,
    text opacity=1,
    text=black,
    font=\tiny,
    align=left
]

\tikzstyle{planning-bg-node}=[
    rectangle,
    rounded corners=2pt,
    fill=planning-orange-light,
    draw=planning-orange,
    inner sep=3pt,
    line width=1pt,
    text opacity=1,
    text=black,
    font=\boxfont,
    align=left
]

\tikzstyle{planning-line-node}=[
    rectangle,
    rounded corners=2pt,
    fill=none,
    fill opacity=.2,
    draw=planning-orange,
    inner sep=3pt,
    line width=0.5pt,
    text opacity=1,
    text=black,
    font=\tiny,
    align=left
]

\tikzstyle{interaction-bg-node}=[
    rectangle,
    rounded corners=2pt,
    fill=interaction-red-light,
    draw=interaction-red,
    inner sep=3pt,
    line width=1pt,
    text opacity=1,
    text=black,
    font=\boxfont,
    align=left
]

\tikzstyle{interaction-line-node}=[
    rectangle,
    rounded corners=2pt,
    fill=none,
    fill opacity=.2,
    draw=interaction-red,
    inner sep=3pt,
    line width=0.5pt,
    text opacity=1,
    text=black,
    font=\tiny,
    align=left
]

\tikzstyle{agentic-bg-node}=[
    rectangle,
    rounded corners=2pt,
    fill=agentic-purple-light,
    draw=agentic-purple,
    inner sep=3pt,
    line width=1pt,
    text opacity=1,
    text=black,
    font=\boxfont,
    align=left
]

\tikzstyle{agentic-line-node}=[
    rectangle,
    rounded corners=2pt,
    fill=none,
    fill opacity=.2,
    draw=agentic-purple,
    inner sep=3pt,
    line width=0.5pt,
    text opacity=1,
    text=black,
    font=\tiny,
    align=left
]


\tikzset{
  perception-line-node/.append style = {font=\boxfont, inner ysep=1.5pt},
  cognition-line-node/.append style  = {font=\boxfont, inner ysep=1.5pt},
  planning-line-node/.append style   = {font=\boxfont, inner ysep=1.5pt},
  interaction-line-node/.append style= {font=\boxfont, inner ysep=1.5pt},
  agentic-line-node/.append style    = {font=\boxfont, inner ysep=1.5pt}
}

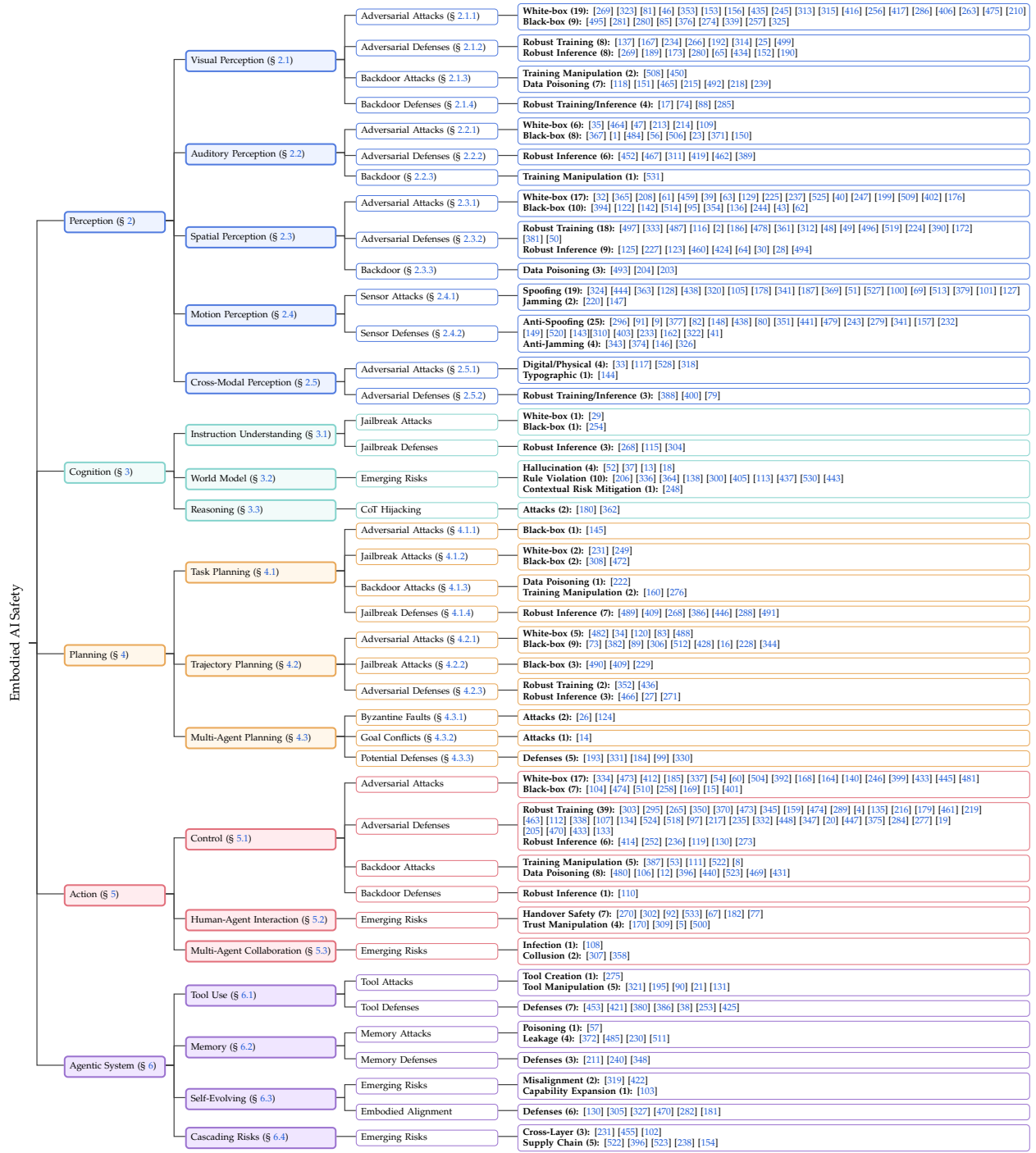
\begin{figure*}[!t]
    \centering
    \resizebox{\textwidth}{!}{%
        \begin{forest}
            forked edges,
            for tree={
                grow=east,
                reversed=true,
                anchor=mid west,
                parent anchor=mid east,
                child anchor=mid west,
                base=left,
                font=\tiny,
                rectangle,
                draw=none,
                sharp corners,
                align=left,
                minimum width=2em,
                edge+={darkgray, line width=0.6pt},
                s sep=2.5pt,
                inner xsep=1.8pt,
                inner ysep=2.5pt,
                line width=0.7pt,
                par/.style={rotate=90, child anchor=north, parent anchor=south, anchor=center},
            },
            where level=0{font=\footnotesize,}{},
            where level=1{text width=\secwidth,font=\tiny,}{},
            where level=2{text width=\subsecwidth,font=\tiny,}{},
            where level=3{text width=\subsubsecwidth,font=\tiny,}{},
            [
                {\EAS}, par
                [
                    Perception (\S~\ref{sec:perception}), perception-bg-node
                    [
                        Visual Perception (\S~\ref{subsec:visual-perception}), perception-bg-node
                        [
                            Adversarial Attacks (\S~\ref{subsubsec:visual-adversarial-attack}), perception-line-node
                            [
                               \textbf{White-box (19):}
                               \cite{melis2017deep} \cite{sitawarin2018darts} \cite{eykholt2018robust} \cite{chen2018shapeshifter} \cite{thys2019fooling} \cite{jia2020robust} \cite{jia2020fooling} \cite{xu2020adversarial} \cite{lovisotto2021slap}
                               \cite{sato2021dirty} \cite{schmalfuss2022pcfa} \cite{wu2023adversarial} \cite{ma2023wip} \cite{wu2024human} \cite{nokabadi2024trackpgd} \cite{wang2024cascaded} \cite{maag2024uncertainty} \cite{zhang2025anyattack} \cite{li2025multi}
                               \\
                               \textbf{Black-box (9):}
                               \cite{zhang2018camou} \cite{nassi2019mobilbye} \cite{nassi2020phantom}
                               \cite{fang2023pso} \cite{wang2024physical} \cite{mohajeransari2024discovering} \cite{sun2024embodied} \cite{ma2024controlloc} \cite{song2025pb}
                               , perception-line-node, text width=\leafwidth
                            ]
                        ]
                        [
                            Adversarial Defenses (\S~\ref{subsubsec:visual-adversarial-defense}), perception-line-node
                            [
                               \textbf{Robust Training (8):}
                               \cite{huang2020dsnet} \cite{kalin2021automating} \cite{liu2022image} \cite{mao2023understanding} \cite{li2023detection} \cite{schlarmann2024robust} \cite{blazevic2025securing} \cite{zhang2025rp}
                               \\
                               \textbf{Robust Inference (8):}
                               \cite{melis2017deep} \cite{li2017aod} \cite{kim2018robust} \cite{nassi2020phantom} \cite{chou2020sentinet} \cite{xu2021model} \cite{jia2024robust} \cite{li2025detecting}
                               , perception-line-node, text width=\leafwidth
                            ]
                        ]
                        [
                            Backdoor Attacks (\S~\ref{subsubsec:visual-backdoor-attack}), perception-line-node
                            [
                               \textbf{Training Manipulation (2):}
                               \cite{zheng2023trojvit} \cite{yang2024swarm}
                               \\
                               \textbf{Data Poisoning (7):}
                               \cite{han2022physical} \cite{jia2022badencoder} \cite{yuan2023badvitvit} \cite{liang2024badclip} \cite{zhang2024towards} \cite{liao2025towards} \cite{liu2025badvisionsslbackdoor}
                               , perception-line-node, text width=\leafwidth
                            ]
                        ]
                        [
                            Backdoor Defenses (\S~\ref{subsubsec:visual-backdoor-defense}), perception-line-node
                            [
                               \textbf{Robust Training/Inference (4):}
                               \cite{bansal2023cleanclip} \cite{doan2023defending} \cite{feng2023decree} \cite{niu2025bdetclip}
                               , perception-line-node, text width=\leafwidth
                            ]
                        ]
                    ]
                    [
                        Auditory Perception (\S~\ref{subsec:auditory-perception}), perception-bg-node
                        [
                            Adversarial Attacks (\S~\ref{subsubsec:auditory-adversarial-attack}), perception-line-node
                            [
                               \textbf{White-box (6):}
                               \cite{carlini2016hidden} \cite{yuan2018commandersong} \cite{chen2020metamorph} \cite{li2020practical} \cite{li2020advpulse} \cite{guo2022specpatch}
                               \\
                               \textbf{Black-box (8):}
                               \cite{vaidya2015cocaine} \cite{abdullah2019practical} \cite{zhang2019activated} \cite{chen2020devil}
                               \cite{zheng2021black} \cite{bilika2023hello} \cite{walker2023barrierbypass} \cite{ji2024watch}
                               , perception-line-node, text width=\leafwidth
                            ]
                        ]
                        [
                            Adversarial Defenses (\S~\ref{subsubsec:auditory-adversarial-defense}), perception-line-node
                            [
                               \textbf{Robust Inference (6):}
                               \cite{yang2018towards} \cite{zeng2019multiversion} \cite{samizade2020adversarial} \cite{wu2023defending} \cite{yu2023antifake} \cite{wang2020differences}
                               , perception-line-node, text width=\leafwidth
                            ]
                        ]
                        [
                            Backdoor (\S~\ref{subsubsec:auditory-backdoor}), perception-line-node
                            [
                               \textbf{Training Manipulation (1):}
                               \cite{zong2023trojanmodel}
                               , perception-line-node, text width=\leafwidth
                            ]
                        ]
                    ]
                    [
                        Spatial Perception (\S~\ref{subsec:spatial-perception}), perception-bg-node
                        [
                            Adversarial Attacks (\S~\ref{subsubsec:spatial-adversarial-attack}), perception-line-node
                            [
                               \textbf{White-box (17):}
                               \cite{cao2019adversarial} \cite{tu2020physically} \cite{li2021fooling} \cite{cheng2021universal} \cite{yoshida2022adversarial} \cite{chawla2022adversarial} \cite{cheng2022physical} \cite{horvath2023targeted}
                               \cite{liu2023slowlidar} \cite{liu2023transferable} \cite{zhu2024ae} \cite{chen2024adversary} \cite{lu2024poison} \cite{li2024adv3d} \cite{zheng2025new} \cite{wang2025imperceptible} \cite{kobayashi2025invisible}
                               \\
                               \textbf{Black-box (10):}
                               \cite{wang2021can} \cite{hau2021object} \cite{ikram2022perceptual} \cite{zhou2022doublestar}
                               \cite{fukunaga2024random} \cite{tian2024adversarial} \cite{huang2024spotattack} \cite{lou2024first} \cite{chen2025lidattack} \cite{cheng2025black}
                               , perception-line-node, text width=\leafwidth
                            ]
                        ]
                        [
                            Adversarial Defenses (\S~\ref{subsubsec:spatial-adversarial-defense}), perception-line-node
                            [
                               \textbf{Robust Training (18):}
                               \cite{zhang2019defense} \cite{sun2021adversarially} \cite{zhang2021acousticfusion} \cite{hahner2021fog} \cite{adamkiewicz2022vision} \cite{lehner20223d} \cite{zhang2022pointcutmix} \cite{tong2022enforcing} \cite{sang2023scene}
                               \cite{chen2024catnips} \cite{chen2024safer} \cite{zhang2024comprehensive} \cite{zhou2024control} \cite{liu2024beyond} \cite{wang2024mobile} \cite{kilic2025lidar}\\\cite{wang2025enhancing} \cite{chen2025splat}
                               \\
                               \textbf{Robust Inference (9):}
                               \cite{heinzler2020cnn} \cite{liu2021pointguard} \cite{hau2021shadow} \cite{you2021temporal} \cite{xiao2023exorcising} \cite{cho2023adopt}
                               \cite{cai2024diffusion} \cite{brunke2025semantically} \cite{zhang2025realtime}
                               , perception-line-node, text width=\leafwidth
                            ]
                        ]
                        [
                            Backdoor (\S~\ref{subsubsec:spatial-backdoor}), perception-line-node
                            [
                               \textbf{Data Poisoning (3):}
                               \cite{zhang2022towards} \cite{li2023badlidet} \cite{li2023towards}
                               , perception-line-node, text width=\leafwidth
                            ]
                        ]
                    ]
                    [
                        Motion Perception (\S~\ref{subsec:motion-perception}), perception-bg-node
                        [
                            Sensor Attacks (\S~\ref{subsubsec:motion-sensor-attack}), perception-line-node
                            [
                               \textbf{Spoofing (19):}
                               \cite{son2015rocking} \cite{yan2016can} \cite{trippel2017walnut} \cite{horton2018development} \cite{xu2018analyzing} \cite{shen2020drift} \cite{gluck2020spoofing} \cite{komissarov2021spoofing}
                               \cite{sun2021control} \cite{lenhart2021relay} \cite{vennam2023mmspoof} \cite{chen2023metawave} \cite{zhu2023tilemask} \cite{gao2023exploring} \cite{dasgupta2024unveiling} \cite{zhong2025analysis} \cite{wang2025practical} \cite{geng2025attacking} \cite{hong2022esp}
                               \\
                               \textbf{Jamming (2):}
                               \cite{lim2018autonomous} \cite{jang2023paralyzing}
                               , perception-line-node, text width=\leafwidth
                            ]
                        ]
                        [
                            Sensor Defenses (\S~\ref{subsubsec:motion-sensor-defense}), perception-line-node
                            [
                               \textbf{Anti-Spoofing (25):}
                               \cite{pozzobon2010anti} \cite{fernandez2016navigation} \cite{anderson2017chips} \cite{wang2017gnss} \cite{falco2018dual} \cite{jansen2018crowd} \cite{xu2018analyzing} \cite{eldosouky2019drones} \cite{tharayil2020sensor}
                               \cite{xue2020deepsim} \cite{zhang2020vanet} \cite{lou2021soundfence} \cite{nallabolu2021frequency} \cite{sun2021control} \cite{jiang2022deeppose} \cite{liu2022traceability}\\
                               \cite{jeong2023rocking} \cite{zhou2023anti} \cite{iqbal2024deep}\cite{sahu2024acoustic} \cite{wang2024vimu} \cite{liu2025gnss} \cite{jin2025spoofing} \cite{singh2025securetrack} \cite{chen2025safety}
                               \\
                               \textbf{Anti-Jamming (4):}
                               \cite{swinney2021gnss} \cite{wang2021machine} \cite{islam2022combating} \cite{spanghero2025gnss}
                               , perception-line-node, text width=\leafwidth
                            ]
                        ]
                    ]
                    [
                        Cross-Modal Perception (\S~\ref{subsec:cross-modal-perception}), perception-bg-node
                        [
                            Adversarial Attacks (\S~\ref{subsubsec:cross-modal-adversarial-attack}), perception-line-node
                            [
                               \textbf{Digital/Physical (4):}
                               \cite{cao_invisible_2021} \cite{hallyburton_frustum_2022} \cite{li_malicious_2024} \cite{hou_dejavu_2025}
                               \\
                               \textbf{Typographic (1):}
                               \cite{iranmanesh2026typographic}
                               , perception-line-node, text width=\leafwidth
                            ]
                        ]
                        [
                            Adversarial Defenses (\S~\ref{subsubsec:cross-modal-adversarial-defense}), perception-line-node
                            [
                               \textbf{Robust Training/Inference (3):}
                               \cite{wang_adversarial_fusion_2022} \cite{wang_mmcert_2024} \cite{el2026robust}
                               , perception-line-node, text width=\leafwidth
                            ]
                        ]
                    ]
                ]
                [
                    Cognition (\S~\ref{sec:cognition}), cognition-bg-node
                    [
                        Instruction Understanding (\S~\ref{subsec:instruction-understanding}), cognition-bg-node
                        [
                            Jailbreak Attacks, cognition-line-node
                            [
                               \textbf{White-box (1):}
                               \cite{burbano2025chai}
                               \\
                               \textbf{Black-box (1):}
                               \cite{bai2025badnaver}
                               , cognition-line-node, text width=\leafwidth
                            ]
                        ]
                        [
                            Jailbreak Defenses, cognition-line-node
                            [
                               \textbf{Robust Inference (3):}
                               \cite{marchiori2025preventing} \cite{abuduweili2025safe} \cite{ravichandran2026safety}
                               , cognition-line-node, text width=\leafwidth
                            ]
                        ]
                    ]
                    [
                        World Model (\S~\ref{subsec:world-model}), cognition-bg-node
                        [
                            Emerging Risks,cognition-line-node
                            [
                               \textbf{Hallucination (4):}
                               \cite{chen2024multiobject} \cite{chakraborty_heal_2025} \cite{bae2025mash} \cite{baraldi2025safety}
                               \\
                               \textbf{Rule Violation (10):}
                               \cite{li2025worldmodelsurvey} \cite{sun2024hrssm} \cite{wen2025scalable} \cite{huang2024safedreamer} \cite{Qu2025VLMSAFEVM} \cite{wang2024drivewm} \cite{guo2026physcond} \cite{xu2026ctrlattack} \cite{zollicoffer2025surprise} \cite{yan2026safedream}
                               \\
                               \textbf{Contextual Risk Mitigation (1):}
                               \cite{lu2026homeguard}
                               , cognition-line-node, text width=\leafwidth
                            ]
                        ]
                    ]
                    [
                        Reasoning (\S~\ref{subsec:reasoning}), cognition-bg-node
                        [
                            CoT Hijacking, cognition-line-node
                            [
                               \textbf{Attacks (2):}
                               \cite{kuo2025hcot} \cite{trinh2026altered}
                               , cognition-line-node, text width=\leafwidth
                            ]
                        ]
                    ]
                ]
                [
                    Planning (\S~\ref{sec:planning}), planning-bg-node
                    [
                        Task Planning (\S~\ref{subsec:Task_Planning}), planning-bg-node
                        [
                            Adversarial Attacks (\S~\ref{subsubsec:Task_Adversarial_Attacks}), planning-line-node
                            [
                               \textbf{Black-box (1):}
                               \cite{islam2024malicious}
                               , planning-line-node, text width=\leafwidth
                            ]
                        ]
                        [
                            Jailbreak Attacks (\S~\ref{subsubsec:Jailbreak_Task_Planners}), planning-line-node
                            [
                               \textbf{White-box (2):}
                               \cite{liu2024exploring} \cite{lu2024poex}
                               \\
                               \textbf{Black-box (2):}
                               \cite{robey2024jailbreaking} \cite{zhang2024badrobot}
                               , planning-line-node, text width=\leafwidth
                            ]
                        ]
                        [
                            Backdoor Attacks (\S~\ref{subsubsec:Backdoor_Task_Planners}), planning-line-node
                            [
                               \textbf{Data Poisoning (1):}
                               \cite{liu2024compromising}
                               \\
                               \textbf{Training Manipulation (2):}
                               \cite{jiao2024can} \cite{nahian2025robotroj}
                               , planning-line-node, text width=\leafwidth
                            ]
                        ]
                        [
                            Jailbreak Defenses (\S~\ref{subsubsec:Task_Planning_Defenses}), planning-line-node
                            [
                               \textbf{Robust Inference (7):}
                               \cite{zhang2024safeembodai} \cite{wen2024secure} \cite{marchiori2025preventing} \cite{wang2025robosafe} \cite{yang2025cee} \cite{obi2025safeplan} \cite{zhang2025enhancing}
                               , planning-line-node, text width=\leafwidth
                            ]
                        ]
                    ]
                    [
                        Trajectory Planning (\S~\ref{subsec:Motion_Planning}), planning-bg-node
                        [
                            Adversarial Attacks (\S~\ref{subsubsec:Motion_Adversarial_Attacks}), planning-line-node
                            [
                               \textbf{White-box (5):}
                               \cite{zhang2022adversarial} \cite{cao2022advdo} \cite{hanselmann2022king} \cite{fan2024adversarial} \cite{zhang2024visual}
                               \\
                               \textbf{Black-box (9):}
                               \cite{ding2020learning} \cite{wang2021advsim} \cite{feng2021intelligent} \cite{rempe2022generating} \cite{zheng2023robustness} \cite{xie2024advdiffuser} \cite{bai2025universal} \cite{liu2025avatar} \cite{szvoren2025adversarial}
                               , planning-line-node, text width=\leafwidth
                            ]
                        ]
                        [
                            Jailbreak Attacks (\S~\ref{subsubsec:Jailbreak_Navigation_Planners}), planning-line-node
                            [
                               \textbf{Black-box (3):}
                               \cite{zhang2024study} \cite{wen2024secure} \cite{liu2026pina}
                               , planning-line-node, text width=\leafwidth
                            ]
                        ]
                        [
                            Adversarial Defenses (\S~\ref{subsubsec:Motion_Planning_Defenses}), planning-line-node
                            [
                               \textbf{Robust Training (2):}
                               \cite{thumm2023reducing} \cite{xu2024robust}
                               \\
                               \textbf{Robust Inference (3):}
                               \cite{yurtsever2019risky} \cite{brudigam2021stochastic} \cite{meng2023vehicle}
                               , planning-line-node, text width=\leafwidth
                            ]
                        ]
                    ]
                    [
                        Multi-Agent Planning (\S~\ref{subsec:Multi_Agent_Planning}), planning-bg-node
                        [
                            Byzantine Faults (\S~\ref{subsubsec:Byzantine_Faults}), planning-line-node
                            [
                               \textbf{Attacks (2):}
                               \cite{blumenkamp_emergence_2021} \cite{he_redteaming_2025}
                               , planning-line-node, text width=\leafwidth
                            ]
                        ]
                        [
                            Goal Conflicts (\S~\ref{subsubsec:Goal_Conflicts}), planning-line-node
                            [
                               \textbf{Attacks (1):}
                               \cite{gao_multirobot_2025}
                               , planning-line-node, text width=\leafwidth
                            ]
                        ]
                        [
                            Potential Defenses (\S~\ref{subsubsec:Multi_Agent_Planning_Defenses}), planning-line-node
                            [
                               \textbf{Defenses (5):}
                               \cite{li_resilient_2020} \cite{strobel_swarms_2023} \cite{lee_distributed_2025} \cite{gandhi_roborebound_2025} \cite{strobel_blockchain_2020}
                               , planning-line-node, text width=\leafwidth
                            ]
                        ]
                    ]
                ]
                [
                    Action (\S~\ref{sec:action}), interaction-bg-node
                    [
                        Control (\S~\ref{subsec:action-control}), interaction-bg-node
                        [
                            Adversarial Attacks, interaction-line-node
                            [
                               \textbf{White-box (17):}
                               \cite{Sun_StealthyEfficientAdversarial_2020} \cite{Zhang_RobustDeepReinforcement_2020} \cite{Weng_EvaluatingRobustnessDeep_2020} \cite{Lee_SpatiotemporallyConstrainedAction_2020} \cite{sun2022strongest} \cite{Chen_DiffusionPolicyAttacker_2024} \cite{Cheng_ManipulationFacingThreats_2024} \cite{Zhao_RethinkingIntermediateFeatures_2024}
                               \cite{Wang_ExploringAdversarialVulnerabilities_2025} \cite{Kalra_HowVulnerableMy_2025} \cite{jones2025adversarial} \cite{huang2025annie} \cite{lu2025robots} \cite{wang2025freezevla} \cite{xu2025model} \cite{yan2025alignment} \cite{zhang2025attention}
                               \\
                               \textbf{Black-box (7):}
                               \cite{Gleave_AdversarialPoliciesAttacking_2020} \cite{Zhang_RobustReinforcementLearning_2021} \cite{Zheng_EvaluatingRobustnessReinforcement_2024} \cite{Ma_SUBPLAYAdversarialPolicies_2024} \cite{karnik_ert_2024} \cite{Bai_RATAdversarialAttacks_2025} \cite{wang2025advedm}
                               , interaction-line-node, text width=\leafwidth
                            ]
                        ]
                        [
                            Adversarial Defenses, interaction-line-node
                            [
                               \textbf{Robust Training (39):}
                               \cite{Rajeswaran_EPOptLearningRobust_2017} \cite{Pinto_RobustAdversarialReinforcement_2017} \cite{Mandlekar_AdversariallyRobustPolicy_2017} \cite{Tessler_ActionRobustReinforcement_2019} \cite{Vinitsky_RobustReinforcementLearning_2020} \cite{Zhang_RobustDeepReinforcement_2020} \cite{Tan_RobustifyingReinforcementLearning_2020} \cite{Jiang_MonotonicRobustPolicy_2021} \cite{Zhang_RobustReinforcementLearning_2021} \cite{Oikarinen_RobustDeepReinforcement_2021} \cite{Ajay_DistributionallyAdaptiveMeta_2022} \cite{Huang_RobustReinforcementLearning_2022} \cite{Liang_EfficientAdversarialTraining_2022}
                               \cite{Kuang_LearningRobustPolicy_2022} \cite{You_UserOrientedRobustReinforcement_2023} \cite{lien2023revisiting} \\\cite{Yuan_RobustMultiAgentCoordination_2023} \cite{guo2023patrol} \cite{sun2023certifiably} \cite{Greenberg_TrainHardFight_2023} \cite{Huang_TradeOffRobustnessRewards_2023} \cite{Zhou_RobustMeanFieldActorCritic_2023} \cite{Zhou_NaturalActorCriticRobust_2023} \cite{Gadot_BringYourOwn_2024} \cite{Liang_GameTheoreticRobustReinforcement_2024a} \cite{Liu_RethinkingAdversarialPolicies_2024}
                               \cite{sun2024breaking} \cite{Yang_RobustOfflineReinforcement_2024} \cite{Tang_AdversariallyRobustDecision_2024} \cite{Belaire_RegretBasedDefenseAdversarial_2024} \cite{Yang_UncertaintyBasedOfflineVariational_2024} \cite{Wang_RobustDeepReinforcement_2025} \cite{Nie_ActionRobustReinforcement_2025} \cite{Nakanishi_OffPolicyActorCriticAdversarial_2025} \cite{Belaire_MinimizingAdversarialCounterfactual_2025} \\\cite{Li_RobustMultiAgentReinforcement_2025} \cite{zhang2025safevla} \cite{xu2025model} \cite{hu2026dream}
                               \\
                               \textbf{Robust Inference (6):}
                               \cite{Wu_CROPCertifyingRobust_2022} \cite{Luu_MitigatingAdversarialPerturbations_2024} \cite{Liu_WorstCaseAttacksRobust_2024} \cite{hancockrun} \cite{hu2025vlsa} \cite{Mishra_AERMANIVLMStructuredPrompting_2025}
                               , interaction-line-node, text width=\leafwidth
                            ]
                        ]
                        [
                            Backdoor Attacks, interaction-line-node
                            [
                               \textbf{Training Manipulation (5):}
                               \cite{Wang_BackdooRLBackdoorAttack_2021} \cite{Chen_MARNetBackdoorAttacks_2023} \cite{Guo_PNActCraftingBackdoor_2025} \cite{zhou2025badvla} \cite{an2026flowhijack}
                               \\
                               \textbf{Data Poisoning (8):}
                               \cite{Li_TooBadRLTriggerOptimization_2025} \cite{gong2024baffle} \cite{Ashcraft_BackdoorsDRLFour_2025} \cite{wang2024trojanrobot} \cite{xu2025tabvla} \cite{zhou_goal_backdoor_2025} \cite{zhan2026beat} \cite{xu2026silentdrift}
                               , interaction-line-node, text width=\leafwidth
                            ]
                        ]
                        [
                            Backdoor Defenses, interaction-line-node
                            [
                               \textbf{Robust Inference (1):}
                               \cite{Guo_PolicyCleanseBackdoorDetection_2023}
                               , interaction-line-node, text width=\leafwidth
                            ]
                        ]
                    ]
                    [
                        Human-Agent Interaction (\S~\ref{subsec:hai}), interaction-bg-node
                        [
                            Emerging Risks,interaction-line-node
                            [
                               \textbf{Handover Safety (7):}
                               \cite{handover_mobile_2024} \cite{trust_handover_biomimetic_2025} \cite{blind_handover_2024} \cite{intention_fuzzy_wearable_2023} \cite{handover_haptic_cues_2021} \cite{handover_adaptive_transport_2023} \cite{handover_survey_2024}
                               \\
                               \textbf{Trust Manipulation (4):}
                               \cite{trust_review_hri_2021} \cite{perceived_safety_survey_2021} \cite{perceived_safety_taxonomy_2023} \cite{zhang2024psysafe}
                               , interaction-line-node, text width=\leafwidth
                            ]
                        ]
                    ]
                    [
                        Multi-Agent Collaboration (\S~\ref{subsec:mac}), interaction-bg-node
                        [
                            Emerging Risks,interaction-line-node
                            [
                               \textbf{Infection (1):}
                               \cite{gu2024agentsmith}
                               \\
                               \textbf{Collusion (2):}
                               \cite{ren2025collusion} \cite{tomasev_distributional_agi_2025}
                               , interaction-line-node, text width=\leafwidth
                            ]
                        ]
                    ]
                ]
                [
                    Agentic System (\S~\ref{sec:agentic}), agentic-bg-node
                    [
                        Tool Use (\S~\ref{subsec:tool-use}), agentic-bg-node
                        [
                            Tool Attacks, agentic-line-node
                            [
                               \textbf{Tool Creation (1):}
                               \cite{yu_robocodex_2024}
                               \\
                               \textbf{Tool Manipulation (5):}
                               \cite{wang_toolhijacker_2025} \cite{li2025stac} \cite{feng2026backdooragent} \cite{belkhiter2026fha} \cite{hu2026maltool}
                               , agentic-line-node, text width=\leafwidth
                            ]
                        ]
                        [
                            Tool Defenses, agentic-line-node
                            [
                               \textbf{Defenses (7):}
                               \cite{brown_safetychip_2024} \cite{ahn_selp_2024} \cite{chen_agentspec_2026} \cite{wang2025robosafe} \cite{chang2026trustboundary} \cite{lv_skill_audit_2026} \cite{xiao_routeguard_2026}
                               , agentic-line-node, text width=\leafwidth
                            ]
                        ]
                    ]
                    [
                        Memory (\S~\ref{subsec:memory}), agentic-bg-node
                        [
                            Memory Attacks, agentic-line-node
                            [
                               \textbf{Poisoning (1):}
                               \cite{chen2024agentpoison}
                               \\
                               \textbf{Leakage (4):}
                               \cite{wang_mextra_2025} \cite{zhang_memoanalyzer_2024} \cite{liu_mama_2025} \cite{zheng_justask_2026}
                               , agentic-line-node, text width=\leafwidth
                            ]
                        ]
                        [
                            Memory Defenses, agentic-line-node
                            [
                               \textbf{Defenses (3):}
                               \cite{hu_memos_2025} \cite{liu2026safeharbor} \cite{tapwal2026prism}
                               , agentic-line-node, text width=\leafwidth
                            ]
                        ]
                    ]
                    [
                        Self-Evolving (\S~\ref{subsec:self-evolving}), agentic-bg-node
                        [
                            Emerging Risks,agentic-line-node
                            [
                               \textbf{Misalignment (2):}
                               \cite{shao_misevolving_2025} \cite{wu2026isc}
                               \\
                               \textbf{Capability Expansion (1):}
                               \cite{selfimproving_efm_2025}
                               , agentic-line-node, text width=\leafwidth
                            ]
                        ]
                        [
                            Embodied Alignment, agentic-line-node
                            [
                               \textbf{Defenses (6):}
                               \cite{hu2025vlsa} \cite{moralanchor_2025} \cite{srikanth2026red} \cite{zhang2025safevla} \cite{nay_law_alignment_2025} \cite{wu_c3ai_2025}
                               , agentic-line-node, text width=\leafwidth
                            ]
                        ]
                    ]
                    [
                        Cascading Risks (\S~\ref{subsec:cascading}), agentic-bg-node
                        [
                            Emerging Risks,agentic-line-node
                            [
                               \textbf{Cross-Layer (3):}
                               \cite{liu2024exploring} \cite{yeke_raven_2025} \cite{geng2026white}
                               \\
                               \textbf{Supply Chain (5):}
                               \cite{zhou2025badvla} \cite{wang2024trojanrobot} \cite{zhou_goal_backdoor_2025} \cite{liu_agent_skills_2026} \cite{jia_skillject_2026}
                               , agentic-line-node, text width=\leafwidth
                            ]
                        ]
                    ]
                ]
            ]
    \end{forest}
    }%
    \caption{The roadmap of this survey.}
    \label{fig:roadmap}
\end{figure*}

Recent years have witnessed rapid advances across the embodied AI pipeline \cite{black2410pi0,kim2025openvla,qu2025spatialvla,wen2025diffusionvla,pertsch2025fast,wen2025tinyvla}. Improvements in perception (e.g., vision, LiDAR, multimodal sensing), cognition (e.g., world modeling, value alignment), planning (e.g., task planning, trajectory optimization), and interaction (e.g., safe control, human–robot collaboration) have expanded the capabilities of embodied agents. However, these capabilities also broaden and complicate the attack surface \cite{li2025attackvla,zhou2025badvla,Wang_ExploringAdversarialVulnerabilities_2025}. Safety challenges that once appeared primarily in digital domains, such as adversarial examples in vision or jailbreak prompts in language models, carry far more severe consequences in physical environments. For instance, small perturbations to a visual sensor may cause an autonomous vehicle to misinterpret a stop sign \cite{eykholt2018robust}, while maliciously poisoned training data may compromise task planning and produce unsafe trajectories \cite{li2025attackvla}. Misaligned or unpredictable human–agent interactions can further generate behaviors that endanger users directly.

Despite their growing importance, the safety challenges unique to embodied AI remain underexamined. Existing surveys in AI safety largely focus on digital-only systems such as vision foundation models \cite{wang2025comprehensive,ma2025safety,ma_safety_report_2026}, large language models (LLMs), multimodal large language models (MLLMs) \cite{ye2025survey}, digital agents \cite{gan2024navigating,deng2025ai}, or Vision-Language-Action models (VLAs) \cite{li2026vlasafety}. While these works offer valuable taxonomies of attacks and defenses, they rarely address embodied settings where perception, cognition, planning, and interaction are tightly coupled and must operate under real-world constraints. A comprehensive treatment of embodied AI safety therefore requires not only synthesizing research within each component but also integrating insights from broader AI safety domains that have direct implications for embodied systems.

\noindent\textbf{Capability--Risk Duality.}
A key organizing principle of this survey is the \emph{capability--risk duality}: each layer of the embodied pipeline represents not merely a functional component but a \emph{capability expansion} that introduces corresponding new vulnerabilities, as illustrated in Figure~\ref{fig:capability_risk}.
Real-world embodied systems vary in the depth of this capability stack.
At the innermost layer, sensor-only devices such as face-recognition access controls represent the simplest embodied systems, where adversaries can only target perceptual inputs.
Adding cognition yields agents like museum guide robots capable of dialogue and scene understanding, opening attack surfaces in reasoning and language comprehension.
Incorporating planning enables navigation and decision-making, as in autonomous vehicles, where adversaries can additionally manipulate route planning and trajectory prediction.
At the action and interaction layer, robotic arms and humanoid robots gain the ability to physically manipulate their environment, exposing control and human--robot interaction to exploitation.
Finally, agentic systems augment all prior capabilities with persistent memory, tool use, and self-evolution, creating the broadest attack surface where compromises at any inner layer can cascade outward.
This duality, \emph{deeper capability entails broader risk}, motivates our layered taxonomy and structures the remainder of this survey: each section addresses both the attacks specific to its capability layer and the pathways through which inner-layer vulnerabilities propagate to outer-layer failures.

To address this need, we conduct a systematic survey of \textbf{safety research in embodied AI}. We propose a multi-level taxonomy that organizes vulnerabilities and defenses across five key components of embodied AI systems: \textbf{perception}, \textbf{cognition}, \textbf{planning}, \textbf{action \& interaction}, and \textbf{agentic systems}. For each component, we categorize attacks and defenses, including adversarial perception, unsafe reasoning, planning under perturbations, unsafe control and interaction, and agentic-level risks such as tool misuse, memory poisoning, and cascading failures, as illustrated in Figure \ref{fig:illustration}. Crucially, we extend our synthesis beyond embodied-specific works to incorporate over \textbf{\numpapers{}} papers from traditional AI safety (vision, language, multimodal foundation models), selecting those with clear embodied relevance. Figure \ref{fig:overview} presents an overview of different attack and defense techniques and their distribution across the pipeline components. This dual perspective situates embodied AI safety within the broader AI safety ecosystem while highlighting the unique risks that emerge when intelligence is deployed in the physical world.

Based on the current literature, we identify and summarize the threats posed by various attacks, as shown in Table \ref{tab:threats}. In \textbf{perception}, \textit{adversarial attacks} introduce subtle perturbations in sensory inputs, such as visual or auditory data, causing misclassifications and leading to incorrect environmental interpretations. \textbf{Backdoor attacks} embed hidden triggers in the model that activate malicious behavior when prompted, while \textbf{sensor attacks}, such as spoofing and jamming, compromise sensor data, resulting in environmental sensing failures or system shutdowns. These vulnerabilities can lead to misinterpretation of objects, failure to detect obstacles, or navigation errors. In \textbf{cognition}, \textit{adversarial attacks} manipulate reasoning processes, causing the system to make unsafe or incorrect decisions, such as faulty spatial understanding or misinterpretation of context. In \textbf{planning}, various attacks, including \textit{adversarial attacks} on task planning and trajectory planning, \textbf{jailbreak attacks}, and \textbf{backdoor attacks}, can manipulate the model’s planned actions, leading to unsafe trajectories, collisions, or failure to follow intended goals. In \textbf{action \& interaction}, \textit{adversarial manipulations} and \textbf{backdoor attacks} can bypass safety mechanisms during human-agent interactions, inducing harmful or unintended behavior, such as violating safety protocols or performing actions that harm users. Finally, in \textbf{agentic systems}, threats arise from the agent's expanded autonomy: \textit{tool misuse} can lead to harmful code execution or unintended physical actions, \textbf{memory poisoning} can corrupt the agent's experience store to cause persistent unsafe behavior, \textbf{memory leakage} can expose private user data and privileged context from agent memory stores, and \textbf{cascading failures} can propagate through inner layers when self-evolving agents erode their own alignment. In Table \ref{tab:threats}, we also categorize the potential real-world dangers caused by these threats.

\begin{table*}[t]
\centering
\caption{Summary of attacks and threats across capability layers of embodied AI.}
\label{tab:threats}
\resizebox{1\textwidth}{!}{
\begin{tabular}{%
    @{}
    p{0.1\textwidth}
    p{0.18\textwidth}
    p{0.32\textwidth}
    p{0.4\textwidth}
    @{}
}
\toprule
\textbf{Capability Layer} & \textbf{Attack} & \textbf{Threat} & \textbf{Real-world Danger} \\
\midrule

\multirow{3}{*}{\centering Perception}
& Adversarial Attack
& Misclassification, misdetection, scene misinterpretation
& Wrong object recognition, traffic sign errors, surveillance failure \\

& Backdoor Attack
& Triggered misperception, hidden model manipulation
& Unsafe behavior activation, safety bypass during deployment \\

& Sensor Attack
& Sensor spoofing, jamming, data corruption
& Navigation failure, loss of situational awareness, system malfunction \\
\midrule

\multirow{1}{*}{\centering Cognition}
& Adversarial Attack
& Faulty reasoning, scene misunderstanding, context errors
& Navigation errors, hazard avoidance failure, unsafe decisions \\
\midrule

\multirow{3}{*}{\centering Planning}
& Adversarial Attack
& Planning perturbation, trajectory errors
& Collision risk, unstable motion, unsafe task execution \\

& Jailbreak Attack
& Safety constraint bypass, unsafe goal generation
& Execution of prohibited actions, violation of safety rules \\

& Backdoor Attack
& Triggered policy manipulation, hidden planning bias
& Malicious plans, safety mechanism bypass \\
\midrule

\multirow{2}{*}{\makecell{Action \&\\ Interaction}}
& Adversarial Attack
& Action manipulation, safety guard evasion
& Unsafe human–robot interaction, physical injury risk \\

& Backdoor Attack
& Triggered harmful actions, hidden interaction flaws
& Malicious responses, loss of control, safety violation \\
\midrule

\multirow{4}{*}{\makecell{Agentic \\ System}}
& Tool / Skill Misuse
& Unsafe tool calls, harmful code execution
& Physical damage, unsafe API or actuator commands, skill-driven misbehavior \\

& Memory Poisoning
& Corrupted memory, unsafe policy update
& Repeated unsafe behavior, long-term reliability loss \\

& Memory Leakage
& Sensitive memory exposure, data extraction
& Privacy breach, leakage of logs or user data \\

& Cascading Failure
& Cross-layer error propagation, alignment drift
& System-wide failure, uncontrolled self-evolution \\
\bottomrule
\end{tabular}}
\end{table*}

\noindent\textbf{Differences from Existing Surveys.}
Prior surveys examine embodied AI safety from complementary perspectives. \cite{xing2026towards} provides an early analysis of vulnerabilities and attack surfaces but offers limited coverage of defensive strategies. \cite{wang2025safety} focuses on robustness issues in navigation but does not extend to cognition, manipulation, or human–robot interaction. \cite{tan2025towards} presents conceptual foundations and system-level safety principles but does not develop detailed attack–defense taxonomies. \cite{baraldi2025safety} analyzes world-model safety, particularly predictive failures, while leaving perception, planning, and interaction risks less explored. \cite{ma_breaks_2026} argues that embodied failures arise from system-level mismatches rather than isolated LLM or CPS flaws, but does not develop component-level attack--defense taxonomies. \cite{wang_closedloop_2026} examines adversarial robustness from a closed-loop propagation perspective, but focuses on adversarial attacks without covering backdoor, jailbreak, or agentic-level threats. \cite{huang_trust_llm_robotics_2026} surveys security threats and defenses for Embodied LLMs, but scopes narrowly to LLM integration without addressing broader perception or interaction layers. \cite{perlo_embodied_risks_2025} provides a policy-oriented risk taxonomy spanning physical, informational, and social dimensions, but does not analyze specific attack mechanisms or defenses. \cite{wang_navigating_2026} surveys embodied AI from an IoT perspective with a dual-brain architecture framework, but treats security and privacy as one component among enabling technologies rather than developing attack--defense taxonomies. In contrast, our survey synthesizes attacks and defenses across the entire embodied pipeline and integrates insights from traditional AI safety to provide a unified, mechanism-oriented understanding of embodied AI safety. Figure \ref{fig:roadmap} provides a roadmap of this survey.

In summary, the main contributions of this work are:
\begin{itemize}
\item We present a systematic survey of \textbf{safety research in embodied AI}, organizing attacks and defenses into a coherent multi-level taxonomy spanning perception, cognition, planning, action \& interaction, and agentic systems.
\item We review over \textbf{\numpapers{}} papers, consolidating embodied-specific research with safety-relevant advances in vision, language, and multimodal foundation models.
\item We identify fundamental challenges, open problems, and future research directions, offering a roadmap for developing embodied agents that are not only capable and autonomous but also safe and trustworthy in real-world environments.
\end{itemize}


\section{Perception}
\label{sec:perception}

Perception forms the innermost layer of embodied AI, granting agents the foundational capability to interpret their environment through multiple sensing modalities. At this layer, the attack surface originates at the sensory boundary: adversaries can corrupt what the agent perceives through adversarial perturbations, sensor spoofing, and backdoor triggers. Because perception underpins all outer layers, errors at this stage propagate and amplify throughout the system: a misclassified object leads to flawed reasoning, unsafe plans, and dangerous actions. This section organizes perception by sensing modality: \textbf{Visual Perception} (Section~\ref{subsec:visual-perception}) addresses vulnerabilities in camera-based perception, with emphasis on modern visual encoders (e.g., CLIP, ViT, and SigLIP) used in vision-language models; \textbf{Auditory Perception} (Section~\ref{subsec:auditory-perception}) covers attacks on speech recognition and speaker verification systems critical for voice-based human-agent interaction; \textbf{Spatial Perception} (Section~\ref{subsec:spatial-perception}) examines threats to 3D understanding including SLAM, depth estimation, pose estimation, and neural scene representations (e.g., NeRF and 3DGS); \textbf{Motion Perception} (Section~\ref{subsec:motion-perception}) addresses vulnerabilities in inertial measurement, GPS, and proprioceptive sensing; and \textbf{Cross-Modal Perception} (Section~\ref{subsec:cross-modal-perception}) discusses attacks on multimodal perception systems and sensor fusion. For each modality, we review attacks, defenses, and evaluation benchmarks. Sensor-level attacks (spoofing and jamming) are integrated within each modality rather than treated separately.


\subsection{Visual Perception}
\label{subsec:visual-perception}

\begin{table}[!tp]
\center
\renewcommand{\arraystretch}{1.15}
\caption{A summary of \textbf{adversarial attacks} for \textbf{visual perception}.}
\resizebox{1\textwidth}{!}{
\rowcolors{2}{tablegray}{white}
{\setlength{\tabcolsep}{0pt}
\begin{tabular}{@{} >{\cellcolor{white}}p{.12\textwidth} p{.16\textwidth} p{.07\textwidth}
               p{.13\textwidth} p{.17\textwidth} p{.20\textwidth} p{.28\textwidth}@{}}

\toprule
\rowcolor{perception-table-blue}
\cellcolor{perception-table-blue}\textbf{Attack} & \textbf{Method} & \textbf{Year} & \textbf{Category} & \textbf{Subcategory} & \textbf{Target Model} & \textbf{Dataset} \\

\midrule

& Melis et al.\cite{melis2017deep} & 2017 & White-box & Digital Attack & Object Classifier & iCubWorld \\

& Thys et al.\cite{thys2019fooling} & 2019 & White-box & Digital Attack & Object Detector & Inria \\
& Adversarial Overlay\cite{wu2023adversarial} & 2023 & White-box & Digital Attack & Object Detector & PASCAL VOC, ROS Gazebo \\
& HitM\cite{wu2024human} & 2024 & White-box & Digital Attack & Object Detector & CARLA, VOC \\

& RTAA\cite{jia2020robust} & 2020 & White-box & Digital Attack & Single-Object Tracker & OTB, UAV, VOT \\
& TrackPGD\cite{nokabadi2024trackpgd} & 2024 & White-box & Digital Attack & Single-Object Tracker & DAVIS, GOT-10k, UAV, VOT \\

& Tracker Hijacking\cite{jia2020fooling} & 2020 & White-box & Digital Attack & Multi-Object Tracker & BDD \\
& Ma et al.\cite{ma2023wip} & 2023 & White-box & Digital Attack & Multi-Object Tracker & BDD \\

& CAA\cite{wang2024cascaded} & 2024 & White-box & Digital Attack & Semantic Segmentation & Cityscapes, RainCityscapes \\
& Uncertainty\cite{maag2024uncertainty} & 2024 & White-box & Digital Attack & Semantic Segmentation & Cityscapes, VOC \\

& AnyAttack\cite{zhang2025anyattack} & 2025 & White-box & Digital Attack & CLIP Image Encoder & LAION-400M \\

& PCFA\cite{schmalfuss2022pcfa} & 2022 & White-box & Digital Attack & Optical Flow Estimator & Sintel, KITTI \\

& DARTS\cite{sitawarin2018darts} & 2018 & White-box & Physical Attack & Object Classifier & GTSDB, GTSRB \\
& RP2\cite{eykholt2018robust} & 2018 & White-box & Physical Attack & Object Classifier & GTSRB, LISA \\

& ShapeShifter\cite{chen2018shapeshifter} & 2018 & White-box & Physical Attack & Object Detector & Real-world data \\
& AdvT\cite{xu2020adversarial} & 2020 & White-box & Physical Attack & Object Detector & Real-world data \\
& SLAP\cite{lovisotto2021slap} & 2021 & White-box & Physical Attack & Object Detector & Real-world data \\
& DRP\cite{sato2021dirty} & 2021 & White-box & Physical Attack & Object Detector & Comma2k19, LGSVL \\

& MFDA\cite{li2025multi} & 2025 & White-box & Physical Attack & Single-Object Tracker & CARLA \\

& AdvTraj\cite{wang2024physical} & 2024 & Black-box & Digital Attack & Multi-Object Tracker & CARLA \\

& Fang et al.\cite{fang2023pso} & 2023 & Black-box & Digital Attack & Object Detector & Carlasc, Comma2k19, CULane \\

& PB-UAP\cite{song2025pb} & 2025 & Black-box & Digital Attack & Semantic Segmentation & VOC, Cityscapes \\

& ELA\cite{sun2024embodied} & 2024 & Black-box & Physical Attack & Object Classifier & CARLA \\

& CAMOU\cite{zhang2018camou} & 2018 & Black-box & Physical Attack & Object Detector & Unreal Engine \\
& MobilBye\cite{nassi2019mobilbye} & 2019 & Black-box & Physical Attack & Object Detector & Real-world data \\
& SSPA\cite{nassi2020phantom} & 2020 & Black-box & Physical Attack & Object Detector & Web data \\
& NS Attack\cite{mohajeransari2024discovering} & 2024 & Black-box & Physical Attack & Object Detector & CARLA \\


& ControlLoc\cite{ma2024controlloc} & 2024 & Black-box & Physical Attack & Multi-Object Tracker & BDD, KITTI \\

\multirow{-29}{*}{\parbox[c]{\linewidth}{\raggedright Adversarial\\Attack}} & Zhu et al.\cite{zhu2026physical} & 2026 & White-box & Physical Attack & Object Detector & LLVIP, FLIR \\

\bottomrule
\end{tabular}
}}
\label{tab:vision_adversarial_attack}
\end{table}

Visual perception encompasses camera-based tasks such as object classification, object detection (including lane detection), object tracking, semantic segmentation, and video understanding (action recognition, optical flow, and video object segmentation), each critical for embodied downstream tasks. Modern visual encoders, including contrastive vision-language models (e.g., CLIP and SigLIP) and Vision Transformers (ViT), serve as shared perception backbones whose vulnerabilities propagate to all downstream systems. This subsection consolidates all visual perception security research: adversarial attacks and defenses that manipulate or protect pixel-level inputs, as well as backdoor attacks and defenses that embed or remove hidden triggers in visual models. We organize the discussion into four parts: \textbf{Adversarial Attacks} (Section~\ref{subsubsec:visual-adversarial-attack}), \textbf{Adversarial Defenses} (Section~\ref{subsubsec:visual-adversarial-defense}), \textbf{Backdoor Attacks} (Section~\ref{subsubsec:visual-backdoor-attack}), and \textbf{Backdoor Defenses} (Section~\ref{subsubsec:visual-backdoor-defense}).


\subsubsection{Adversarial Attacks}\label{subsubsec:visual-adversarial-attack}

Adversarial attacks on vision pipelines typically occur in two domains: in the digital space, where they perturb pixel values, and in the physical world, where they manipulate real-world signals (e.g., road signs and flashlights) to deceive perception systems.

\noindent\textbf{White-box Attacks}.\quad
White-box attacks exploit full model access to craft precise perturbations, organized into digital and physical attack strategies.
Digital attacks manipulate inputs directly in the digital space.
For object classification,
Melis et al.~\cite{melis2017deep} introduced region-constrained perturbations against the iCub robot's vision pipeline.
For object detection,
Thys et al.~\cite{thys2019fooling} used adversarial patches to conceal detections or degrade localization.
Adversarial Overlay~\cite{wu2023adversarial} proposes real-time attacks,
and HitM~\cite{wu2024human} introduces a human-in-the-middle threat model that intercepts camera data before OS processing.
For single-object tracking,
RTAA~\cite{jia2020robust} exploits temporal information by leveraging motion and recent predictions across frames.
TrackPGD~\cite{nokabadi2024trackpgd} targets Transformer-based trackers specifically.
For multi-object tracking,
under tracking-by-detection (TBD) and joint-detection-tracking (JDT) paradigms,
Tracker Hijacking~\cite{jia2020fooling} exploits sparse-frame perturbations to induce long-term tracking failures.
Ma et al.'s attack~\cite{ma2023wip} corrupts the detection stage via adversarial patches.
For semantic segmentation,
CAA~\cite{wang2024cascaded} performs multi-task attacks on joint networks.
Uncertainty~\cite{maag2024uncertainty} applies loss-weighting based on uncertainty.

Modern visual encoders introduce new attack surfaces beyond task-specific models. For contrastive vision-language encoders such as CLIP and SigLIP, whose compromise cascades to all downstream VLMs and embodied agents,
  AnyAttack~\cite{zhang2025anyattack} pre-trains a self-supervised perturbation generator on LAION-400M that produces cross-model attacks against CLIP, BLIP, BLIP2, and commercial systems without label supervision.

Video perception models face temporal adversarial threats absent from single-image models.
PCFA~\cite{schmalfuss2022pcfa} targets optical flow models with global perturbations that shift predicted flow toward attacker-chosen targets.


Physical attacks modify objects or scenes to deceive perception under realistic conditions.
For object classification,
DARTS~\cite{sitawarin2018darts} and RP2~\cite{eykholt2018robust} perform physical attacks by attaching stickers or printing patterns on objects.
For object detection,
ShapeShifter~\cite{chen2018shapeshifter} extended Expectation over Transformation (EoT) to the physical domain.
AdvT~\cite{xu2020adversarial}, SLAP~\cite{lovisotto2021slap} and DRP~\cite{sato2021dirty} perform physical adversarial attacks on objects or environments by printing patterns on garments, projecting textures onto traffic signs, or disguising patches as road stains.
For single-object tracking,
MFDA~\cite{li2025multi} fools SOT models under viewpoint, deformation, and illumination changes.

\noindent\textbf{Black-box Attacks}.\quad
Black-box attacks operate without model access, relying on transferability, query-based optimization, or surrogate models, organized into digital and physical attack strategies.
In the digital domain,
for multi-object tracking,
AdvTraj~\cite{wang2024physical} confuses the association phase by swapping the attacker's ID with a target's ID.
For object detection,
Fang et al.~\cite{fang2023pso} used Particle Swarm Optimization to perform heuristic searches on lane-like perturbations for lane detection.
For semantic segmentation,
PB-UAP~\cite{song2025pb} generates black-box universal perturbations transferable across models.

In the physical domain,
for object classification,
ELA~\cite{sun2024embodied} predicts traffic-sign poses and trains an RL agent to project adversarial laser patterns in real time.
For object detection,
CAMOU~\cite{zhang2018camou} perturbs vehicle appearances,
while MobilBye~\cite{nassi2019mobilbye} and SSPA~\cite{nassi2020phantom} employ optical-based attacks via projecting phantom objects.
NS Attack~\cite{mohajeransari2024discovering} perturbs road appearances to evade detection.
For multi-object tracking,
ControlLoc~\cite{ma2024controlloc} searches for optimal patch placements to manipulate objects' positions and shapes.
For visible-thermal detection, Zhu et al.~\cite{zhu2026physical} craft a single physical garment with non-overlapping RGB and thermal patterns to fool both modalities at once.


\subsubsection{Adversarial Defenses}\label{subsubsec:visual-adversarial-defense}

Visual defenses protect object classification, object detection (including lane detection), tracking, and segmentation pipelines from adversarial manipulation through robust training and robust inference strategies.

\noindent\textbf{Robust Training}.\quad
Robust training hardens models by incorporating adversarial examples, augmented data, or feature recovery during training.
For object classification,
Kalin et al.~\cite{kalin2021automating} retrained models with visible-light and infrared imagery, guided by adversarial-surface analysis.
For object detection,
DSNet~\cite{huang2020dsnet}, IA-YOLO~\cite{liu2022image}, and BAD-Net~\cite{li2023detection} jointly learn visibility enhancement, feature restoration, or dehazing with detection,
and Blazevic et al.~\cite{blazevic2025securing} trained robust lane detection models against adversarial perturbations.
For semantic segmentation,
RP-PGD~\cite{zhang2025rp} employs adversarial training to enhance model robustness.

\noindent\textbf{Robust Inference}.\quad
Robust inference defends models through input or output moderation.
Input moderation detects anomalous inputs, preprocesses signals, restores degraded data, or fuses multimodal information.
For object classification,
Melis et al.~\cite{melis2017deep} detected and filtered inputs deviating from training distributions in deep feature space.
For object detection,
AOD-Net~\cite{li2017aod} provides lightweight dehazing to restore visibility.
DGFN~\cite{kim2018robust} fuses camera and LiDAR data with gating mechanisms.
GhostBusters~\cite{nassi2020phantom} deploys four specialized CNNs analyzing various visual features.
For single-object tracking,
Jia et al.~\cite{jia2024robust} detected attacks using similarity differences in the feature space.
Modern visual encoders also require architecture-aware defenses. For CLIP-family encoders,
TeCoA~\cite{mao2023understanding} introduces contrastive adversarial fine-tuning that improves zero-shot robustness. Robust CLIP~\cite{schlarmann2024robust} proposes unsupervised adversarial fine-tuning to ensure that downstream tasks inherit encoder-level robustness.
Output moderation verifies model predictions by analyzing output behavior or comparing predictions across transformations.
For object detection,
SentiNet~\cite{chou2020sentinet} employs output-based detection by localizing suspicious regions,
while Xu et al.~\cite{xu2021model} applied a CNN to the detected lanes to classify them as real or fake.
For single-object tracking,
Li et al.~\cite{li2025detecting} compared full- and low-frequency tracking, using the low-frequency branch as a stable reference.

\begin{table}[!tp]
\center
\renewcommand{\arraystretch}{1.15}
\caption{A summary of \textbf{adversarial defenses} for \textbf{visual perception}.}
\resizebox{1\textwidth}{!}{
\rowcolors{2}{tablegray}{white}
{\setlength{\tabcolsep}{0pt}
\begin{tabular}{@{} >{\cellcolor{white}}p{.15\textwidth} p{.16\textwidth} p{.05\textwidth}
               p{.16\textwidth} p{.20\textwidth} p{.15\textwidth} p{.28\textwidth}@{}}

\toprule
\rowcolor{perception-table-blue}
\cellcolor{perception-table-blue}\textbf{Defense} & \textbf{Method} & \textbf{Year} & \textbf{Category} & \textbf{Subcategory} & \textbf{Target Model} & \textbf{Dataset} \\

\midrule

& Kalin et al.\cite{kalin2021automating} & 2021 & Robust Training & Adversarial Training & Object Classifier & VEDAI \\
& DSNet\cite{huang2020dsnet} & 2020 & Robust Training & Adversarial Training & Object Detector & FOD, Foggy Driving \\
& IA-YOLO\cite{liu2022image} & 2022 & Robust Training & Adversarial Training & Object Detector & VOC\_Foggy, RTTS \\
& BAD-Net\cite{li2023detection} & 2023 & Robust Training & Adversarial Training & Object Detector & RTTS, VOChaze \\
& Blazevic et al.\cite{blazevic2025securing} & 2025 & Robust Training & Adversarial Training & Object Detector & MetaDrive \\
& RP-PGD\cite{zhang2025rp} & 2025 & Robust Training & Adversarial Training & Semantic Segmentation & ADE20K, VOC, Cityscapes \\

& TeCoA\cite{mao2023understanding} & 2023 & Robust Training & Adversarial Training & CLIP Image Encoder & ImageNet, 15 ZS datasets \\
& Robust CLIP\cite{schlarmann2024robust} & 2024 & Robust Training & Adversarial Training & CLIP Image Encoder & ImageNet, COCO \\
& Melis et al.\cite{melis2017deep} & 2017 & Robust Inference & Input Moderation & Object Classifier & iCubWorld \\
& AOD-Net\cite{li2017aod} & 2017 & Robust Inference & Input Moderation & Object Detector & Middlebury, Real-world data \\
& DGFN\cite{kim2018robust} & 2018 & Robust Inference & Input Moderation & Object Detector & KITTI \\

& GhostBusters\cite{nassi2020phantom} & 2020 & Robust Inference & Input Moderation & Object Detector & Real-world data \\
& Jia et al.\cite{jia2024robust} & 2024 & Robust Inference & Input Moderation & Single-Object Tracker & UAV \\

& SentiNet\cite{chou2020sentinet} & 2020 & Robust Inference & Output Moderation & Object Detector & ImageNet, LFW, LISA, VGG-Face \\
& Xu et al.\cite{xu2021model} & 2021 & Robust Inference & Output Moderation & Object Detector & TuSimple \\
\multirow{-16}{*}{\parbox[c]{\linewidth}{\raggedright Adversarial\\Defense}} & Li et al.\cite{li2025detecting} & 2025 & Robust Inference & Output Moderation & Single-Object Tracker & LaSOT, OTB, UAV \\

\bottomrule
\end{tabular}
}}
\label{tab:vision_adversarial_defense}
\end{table}


\subsubsection{Backdoor Attacks}\label{subsubsec:visual-backdoor-attack}

Backdoor attacks on visual perception embed hidden triggers during model training so that the system behaves normally on benign inputs but executes attacker-chosen behaviors when the trigger appears. These attacks target visual models through training manipulation or data poisoning.

\noindent\textbf{Training Manipulation}.\quad
Training manipulation attacks embed backdoors by manipulating training objectives or procedures.
For ViTs,
  TrojViT~\cite{zheng2023trojvit} injects trojans via RowHammer-based bit-flipping without training-time poisoning,
and SWARM~\cite{yang2024swarm} targets prompt-tuned ViTs with a switchable backdoor.

\noindent\textbf{Data Poisoning}.\quad
Data poisoning attacks embed backdoors by injecting triggered samples into training data.
Han et al.~\cite{han2022physical} inserted physical objects as triggers using poison-annotation strategies for object detectors.
BadLANE~\cite{zhang2024towards} and DBALD~\cite{liao2025towards} embed visual pattern triggers via meta-learning or diffusion-based synthesis for object detectors.


  Modern visual encoders are also vulnerable to backdoor attacks. For vision-language encoders, a single compromised encoder propagates backdoor behavior to all downstream tasks:
  BadEncoder~\cite{jia2022badencoder} first demonstrates this supply-chain threat on pre-trained encoders including CLIP,
  BadCLIP~\cite{liang2024badclip} optimizes triggers via dual-embedding guided Bayesian reasoning,
  and BadVision~\cite{liu2025badvisionsslbackdoor} exploits SSL encoder backdoors to induce visual hallucinations in LVLMs.
  For ViTs,
  BadViT~\cite{yuan2023badvitvit} shows that self-attention makes ViTs more sensitive to patch-wise triggers than CNNs.


\subsubsection{Backdoor Defenses}\label{subsubsec:visual-backdoor-defense}

Backdoor defenses for visual perception aim to detect or remove hidden visual triggers implanted during training. Current defense research for task-specific visual backdoor attacks remains limited, highlighting an important open challenge for securing visual perception in embodied systems.

For modern visual encoders, backdoor defenses must operate at encoder level.
  For ViTs, Doan et al.~\cite{doan2023defending} exploited patch-transformation responses to detect triggers without training data access.
  For CLIP-family encoders,
  CleanCLIP~\cite{bansal2023cleanclip} re-aligns modality representations via multimodal contrastive fine-tuning to weaken backdoor associations,
  DECREE~\cite{feng2023decree} detects backdoors in pre-trained encoders without classifier headers or input labels,
and BDetCLIP~\cite{niu2025bdetclip} enables efficient test-time backdoor detection via contrastive prompting.


\subsection{Auditory Perception}
\label{subsec:auditory-perception}

\begin{table}[!tp]
\center
\renewcommand{\arraystretch}{1.15}
\caption{A summary of \textbf{adversarial} attacks and defenses for \textbf{auditory perception}.}
\resizebox{1\textwidth}{!}{

\rowcolors{2}{tablegray}{white}
{\setlength{\tabcolsep}{0pt}
\begin{tabular}{@{} >{\cellcolor{white}}p{.15\textwidth} p{.20\textwidth} p{.06\textwidth}
               p{.18\textwidth} p{.18\textwidth} p{.16\textwidth} p{.25\textwidth}@{}}

\toprule
\rowcolor{perception-table-blue}
\cellcolor{perception-table-blue}\textbf{Attack/Defense} &
\cellcolor{perception-table-blue}\textbf{Method} &
\cellcolor{perception-table-blue}\textbf{Year} &
\cellcolor{perception-table-blue}\textbf{Category} &
\cellcolor{perception-table-blue}\textbf{Subcategory} &
\cellcolor{perception-table-blue}\textbf{Target Model} &
\cellcolor{perception-table-blue}\textbf{Dataset} \\
\midrule

& Carlini et al.\cite{carlini2016hidden} & 2016 & White-box & Physical Attack & Speech Recognizer & Custom dataset \\
& CommanderSong\cite{yuan2018commandersong} & 2018 & White-box & Physical Attack & Speech Recognizer & Custom dataset \\
& Metamorph\cite{chen2020metamorph} & 2020 & White-box & Physical Attack & Speech Recognizer & AIR, Common Voice, MARDY \\
& SpecPatch\cite{guo2022specpatch} & 2022 & White-box & Physical Attack & Speech Recognizer & TIMIT \\
& Li et al.\cite{li2020practical} & 2020 & White-box & Physical Attack & Speech Verifier & CSTR VCTK Corpus \\
& AdvPulse\cite{li2020advpulse} & 2020 & White-box & Physical Attack & Speech Rec./Ver. & CSTR VCTK Corpus, Voice Commands \\

& TSMAE\cite{ji2024watch} & 2024 & Black-box & Digital Attack & Speech Recognizer & CSTR VCTK Corpus, Custom dataset \\
& Occam\cite{zheng2021black} & 2021 & Black-box & Digital Attack & Speech Rec./Ver. & Common Voice, LibriSpeech, VoxCeleb \\

& Cocaine Noodles\cite{vaidya2015cocaine} & 2015 & Black-box & Physical Attack & Speech Recognizer & Custom dataset \\
& Wang et al.\cite{wang2020differences} & 2020 & Black-box & Physical Attack & Speech Recognizer & Custom dataset \\
& Devil's Whisper\cite{chen2020devil} & 2020 & Black-box & Physical Attack & Speech Recognizer & CommanderSong dataset \\
& BarrierBypass\cite{walker2023barrierbypass} & 2023 & Black-box & Physical Attack & Speech Recognizer & Custom dataset \\
& Vaspy\cite{zhang2019activated} & 2019 & Black-box & Physical Attack & Speech Verifier & Custom dataset \\
& Bilika et al.\cite{bilika2023hello} & 2023 & Black-box & Physical Attack & Speech Verifier & Custom dataset \\
\multirow{-15}{*}{\parbox[c]{\linewidth}{\raggedright Adversarial\\Attack}} & Abdullah et al.\cite{abdullah2019practical} & 2019 & Black-box & Physical Attack & Speech Rec./Ver. & LibriSpeech, TIMIT \\

\midrule

& Yang et al.\cite{yang2018towards} & 2018 & Robust Inference & Input Moderation & Speech Recognizer & Common Voice, LIBRIS \\
& Samizade et al.\cite{samizade2020adversarial} & 2020 & Robust Inference & Input Moderation & Speech Recognizer & Common Voice, Speech Commands \\
& AudioPure\cite{wu2023defending} & 2023 & Robust Inference & Input Moderation & Speech Recognizer & Qualcomm Keyword Speech \\
& AntiFake\cite{yu2023antifake} & 2023 & Robust Inference & Input Moderation & Speech Verifier & CSTR VCTK, LibriSpeech, TIMIT \\

& Yang et al.\cite{yang2018towards} & 2018 & Robust Inference & Output Moderation & Speech Recognizer & Common Voice, LIBRIS \\
\multirow{-6}{*}{\parbox[c]{\linewidth}{\raggedright Adversarial\\Defense}} & MVP-EARS\cite{zeng2019multiversion} & 2019 & Robust Inference & Output Moderation & Speech Recognizer & Common Voice, Custom dataset \\

\bottomrule
\end{tabular}
}}
\label{tab:auditory_perception_attack_and_defense_part2}
\end{table}

Auditory perception supports human-robot interaction and voice-based control through speech recognizers, which convert spoken language into text, and speaker verifiers, which authenticate speaker identity based on voice characteristics. This subsection consolidates all auditory perception security research, organized into three parts: \textbf{Adversarial Attacks} (Section~\ref{subsubsec:auditory-adversarial-attack}), \textbf{Adversarial Defenses} (Section~\ref{subsubsec:auditory-adversarial-defense}), and \textbf{Backdoor Attacks and Defenses} (Section~\ref{subsubsec:auditory-backdoor}).


\subsubsection{Adversarial Attacks}\label{subsubsec:auditory-adversarial-attack}

Adversarial attacks on auditory perception operate in the digital space by perturbing audio waveforms and in the physical space by manipulating signals (e.g., voice injection and speaker spoofing).

\noindent\textbf{White-box Attacks}.\quad
In the physical space, for speech recognizers,
Carlini et al.~\cite{carlini2016hidden} converted audio commands into forms unintelligible to humans.
CommanderSong~\cite{yuan2018commandersong} embeds perturbations into songs.
Metamorph~\cite{chen2020metamorph} employs background-like audio perturbations,
while SpecPatch~\cite{guo2022specpatch} employs audio spectrogram patches.
For speaker verifiers,
Li et al.~\cite{li2020practical} incorporated Room Impulse Response to maintain effectiveness under over-the-air playback.
For the joint speech recognizers and speaker verifiers,
AdvPulse~\cite{li2020advpulse} designs subsecond perturbations.

\noindent\textbf{Black-box Attacks}.\quad
In the digital space,
TSMAE~\cite{ji2024watch} adjusts playback speed to attack speech recognizers.
Occam~\cite{zheng2021black} introduces decision-only adversarial examples for cloud APIs.
In the physical space, for speech recognizers,
BarrierBypass~\cite{walker2023barrierbypass} injects commands through physical barriers.
Cocaine Noodles~\cite{vaidya2015cocaine} and Abdullah et al.~\cite{abdullah2019practical} generated mangled inputs that are unintelligible to humans but still recognized by machines.
Devil's Whisper~\cite{chen2020devil} crafts adversarial examples by pairing a query-based substitute with a stronger white-box recognizer. Wang et al.~\cite{wang2020differences} modulated the signals to compensate for the distortion in the frequency domain.
For speaker verifiers,
Vaspy~\cite{zhang2019activated} synthesizes activation keywords via speech recognition and voice cloning.
Bilika et al.~\cite{bilika2023hello} demonstrated practical feasibility in real-world scenarios.


\subsubsection{Adversarial Defenses}\label{subsubsec:auditory-adversarial-defense}


Auditory defenses protect speech recognizer and speaker verifier systems from adversarial perturbations through robust inference strategies.

\noindent\textbf{Robust Inference}.\quad
For input moderation of speech recognizers,
Samizade et al.~\cite{samizade2020adversarial} extracted MFCC features and classified benign and adversarial samples using CNNs.
Yang et al.~\cite{yang2018towards} explored input preprocessing techniques, including perturbation, compression, quantization, smoothing, reconstruction, and downsampling.
AudioPure~\cite{wu2023defending} employs diffusion models to purify and restore input signals.
For speaker verifiers,
AntiFake~\cite{yu2023antifake} embeds protective perturbations into audio to prevent voice cloning and forgery.
For output moderation of speech recognizers,
Yang et al.'s approach~\cite{yang2018towards} compares transcription consistency between audio segments and complete audio.
MVP-EARS~\cite{zeng2019multiversion} uses multiple models for cross-verification.


\subsubsection{Backdoor Attacks and Defenses}\label{subsubsec:auditory-backdoor}

Backdoor research on auditory perception remains nascent.
TrojanModel~\cite{zong2023trojanmodel} inserts backdoors into acoustic models for speech recognizers via training manipulation, but no dedicated defense has been proposed, highlighting an important open challenge for securing speech-based embodied systems.


\subsection{Spatial Perception}
\label{subsec:spatial-perception}

Spatial perception enables embodied agents to build, maintain, and reason about 3D representations of their environment for navigation, manipulation, and obstacle avoidance. It encompasses point cloud classification, 3D object detection and tracking, trajectory prediction, depth and pose estimation, SLAM, and neural scene representations (NeRF, 3DGS). This subsection consolidates all spatial perception security research, organized into three parts: \textbf{Adversarial Attacks} (Section~\ref{subsubsec:spatial-adversarial-attack}), \textbf{Adversarial Defenses} (Section~\ref{subsubsec:spatial-adversarial-defense}), and \textbf{Backdoor Attacks and Defenses} (Section~\ref{subsubsec:spatial-backdoor}).

\begin{table}[!tp]
\center
\renewcommand{\arraystretch}{1.15}
\caption{A summary of \textbf{adversarial attacks} for \textbf{spatial perception}.}
\resizebox{1\textwidth}{!}{

\rowcolors{2}{tablegray}{white}
{\setlength{\tabcolsep}{0pt}
\begin{tabular}{@{} >{\cellcolor{white}}p{.12\textwidth} p{.20\textwidth} p{.06\textwidth}
               p{.18\textwidth} p{.18\textwidth} p{.16\textwidth} p{.25\textwidth}@{}}

\toprule
\rowcolor{perception-table-blue}
\cellcolor{perception-table-blue}\textbf{Attack} & \textbf{Method} & \textbf{Year} & \textbf{Category} & \textbf{Subcategory} & \textbf{Target Model} & \textbf{Dataset} \\

\midrule


& FLAT\cite{li2021fooling} & 2021 & White-box & Digital Attack & 3D Object Detector & nuScenes \\
& SlowLiDAR\cite{liu2023slowlidar} & 2023 & White-box & Digital Attack & 3D Object Detector & KITTI \\
& Zheng et al.\cite{zheng2025new} & 2025 & White-box & Digital Attack & 3D Object Detector & KITTI, nuScenes \\
& Wang et al.\cite{wang2025imperceptible} & 2025 & White-box & Digital Attack & 3D Object Detector & KITTI, Waymo \\

& Cheng et al.\cite{cheng2021universal} & 2021 & White-box & Digital Attack & 3D Object Tracker & KITTI \\
& TAN\cite{liu2023transferable} & 2023 & White-box & Digital Attack & 3D Object Tracker & KITTI \\

& Yoshida et al.~\cite{yoshida2022adversarial} & 2022 & White-box & Digital Attack & SLAM (LiDAR) & Self-constructed data \\
& Horv\'{a}th and J\'{o}zsa~\cite{horvath2023targeted} & 2023 & White-box & Digital Attack & NeRF Navigator & LLFF \\
& Poison-Splat~\cite{lu2024poison} & 2024 & White-box & Digital Attack & 3DGS Navigator & NeRF-Synthetic, Mip-NeRF360 \\

& LiDAR-Adv\cite{cao2019adversarial} & 2019 & White-box & Physical Attack & 3D Object Detector & Real-world data \\
& Tu et al.\cite{tu2020physically} & 2020 & White-box & Physical Attack & 3D Object Detector & KITTI \\
& AE-Morpher\cite{zhu2024ae} & 2024 & White-box & Physical Attack & 3D Object Detector & Custom dataset, SVL simulator \\
& Adv3D~\cite{li2024adv3d} & 2024 & White-box & Physical Attack & 3D Object Detector & nuScenes \\
& ShadowHack\cite{kobayashi2025invisible} & 2025 & White-box & Physical Attack & 3D Object Detector & AWSIM simulator \\

& Cheng et al.~\cite{cheng2022physical} & 2022 & White-box & Physical Attack & Depth Estimator & KITTI \\
& Chawla et al.~\cite{chawla2022adversarial} & 2022 & White-box & Physical Attack & Pose Estimator & KITTI odometry \\
& AoR~\cite{chen2024adversary} & 2024 & White-box & Physical Attack & SLAM (Visual) & KITTI, Oxford RobotCar, 4Seasons \\


& Hau et al.\cite{hau2021object} & 2021 & Black-box & Digital Attack & 3D Object Detector & KITTI \\

& TAPG\cite{tian2024adversarial} & 2024 & Black-box & Digital Attack & 3D Object Tracker & KITTI Tracking, nuScenes \\
& Cheng et al.\cite{cheng2025black} & 2025 & Black-box & Digital Attack & 3D Object Tracker & KITTI, nuScenes, Waymo \\

& Wang et al.~\cite{wang2024benchmarking} & 2024 & Black-box & Digital Attack & NeRF Navigator & LLFF-C, Blender-C \\

& SpotAttack\cite{huang2024spotattack} & 2024 & Black-box & Physical Attack & 3D Object Detector & MATLAB simulator \\
& LiDAttack\cite{chen2025lidattack} & 2025 & Black-box & Physical Attack & 3D Object Detector & Custom dataset, KITTI, nuScenes \\
& ICSL Attack~\cite{wang2021can} & 2021 & Black-box & Physical Attack & Camera & Bosch Night, KITTI \\
& DoubleStar~\cite{zhou2022doublestar} & 2022 & Black-box & Physical Attack & Stereo Depth & Self-constructed data \\
& Ikram et al.~\cite{ikram2022perceptual} & 2022 & Black-box & Physical Attack & SLAM (Visual) & Manhattan, Intel, MIT, Garage \\
& Fukunaga et al.~\cite{fukunaga2024random} & 2024 & Black-box & Physical Attack & LiDAR SLAM & Self-constructed data \\
\multirow{-28}{*}{\parbox[c]{\linewidth}{\raggedright Adversarial\\Attack}} & Lou et al.~\cite{lou2024first} & 2024 & Black-box & Physical Attack & Trajectory Predictor & nuScenes, Real world \\

\bottomrule
\end{tabular}
}}
\label{tab:spatial_perception_attack}
\end{table}


\subsubsection{Adversarial Attacks}\label{subsubsec:spatial-adversarial-attack}

Adversarial attacks on spatial perception models perturb point clouds, depth maps, or neural scene representations in the digital space, or manipulate 3D object geometries, LiDAR signals, or camera inputs in the physical space to disrupt detection, localization, and navigation.

\noindent\textbf{White-box Attacks}.\quad
White-box attacks exploit full model access to craft precise perturbations, organized into digital and physical attack strategies.
In the digital domain,
for 3D object detectors,
FLAT~\cite{li2021fooling} manipulates the vehicle trajectory to affect LiDAR motion compensation.
SlowLiDAR~\cite{liu2023slowlidar} introduces slow-acting perturbations, causing delayed failures.
Zheng et al.~\cite{zheng2025new} leveraged smoke-like perturbations to interfere with sensing,
and Wang et al.~\cite{wang2025imperceptible} utilized saliency maps to identify critical points before optimizing perturbations.
For 3D object trackers,
Cheng et al.'s method~\cite{cheng2021universal} crafts universal perturbations,
and TAN~\cite{liu2023transferable} crafts transferable perturbations.
For SLAM systems, Yoshida et al.~\cite{yoshida2022adversarial} applied small, well-timed perturbations to LiDAR point clouds to disrupt scan matching and degrade map consistency, leading to accumulating localization errors.
For scene representation models, Horv\'{a}th and J\'{o}zsa~\cite{horvath2023targeted} demonstrated that NeRFs can be subverted by carefully perturbed input views to render photorealistic but falsified scenes,
and Poison-Splat~\cite{lu2024poison} introduces a data-poisoning attack on 3DGS that corrupts adaptive density control, causing denial-of-service through excessive memory and compute usage.
In the physical domain,
for 3D object detectors,
LiDAR-Adv~\cite{cao2019adversarial} and
Tu et al.'s attack~\cite{tu2020physically} optimize adversarial 3D mesh geometries under physical constraints.
AE-Morpher~\cite{zhu2024ae} generates adversarial meshes with morphing constraints.
Adv3D~\cite{li2024adv3d} embeds adversarial objects directly as NeRFs, enabling transferable and contact-free perturbations.
ShadowHack~\cite{kobayashi2025invisible} exploits optimized planar materials to manipulate shadow patterns.
For depth estimation, Cheng et al.~\cite{cheng2022physical} optimized physical adversarial patches that bias monocular depth estimators.
For pose estimation, Chawla et al.~\cite{chawla2022adversarial} crafted patches that yield large trajectory errors.
For visual SLAM, AoR~\cite{chen2024adversary} uses adversarial patches to trigger false loop closures, producing severe localization drift.

\noindent\textbf{Black-box Attacks}.\quad
In the digital space,
for 3D object detectors,
Hau et al.~\cite{hau2021object} forced the sensor to record injected points, thereby pushing the genuine points outside the object's bounding box.
For 3D object trackers,
TAPG~\cite{tian2024adversarial} and Cheng et al.~\cite{cheng2025black} optimized black-box attacks via transfer-based and explainability-guided methods, respectively.
For NeRF-based navigation, Wang et al.~\cite{wang2024benchmarking} benchmarked NeRF navigators under visual corruptions.
In the physical domain,
for 3D object detectors,
SpotAttack~\cite{huang2024spotattack} employs a genetic algorithm-based global search to optimize non-reflective adversarial spots,
while LiDAttack~\cite{chen2025lidattack} combines global search with local refinement for covert attacks.
For camera-based systems, ICSL Attack~\cite{wang2021can} creates ghost traffic signals using infrared projection invisible to humans.
For stereo depth estimation, DoubleStar~\cite{zhou2022doublestar} uses long-range, synchronized light patterns to exploit stereo-matching artifacts and fabricate obstacles for drones.
For visual SLAM, Ikram et al.~\cite{ikram2022perceptual} employed duplicated textured regions to trigger perceptual aliasing and induce localization drift.
For LiDAR SLAM, Fukunaga et al.~\cite{fukunaga2024random} injected simple, well-timed points into LiDAR streams to disrupt scan matching.
For trajectory predictors,
Lou et al.~\cite{lou2024first} introduced the first physical-world attack on trajectory prediction via LiDAR-induced deceptions, employing a two-stage framework that identifies velocity-insensitive state perturbations and matches them to feasible object locations.


\subsubsection{Adversarial Defenses}\label{subsubsec:spatial-adversarial-defense}

\begin{table}[!tp]
\center
\renewcommand{\arraystretch}{1.15}
\caption{A summary of \textbf{adversarial defenses} for \textbf{spatial perception}.}
\resizebox{1\textwidth}{!}{

\rowcolors{2}{tablegray}{white}
{\setlength{\tabcolsep}{0pt}
\begin{tabular}{@{} >{\cellcolor{white}}p{.12\textwidth} p{.20\textwidth} p{.06\textwidth}
               p{.18\textwidth} p{.18\textwidth} p{.16\textwidth} p{.25\textwidth}@{}}

\toprule
\rowcolor{perception-table-blue}
\cellcolor{perception-table-blue}\textbf{Defense} & \textbf{Method} & \textbf{Year} & \textbf{Category} & \textbf{Subcategory} & \textbf{Target Model} & \textbf{Dataset} \\

\midrule

& Defense-PointNet\cite{zhang2019defense} & 2019 & Robust Training & Adversarial Training & Point Cloud Classifier & ShapeNet \\
& Sun et al.\cite{sun2021adversarially} & 2021 & Robust Training & Adversarial Training & Point Cloud Classifier & ModelNet, ScanObjectNN \\
& PointCutMix\cite{zhang2022pointcutmix} & 2022 & Robust Training & Adversarial Training & Point Cloud Classifier & ModelNet, ScanObjectNN\\

& Hahner et al.\cite{hahner2021fog} & 2021 & Robust Training & Adversarial Training & 3D Object Detector & KITTI \\
& 3D-VField\cite{lehner20223d} & 2022 & Robust Training & Adversarial Training & 3D Object Detector & CrashD, KITTI, Waymo \\
& BAFT\cite{zhang2024comprehensive} & 2024 & Robust Training & Adversarial Training & 3D Object Detector & KITTI, Waymo \\
& DART\cite{wang2025enhancing} & 2025 & Robust Training & Adversarial Training & 3D Object Detector & KITTI \\
& LISA\cite{kilic2025lidar} & 2025 & Robust Training & Adversarial Training & 3D Object Detector & KITTI, Waymo \\

& AcousticFusion~\cite{zhang2021acousticfusion} & 2021 & Robust Training & Multi-Modal Fusion & SLAM & Azure Kinect audio and RGB-D \\
& Wang et al.~\cite{wang2024mobile} & 2024 & Robust Training & Multi-Modal Fusion & Safety System & Self-constructed data \\

& Sang et al.~\cite{sang2023scene} & 2023 & Robust Training & Scene Augmentation & Scene Understanding & iGibson \\

& Adamkiewicz et al.~\cite{adamkiewicz2022vision} & 2022 & Robust Training & Environment Augmentation & NeRF Navigator & Self-constructed data \\
& Splat-Nav~\cite{chen2025splat} & 2025 & Robust Training & Environment Augmentation & 3DGS Navigator & Stonehenge, Statues, Flightroom \\

& Tong et al.~\cite{tong2022enforcing} & 2022 & Robust Training & Formal Safety Methods & NeRF Navigator & Replica Dataset \\
& CATNIPS~\cite{chen2024catnips} & 2024 & Robust Training & Formal Safety Methods & Navigation & Stonehenge, Statues, Flightroom \\
& Zhou et al.~\cite{zhou2024control} & 2024 & Robust Training & Formal Safety Methods & SLAM + Control & Self-constructed data \\
& SAFER-Splat~\cite{chen2024safer} & 2024 & Robust Training & Formal Safety Methods & 3DGS Navigator & Self-constructed data \\

& RaEM~\cite{liu2024beyond} & 2024 & Robust Training & Risk-Aware Planning & Active Perception & Matterport3D \\

& PointGuard\cite{liu2021pointguard} & 2021 & Robust Inference & Input Moderation & Point Cloud Classifier & ModelNet40, ScanNet \\

& WeatherNet\cite{heinzler2020cnn} & 2020 & Robust Inference & Input Moderation & 3D Object Detector & Chamber \& road(custom) \\
& Shadow-Catcher\cite{hau2021shadow} & 2021 & Robust Inference & Input Moderation & 3D Object Detector & KITTI \\
& LOP\cite{xiao2023exorcising} & 2023 & Robust Inference & Input Moderation & 3D Object Detector & KITTI, LGSVL simulator \\
& ADoPT\cite{cho2023adopt} & 2023 & Robust Inference & Input Moderation & 3D Object Detector & nuScenes \\
& LiDARPure\cite{cai2024diffusion} & 2024 & Robust Inference & Input Moderation & 3D Object Detector & KITTI \\
& Zhang et al.~\cite{zhang2025realtime} & 2025 & Robust Inference & Input Moderation & 3D Object Detector & nuScenes, KITTI \\

& Brunke et al.~\cite{brunke2025semantically} & 2025 & Robust Inference & Input Moderation & Scene Understanding & ScanNet200, Self-constructed data \\


& 3D-TC2\cite{you2021temporal} & 2021 & Robust Inference & Output Moderation & 3D Object Detector & nuScenes \\

\multirow{-28}{*}{\parbox[c]{\linewidth}{\raggedright Adversarial\\Defense}} & ViewFool~\cite{dong2022viewfool} & 2022 & Robust Inference & Output Moderation & Object Classifier & BlenderKit, Objectron \\

\bottomrule
\end{tabular}
}}
\label{tab:spatial_perception_defense}
\end{table}


Spatial defenses protect point cloud classifiers, 3D object detectors, SLAM systems, and neural scene representations from adversarial perturbations through robust training and robust inference strategies.

\noindent\textbf{Robust Training}.\quad
Robust training strengthens spatial perception models by exposing them to adversarial examples, augmented data, or physical constraints during the training process.
We refer to adversarial training as a framework where training data is augmented by natural perturbations (e.g., weather, fog, crash simulations) or adversarial perturbations (e.g., gradient-based attacks) to improve model robustness under distribution shift and adversarial manipulation.
For point cloud classifiers,
Defense-PointNet~\cite{zhang2019defense}, Sun et al.'s work~\cite{sun2021adversarially}, and PointCutMix~\cite{zhang2022pointcutmix} employ robust training with data augmentation or perturbation strategies.
For 3D object detectors,
Hahner et al.~\cite{hahner2021fog} introduced LiDAR fog simulation and augmentation techniques.
3D-VField~\cite{lehner20223d} learns vector fields for adversarial augmentation,
BAFT~\cite{zhang2024comprehensive} and DART~\cite{wang2025enhancing} use optimized perturbation strategies,
and LISA~\cite{kilic2025lidar} improves robustness via LiDAR-specific augmentation.

Beyond adversarial training, spatial perception benefits from multi-modal fusion, scene augmentation, environment augmentation, formal safety methods, and risk-aware planning.
AcousticFusion~\cite{zhang2021acousticfusion} fuses auditory cues with visual SLAM to stabilize localization under motion and scene changes,
and Wang et al.~\cite{wang2024mobile} developed a cooperative safety system fusing multi-camera 3D inputs for dynamic human detection and real-time safety-zone enforcement.
Sang et al.~\cite{sang2023scene} proposed systematic scene augmentation that procedurally varies layouts, objects, and states to improve generalization.
Adamkiewicz et al.~\cite{adamkiewicz2022vision} achieved collision-free navigation in NeRF-modeled environments,
and Splat-Nav~\cite{chen2025splat} demonstrate safe navigation using Gaussian Splatting.
Tong et al.~\cite{tong2022enforcing} paired predictive NeRF rendering with CBF filtering,
CATNIPS~\cite{chen2024catnips} reinterprets NeRF densities for collision-probability estimation,
Zhou et al.~\cite{zhou2024control} coupled visual-inertial SLAM with control barrier functions (CBFs),
and SAFER-Splat~\cite{chen2024safer} embeds CBFs over Gaussian Splatting primitives.
RaEM~\cite{liu2024beyond} introduces risk-aware view acquisition that prioritizes safety-critical regions.

\noindent\textbf{Robust Inference}.\quad
Input moderation for spatial perception models detects anomalous inputs, preprocesses point clouds, or restores degraded data before inference.
For point cloud classifiers,
PointGuard~\cite{liu2021pointguard} provides certified robustness guarantees.
For 3D object detectors,
WeatherNet~\cite{heinzler2020cnn} is trained to remove noise induced by bad weather.
Shadow-Catcher~\cite{hau2021shadow} and LOP~\cite{xiao2023exorcising} detect adversarial point clouds via physical invariants such as 3D shadows and depth-density relations.
ADoPT~\cite{cho2023adopt} leverages temporal consistency for abnormal input detection.
LiDARPure~\cite{cai2024diffusion} employs diffusion models to purify input point clouds.
Zhang et al.~\cite{zhang2025realtime} proposed the first real-time defense against object-based LiDAR attacks, employing a generative model positioned between sensing and perception to identify and remove adversarial points from suspicious regions in point clouds.
For scene understanding, Brunke et al.~\cite{brunke2025semantically} introduced a semantic safety filter integrating 3D semantic maps with LLM reasoning, compiling abstract language-grounded rules into CBFs that enforce both geometric and semantic safety.
For output moderation,
3D-TC2~\cite{you2021temporal} verifies predictions of 3D object detectors via temporal consistency across frames.
ViewFool~\cite{dong2022viewfool} employs NeRF to systematically identify failure-inducing viewpoints for robustness assessment of object classifiers.


\subsubsection{Backdoor Attacks and Defenses}\label{subsubsec:spatial-backdoor}

\begin{table}[!tp]
\center
\renewcommand{\arraystretch}{1.15}
\caption{A summary of \textbf{backdoor} attacks and defenses for \textbf{embodied perception}.}
\resizebox{1\textwidth}{!}{

\rowcolors{2}{tablegray}{white}
{\setlength{\tabcolsep}{0pt}
\begin{tabular}{@{} >{\cellcolor{white}}p{.12\textwidth} p{.20\textwidth} p{.06\textwidth}
               p{.18\textwidth} p{.18\textwidth} p{.16\textwidth} p{.25\textwidth}@{}}

\toprule
\rowcolor{perception-table-blue}
\cellcolor{perception-table-blue}\textbf{Attack/Defense} & \textbf{Method} & \textbf{Year} & \textbf{Category} & \textbf{Subcategory} & \textbf{Target Model} & \textbf{Dataset} \\

\midrule


& TrojViT\cite{zheng2023trojvit} & 2023 & Training Manipulation & Bit-Flipping & Vision Transformer & ImageNet \\
& SWARM\cite{yang2024swarm} & 2024 & Training Manipulation & Prompt Poisoning & Vision Transformer & CIFAR-100, ImageNet \\

& TrojanModel\cite{zong2023trojanmodel} & 2023 & Training Manipulation & Trigger Injection & Speech Recognizer & Google Speech Commands \\

& Han et al.\cite{han2022physical} & 2022 & Data Poisoning & Physical Object & Object Detector & TuSimple \\
& BadLANE\cite{zhang2024towards} & 2024 & Data Poisoning & Visual Pattern & Object Detector & CULane, TuSimple \\
& DBALD\cite{liao2025towards} & 2025 & Data Poisoning & Visual Pattern & Object Detector & CULane, TuSimple \\

& BadEncoder\cite{jia2022badencoder} & 2022 & Data Poisoning & Embedding Poisoning & CLIP Image Encoder & CIFAR-10, STL-10, SVHN \\
& BadCLIP\cite{liang2024badclip} & 2024 & Data Poisoning & Embedding Poisoning & CLIP Image Encoder & ImageNet, 11 ZS datasets \\
& BadVision\cite{liu2025badvisionsslbackdoor} & 2025 & Data Poisoning & Embedding Poisoning & CLIP Image Encoder & COCO, VQA-v2 \\

& BadViT\cite{yuan2023badvitvit} & 2023 & Data Poisoning & Attention Manipulation & Vision Transformer & CIFAR-10, ImageNet \\

& Zhang et al.\cite{zhang2022towards} & 2022 & Data Poisoning & BEV Trigger & 3D Object Detector & KITTI \\
& BadLiDet\cite{li2023badlidet} & 2023 & Data Poisoning & Point Perturbation & 3D Object Detector & KITTI, nuScenes \\

\multirow{-13}{*}{\parbox[c]{\linewidth}{\raggedright Backdoor\\Attack}} & BadLiSeg\cite{li2023towards} & 2023 & Data Poisoning & Spoofed Pattern & 3D Segmentation & SemanticKITTI \\

\midrule


& Doan et al.\cite{doan2023defending} & 2023 & Robust Inference & Patch Processing & Vision Transformer & CIFAR-10, ImageNet \\

& CleanCLIP\cite{bansal2023cleanclip} & 2023 & Robust Training & Fine-Tuning & CLIP Image Encoder & ImageNet, 8 ZS datasets \\
& DECREE\cite{feng2023decree} & 2023 & Robust Inference & Detection & CLIP Image Encoder & CIFAR-10, ImageNet, STL-10 \\
\multirow{-4}{*}{\parbox[c]{\linewidth}{\raggedright Backdoor\\Defense}} & BDetCLIP\cite{niu2025bdetclip} & 2025 & Robust Inference & Test-Time Detection & CLIP Image Encoder & ImageNet, 11 ZS datasets \\

\bottomrule
\end{tabular}
}}
\label{tab:backdoor_attack_and_defense}
\end{table}

Backdoor attacks on spatial perception target 3D object detectors and LiDAR segmentation through data poisoning.
Zhang et al.~\cite{zhang2022towards} and BadLiDet~\cite{li2023badlidet} demonstrate pixel-level trigger injection in bird's-eye-view representations and imperceptible point-level perturbations for 3D object detectors, while BadLiSeg~\cite{li2023towards} embeds backdoors in LiDAR segmentation via crafted spoofed patterns.
No dedicated defense has been proposed for spatial backdoors, an important open challenge for securing 3D perception in embodied systems.


\subsection{Motion Perception}
\label{subsec:motion-perception}

Motion perception enables embodied agents to estimate pose, velocity, and trajectory through inertial measurement units (IMUs), visual and LiDAR odometry, Global Navigation Satellite System (GNSS) receivers, and simultaneous localization and mapping (SLAM) pipelines. Compromised motion perception leads to dangerous physical behavior: erroneous position estimates cause collision, drift, or loss of navigation, while corrupted pose estimation destabilizes control loops in drones, ground robots, and autonomous vehicles.
We organize the discussion into two parts: \textbf{Sensor Attacks} (Section~\ref{subsubsec:motion-sensor-attack}) cover IMU-based perception attacks, localization and odometry attacks, and sensor-level spoofing and jamming of GNSS, IMU, ultrasonic, and mmWave radar systems; and \textbf{Sensor Defenses} (Section~\ref{subsubsec:motion-sensor-defense}) protect motion estimation through anomaly detection, cross-sensor verification, robust state estimation, and anti-spoofing/anti-jamming mechanisms.


\subsubsection{Sensor Attacks}\label{subsubsec:motion-sensor-attack}

Sensor attacks on motion perception manipulate raw sensor signals before data reaches perception algorithms, exploiting hardware vulnerabilities, signal-processing pipelines, or physical-layer characteristics. These attacks compromise the integrity of raw measurements through physical injection and interference to achieve spoofing (deception) or jamming (disruption) goals.

\begin{table}[!tp]
\center
\renewcommand{\arraystretch}{1.15}
\caption{A summary of \textbf{sensor attacks} for \textbf{motion perception}. RF: Radio Frequency; EM: Electromagnetic.}
\resizebox{1\textwidth}{!}{

\rowcolors{2}{tablegray}{white}
{\setlength{\tabcolsep}{0pt}
\begin{tabular}{@{} >{\cellcolor{white}}p{.08\textwidth} p{.20\textwidth} p{.06\textwidth}
               p{.12\textwidth} p{.23\textwidth} p{.16\textwidth} p{.25\textwidth}@{}}

\toprule
\rowcolor{perception-table-blue}
\cellcolor{perception-table-blue}\textbf{Attack} & \textbf{Method} & \textbf{Year} & \textbf{Category} & \textbf{Subcategory} & \textbf{Target Sensor} & \textbf{Dataset} \\

\midrule

& Lenhart et al.\cite{lenhart2021relay} & 2021 & Spoofing & Replay Spoofing & GNSS & real-world data \\
& Wang et al.\cite{wang2025practical} & 2025 & Spoofing & Replay Spoofing & GNSS & ESA, real-world data \\

& Horton et al.\cite{horton2018development} & 2018 & Spoofing & Generative Spoofing & GNSS & real-world data \\
& FusionRipper\cite{shen2020drift} & 2020 & Spoofing & Generative Spoofing & GNSS & Apollo Data, KAIST Complex Urban \\
& Dasgupta et al.\cite{dasgupta2024unveiling} & 2024 & Spoofing & Generative Spoofing & GNSS & custom dataset \\
& Zhong et al.\cite{zhong2025analysis} & 2025 & Spoofing & Generative Spoofing & GNSS & real-world data \\

& Son et al.\cite{son2015rocking} & 2015 & Spoofing & Acoustic Injection & IMU & real-world data \\
& WALNUT\cite{trippel2017walnut} & 2017 & Spoofing & Acoustic Injection & IMU & real-world data \\
& KITE\cite{gao2023exploring} & 2023 & Spoofing & Acoustic Injection & IMU & custom dataset, real-world data \\

& Yan et al.\cite{yan2016can} & 2016 & Spoofing & Acoustic Injection & Ultrasonic Ranging & real-world data \\
& Xu et al.\cite{xu2018analyzing} & 2018 & Spoofing & Acoustic Injection & Ultrasonic Ranging & real-world data \\
& Gluck et al.\cite{gluck2020spoofing} & 2020 & Spoofing & Acoustic Injection & Ultrasonic Ranging & real-world data \\

& Sun et al.\cite{sun2021control} & 2021 & Spoofing & RF Injection & mmWave Radar & real-world data \\
& Komissarov et al.\cite{komissarov2021spoofing} & 2021 & Spoofing & RF Injection & mmWave Radar & real-world data \\
& mmSpoof\cite{vennam2023mmspoof} & 2023 & Spoofing & RF Injection & mmWave Radar & real-world data \\

& MetaWave\cite{chen2023metawave} & 2023 & Spoofing & Physical Manipulation & mmWave Radar & real-world data \\
& TileMask\cite{zhu2023tilemask} & 2023 & Spoofing & Physical Manipulation & mmWave Radar & real-world data \\
& mmHide\cite{geng2025attacking} & 2025 & Spoofing & Physical Manipulation & mmWave Radar & real-world data \\

& Lim et al.\cite{lim2018autonomous} & 2018 & Jamming & Acoustic Interference & Ultrasonic Ranging & real-world data \\

\multirow{-20}{*}{\parbox[c]{\linewidth}{\raggedright Sensor\\Attack}} & Jang et al.\cite{jang2023paralyzing} & 2023 & Jamming & EM Interference & IMU & custom dataset, real-world data \\

\bottomrule
\end{tabular}
}}
\label{tab:sensor_attack}
\end{table}

\noindent\textbf{Spoofing}.\quad
Spoofing covers adversarial techniques that manipulate sensor inputs by injecting or presenting false, yet physically plausible, signals, inducing erroneous measurements or misleading perception.

For GNSS, replay spoofing captures and retransmits authentic signals with intentional delay. Lenhart et al.~\cite{lenhart2021relay} demonstrated long-range real-time relay systems using commercial software-defined radios (SDRs), and Wang et al.~\cite{wang2025practical} exposed vulnerabilities in Galileo's OSNMA via manipulated time synchronization.

Generative spoofing synthesizes counterfeit yet realistic GNSS signals. Horton and Ranganathan~\cite{horton2018development} leveraged low-cost SDR platforms to generate spoofing signals; Shen et al.~\cite{shen2020drift} crafted fusion-aware perturbations to mislead integrated navigation systems; Dasgupta et al.~\cite{dasgupta2024unveiling} devised slow-drift attacks; and Zhong et al.~\cite{zhong2025analysis} analyzed perturbations against integrated navigation.

Acoustic injection exploits resonance in inertial sensors by emitting ultrasonic or audible signals. For IMUs, Son et al.~\cite{son2015rocking} showed that single-tone acoustic excitation can induce gyroscope deviations. WALNUT~\cite{trippel2017walnut} combines acoustic induction with ADC aliasing to trigger false readings, and KITE~\cite{gao2023exploring} refines resonance-response modeling for precise control.
For ultrasonic ranging systems, Yan et al.~\cite{yan2016can}, Xu et al.~\cite{xu2018analyzing}, and Gluck et al.~\cite{gluck2020spoofing} showed that crafted acoustic echoes can spoof distance measurements.

RF injection targets radio-frequency sensors such as mmWave radar by introducing synchronized or tailored electromagnetic signals. Sun et al.~\cite{sun2021control} and Komissarov and Wool~\cite{komissarov2021spoofing} injected synchronized RF signals to manipulate radar point clouds, while Vennam et al.~\cite{vennam2023mmspoof} synthesized customized spoofing waveforms to generate deceptive objects.

Physical manipulation involves placing adversarial objects or engineered surfaces in the environment to passively spoof sensors. For mmWave radar, MetaWave~\cite{chen2023metawave}, TileMask~\cite{zhu2023tilemask}, and mmHide~\cite{geng2025attacking} employ metamaterial surface patterns to control radar reflections.

\noindent\textbf{Jamming}.\quad
Jamming attacks disrupt or block legitimate sensor signals through interference, thereby degrading or completely disabling sensor functionality. These attacks exploit physical-layer vulnerabilities by overwhelming or corrupting the sensing modality with high-energy or carefully crafted noise across acoustic, electromagnetic, or optical domains.

Acoustic interference employs high-intensity sound to jam ultrasonic sensors. For ultrasonic ranging, Lim et al.~\cite{lim2018autonomous} demonstrated that high-power acoustic jamming can cause missed detections and destabilize autonomous control policies. Electromagnetic interference disrupts sensor operation through radiated or conducted RF noise.
For IMUs, Jang et al.~\cite{jang2023paralyzing} presented a remote electromagnetic injection attack that corrupts communication between the IMU and flight controller, leading to system failure.


\subsubsection{Sensor Defenses}\label{subsubsec:motion-sensor-defense}

\begin{table}[!tp]
\center
\renewcommand{\arraystretch}{1.15}
\caption{A summary of \textbf{sensor defenses} for \textbf{motion perception}.}
\resizebox{1\textwidth}{!}{

\rowcolors{2}{tablegray}{white}
{\setlength{\tabcolsep}{0pt}
\begin{tabular}{@{} >{\cellcolor{white}}p{.12\textwidth} p{.20\textwidth} p{.06\textwidth}
               p{.18\textwidth} p{.18\textwidth} p{.16\textwidth} p{.25\textwidth}@{}}

\toprule
\rowcolor{perception-table-blue}
\cellcolor{perception-table-blue}\textbf{Defense} & \textbf{Method} & \textbf{Year} & \textbf{Category} & \textbf{Subcategory} & \textbf{Target Sensor} & \textbf{Dataset} \\

\midrule

& Xu et al.\cite{xu2018analyzing} & 2018 & Anti-Spoofing & Detection & Ultrasonic Ranging & real-world data \\
& SoundFence\cite{lou2021soundfence} & 2021 & Anti-Spoofing & Detection & Ultrasonic Ranging & real-world data \\
& SecureTrack\cite{singh2025securetrack} & 2025 & Anti-Spoofing & Detection & Ultrasonic Ranging & real-world data \\
& Sun et al.\cite{sun2021control} & 2021 & Anti-Spoofing & Detection & mmWave Radar & real-world data \\
& Nallabolu et al.\cite{nallabolu2021frequency} & 2021 & Anti-Spoofing & Detection & mmWave Radar & real-world data \\
& Falco et al.\cite{falco2018dual} & 2018 & Anti-Spoofing & Detection & GNSS & Simulation data \\
& Crowd-GPS-Sec\cite{jansen2018crowd} & 2018 & Anti-Spoofing & Detection & GNSS & Real-world data, Simulation data \\

& DeepSIM\cite{xue2020deepsim} & 2020 & Anti-Spoofing & Detection & GNSS & SatUAV(custom) \\
& DeepPOSE\cite{jiang2022deeppose} & 2022 & Anti-Spoofing & Detection & GNSS & BDD-100K, Custom dataset \\
& Iqbal et al.\cite{iqbal2024deep} & 2024 & Anti-Spoofing & Detection & GNSS & TEXBAT \\

& PADS\cite{liu2025gnss} & 2025 & Anti-Spoofing & Detection & GNSS & Jammertest \\
& Jin et al.\cite{jin2025spoofing} & 2025 & Anti-Spoofing & Detection & GNSS & Real-world data \\
& SDI\cite{tharayil2020sensor} & 2020 & Anti-Spoofing & Detection & IMU & Custom dataset, Real-world data \\

& Liu et al.\cite{liu2022traceability} & 2022 & Anti-Spoofing & Detection & IMU & Custom dataset, Real-world data \\
& CPD-MhIMU\cite{sahu2024acoustic} & 2024 & Anti-Spoofing & Detection & IMU & Custom dataset, Real-world data \\

& SAS\cite{pozzobon2010anti} & 2010 & Anti-Spoofing & Authentication & GNSS & Simulation data \\
& NMA\cite{fernandez2016navigation} & 2016 & Anti-Spoofing & Authentication & GNSS & Real-world data, Simulation data \\
& Chimera\cite{anderson2017chips} & 2017 & Anti-Spoofing & Authentication & GNSS & Real-world data, Simulation data \\

& Wang et al.\cite{wang2017gnss} & 2017 & Anti-Spoofing & Mitigation & GNSS & Simulation data, TEXBAT \\
& Eldosouky et al.\cite{eldosouky2019drones} & 2019 & Anti-Spoofing & Mitigation & GNSS & Simulation data \\
& Zhou et al.\cite{zhou2023anti} & 2023 & Anti-Spoofing & Mitigation & GNSS & TEXBAT \\

& Hong et al.\cite{hong2022esp} & 2022 & Anti-Spoofing & Mitigation & IMU & Simulation data \\
& UNROCKER\cite{jeong2023rocking} & 2023 & Anti-Spoofing & Mitigation & IMU & Real-world data, Simulation data \\
& VIMU\cite{wang2024vimu} & 2024 & Anti-Spoofing & Mitigation & IMU & Real-world data, Simulation data \\

& Zhang et al.\cite{zhang2020vanet} & 2020 & Anti-Spoofing & Mitigation & mmWave Radar & real-world data \\

& Chen et al.~\cite{chen2025safety} & 2025 & Anti-Spoofing & Mitigation & Camera/ADAS & openpilot, CARLA \\

& Swinney et al.\cite{swinney2021gnss} & 2021 & Anti-Jamming & Detection & GNSS & Custom dataset \\
& Spanghero et al.\cite{spanghero2025gnss} & 2025 & Anti-Jamming & Detection & GNSS & Jammertest, Real-world data \\

& Wang et al.\cite{wang2021machine} & 2021 & Anti-Jamming & Mitigation & GNSS & Simulation data \\
\multirow{-30}{*}{\parbox[c]{\linewidth}{\raggedright Sensor\\Defense}} & MFMC\cite{islam2022combating} & 2022 & Anti-Jamming & Mitigation & GNSS & Custom dataset \\

\bottomrule
\end{tabular}
}}
\label{tab:sensor_defense}
\end{table}

Defenses for motion perception apply anti-spoofing and anti-jamming mechanisms to detect and recover from corrupted sensor signals.

\noindent\textbf{Anti-Spoofing Defenses}.\quad
Anti-spoofing defenses detect, authenticate, and mitigate forged signals through three strategies: Detection, Authentication, and Mitigation. Detection identifies malicious signals or anomalous sensor readings by analyzing inconsistencies in signal characteristics, sensor outputs, or cross-modal correlations.

For GNSS, Falco et al.~\cite{falco2018dual} used dual-antenna double-difference dispersion to detect spatially inconsistent signals.
Crowd-GPS-Sec~\cite{jansen2018crowd} analyzes temporal and spatial inconsistencies across a crowd of devices to identify outliers.
DeepSIM~\cite{xue2020deepsim} uses Siamese networks to match ground-level imagery with satellite maps for consistency verification.
DeepPOSE~\cite{jiang2022deeppose} reconstructs vehicle speed and trajectory using ConvLSTM to detect implausible motion patterns.
Iqbal et al.~\cite{iqbal2024deep} introduced a VAE-WGAN framework for zero-day spoofing detection.
PADS~\cite{liu2025gnss} fuses GNSS with Wi-Fi and cellular data to improve robustness.
Jin et al.~\cite{jin2025spoofing} employed anti-jamming antenna arrays coupled with LightGBM for real-time spoofing classification.

For IMU,
SDI~\cite{tharayil2020sensor} introduces cross-sensor consistency checks,
Liu et al.~\cite{liu2022traceability} localized acoustic sources via MLPs,
and CPD-MhIMU~\cite{sahu2024acoustic} deploys heterogeneous IMUs with adaptive EKF fusion.
For ultrasonic ranging,
Xu et al.~\cite{xu2018analyzing} introduced physical shift authentication. SoundFence~\cite{lou2021soundfence} randomizes pulse periods, and SecureTrack~\cite{singh2025securetrack} incorporates EMI monitoring.
For mmWave radar,
Sun et al.~\cite{sun2021control} proposed challenge-response mechanisms,
and Nallabolu and Li~\cite{nallabolu2021frequency} designed hybrid-slope chirps.

Authentication cryptographically verifies the authenticity and integrity of sensor signals to ensure they originate from legitimate sources. For GNSS, SAS~\cite{pozzobon2010anti} introduces encrypted authentication sequences to bind signals to their true source.
NMA~\cite{fernandez2016navigation} demonstrates navigation message authentication for Galileo based on the TESLA protocol.
Chimera~\cite{anderson2017chips} proposes time-binding tags that link signal transmission to precise time slots, preventing replay.

Mitigation actively counteracts or reduces attack effects.
For GNSS,
Wang et al.~\cite{wang2017gnss} introduced MLE-based localization and cancellation,
Eldosouky et al.~\cite{eldosouky2019drones} used cross-UAV localization,
and Zhou et al.~\cite{zhou2023anti} proposed VTL-based correction pipelines.
For IMU,
Hong et al.'s work~\cite{hong2022esp} combines LSTM prediction with CUSUM monitoring,
UNROCKER~\cite{jeong2023rocking} applies denoising autoencoders,
and VIMU~\cite{wang2024vimu} integrates physical modeling with anomaly detection.
For mmWave radar,
Zhang et al.~\cite{zhang2020vanet} introduced VANET-based coordination.
For camera-based ADAS,
Chen et al.~\cite{chen2025safety} evaluated automated and human-driver safety interventions in open-source ADAS against adversarial patch attacks, analyzing intervention conflicts and their resolution to enhance system resilience.

\noindent\textbf{Anti-Jamming Defenses}.\quad
Anti-jamming defenses enhance receiver resilience against intentional or unintentional interference through two strategies: Detection, which identifies jamming signals or interference patterns, and Mitigation, which suppresses or filters interference to restore signal integrity.

For detection of jamming in GNSS, Swinney and Woods~\cite{swinney2021gnss} fused frequency- and time-domain representations using VGG16 with transfer learning to detect jamming.
Spanghero et al.~\cite{spanghero2025gnss} leveraged VTOL UAVs to localize jammers by analyzing spatial signal degradation patterns.

For mitigation of jamming in GNSS, Wang et al.~\cite{wang2021machine} demonstrated that reservoir computing and LSTM networks can reconstruct GPS signals corrupted by jamming.
MFMC~\cite{islam2022combating} develops multi-frequency, multi-constellation receivers to exploit signal diversity and reduce vulnerability.


\subsection{Cross-Modal Perception}
\label{subsec:cross-modal-perception}

Modern embodied systems increasingly rely on multi-sensor fusion (combining cameras, LiDAR, and radar) to achieve robust perception. Cross-modal perception introduces unique vulnerabilities absent from single-modality systems: attackers can exploit inconsistencies between modalities, corrupt fusion mechanisms, or target the weakest channel to compromise the entire perception pipeline. These attacks are particularly dangerous because they can bypass defenses designed for individual modalities. This subsection reviews vulnerabilities and defenses in multi-modal perception models, organized into two parts: \textbf{Adversarial Attacks} (Section~\ref{subsubsec:cross-modal-adversarial-attack}) target sensor fusion pipelines through cross-channel perturbations and temporal misalignment; and \textbf{Adversarial Defenses} (Section~\ref{subsubsec:cross-modal-adversarial-defense}) protect fusion systems through certified robustness, adversarial training, and modality-specific sanitization.

\begin{table}[!tp]
\center
\renewcommand{\arraystretch}{1.15}
\caption{A summary of \textbf{adversarial} attacks and defenses for \textbf{cross-modal perception}.}
\resizebox{1\textwidth}{!}{
\rowcolors{2}{tablegray}{white}
{\setlength{\tabcolsep}{0pt}
\begin{tabular}{@{} >{\cellcolor{white}}p{.12\textwidth} p{.20\textwidth} p{.06\textwidth}
               p{.18\textwidth} p{.18\textwidth} p{.16\textwidth} p{.25\textwidth}@{}}

\toprule
\rowcolor{perception-table-blue}
\cellcolor{perception-table-blue}\textbf{Attack/Defense} & \textbf{Method} & \textbf{Year} & \textbf{Category} & \textbf{Subcategory} & \textbf{Target Model} & \textbf{Dataset} \\

\midrule

& DejaVu\cite{hou_dejavu_2025} & 2025 & Digital Attack & Temporal Misalignment & Fusion 3D Detector & nuScenes \\

& Cao et al.\cite{cao_invisible_2021} & 2021 & Physical Attack & Adversarial Object & Fusion 3D Detector & KITTI, nuScenes \\
& Hallyburton et al.\cite{hallyburton_frustum_2022} & 2022 & Physical Attack & LiDAR Spoofing & Fusion 3D Detector & KITTI, nuScenes \\
& Li et al.\cite{li_malicious_2024} & 2024 & Physical Attack & Adversarial Object & Fusion 3D Detector & nuScenes \\
\multirow{-5}{*}{\parbox[c]{\linewidth}{\raggedright Adversarial\\Attack}} & Iranmanesh et al.\cite{iranmanesh2026typographic} & 2026 & Physical Attack & Typographic & VLM & HomeRobot (Habitat) \\

\midrule

& Wang et al.\cite{wang_adversarial_fusion_2022} & 2022 & Robust Training & Adversarial Training & Fusion 3D Detector & KITTI \\

& MMCert\cite{wang_mmcert_2024} & 2024 & Robust Inference & Certified Defense & Multi-Modal Model & Kinetics-400, Food-101 \\
& El-Fatyany~\cite{el2026robust} & 2026 & Robust Inference & Input Moderation & Fusion 3D Detector & nuScenes, KITTI \\
\multirow{-4}{*}{\parbox[c]{\linewidth}{\raggedright Adversarial\\Defense}} & BlueSuffix\cite{zhao2025bluesuffix} & 2025 & Robust Inference & Input Moderation & Vision-Language Model & MM-SafetyBench, RedTeam-2K \\

\bottomrule
\end{tabular}
}}
\label{tab:cross_modal_attack_and_defense}
\end{table}


\subsubsection{Adversarial Attacks}\label{subsubsec:cross-modal-adversarial-attack}

Multi-sensor fusion combines modalities to improve perception accuracy and robustness, but the fusion process itself introduces attack surfaces at the feature alignment, projection, and decision stages.

\noindent\textbf{Digital Attacks}.\quad
  Cao et al.~\cite{cao_invisible_2021} showed that adversarial perturbations can deceive multi-sensor fusion (camera + LiDAR) perception in autonomous driving by exploiting the cross-modal alignment between the two channels.
  DejaVu~\cite{hou_dejavu_2025} exploits synchronization dependencies between sensors: a single-frame LiDAR delay causes 88.5\% mAP degradation in 3D detection, while three-frame camera delays reduce MOT performance by 73\% MOTA, revealing that temporal misalignment is a critical but underprotected attack surface.
  UMAB~\cite{li2026umab} crafts untargeted adversarial perturbations that break image-text modality alignment in large vision-language models.

\noindent\textbf{Physical Attacks}.\quad
  Physical attacks on cross-modal perception exploit the geometric correspondence between 2D images and 3D point clouds.
  Extending the digital fusion attack introduced above, Cao et al.~\cite{cao_invisible_2021} presented the first physical-world attack that simultaneously fools both camera and LiDAR channels using 3D-printed adversarial objects, demonstrating that physically realizable perturbations can evade all tested fusion-based detectors.
  Hallyburton et al.~\cite{hallyburton_frustum_2022} proposed the frustum attack that compromises all eight widely-used perception algorithms (both LiDAR-only and camera-LiDAR fusion) through black-box LiDAR spoofing, while remaining stealthy to existing defenses.
Li et al.~\cite{li_malicious_2024} conducted the first comprehensive study attacking all three sensing modalities simultaneously (camera, LiDAR, and radar) using a single adversarial object that exploits cross-modal geometric constraints.
Iranmanesh and Liu~\cite{iranmanesh2026typographic} extend typographic attacks to robot manipulation: printed scene text overrides CLIP-based VLM perception and propagates through HomeRobot's sense-plan-act pipeline.


\subsubsection{Adversarial Defenses}\label{subsubsec:cross-modal-adversarial-defense}

Defenses for cross-modal perception systems exploit redundancy between modalities and enforce consistency constraints to detect or mitigate multi-channel attacks.

\noindent\textbf{Robust Training}.\quad
  Adversarial training for fusion models must account for cross-channel externalities.
  Wang et al.~\cite{wang_adversarial_fusion_2022} discovered that single-channel adversarial training can reduce robustness to attacks on other channels (a cross-channel externality) and propose multi-channel adversarial training as a countermeasure.

\noindent\textbf{Robust Inference}.\quad
  Wang et al.~\cite{wang_mmcert_2024} proposed MMCert, the first certified defense against adversarial attacks on multi-modal models, deriving provable robustness bounds through randomized smoothing across modalities.
El-Fatyany~\cite{el2026robust} applied modality-specific input sanitization before sensor fusion, preventing adversarial patches on one modality from corrupting the joint representation.
BlueSuffix~\cite{zhao2025bluesuffix} pairs visual and textual purifiers with a reinforcement-trained suffix generator to defend vision-language models against jailbreak attacks while preserving cross-modal alignment.


\section{Cognition}
\label{sec:cognition}

Cognition forms the second layer, encompassing perception while adding semantic interpretation and logical inference, expanding the agent's capability from sensing to understanding. This expansion introduces new attack surfaces beyond perceptual corruption: adversaries can now manipulate how agents interpret instructions, exploit world model hallucinations, and hijack reasoning chains. With LLMs and VLMs increasingly serving as the ``brain'' of embodied systems, the cognitive layer becomes both the engine of intelligent behavior and a high-value target for adversarial manipulation. \textbf{Instruction Understanding} (Section~\ref{subsec:instruction-understanding}) addresses jailbreak attacks that manipulate natural language instructions to bypass safety constraints and induce harmful intent understanding, along with corresponding defenses and benchmarks; \textbf{World Model} (Section~\ref{subsec:world-model}) examines hallucination in scene understanding and rule violations in predictive models; and \textbf{Reasoning} (Section~\ref{subsec:reasoning}) covers chain-of-thought hijacking attacks that corrupt reasoning.

\begin{table}[!tp]
\center
\renewcommand{\arraystretch}{1.15}
\caption{A summary of \textbf{cognitive attacks and defenses} for \textbf{embodied cognition}.}
\resizebox{1\textwidth}{!}{

\rowcolors{2}{tablegray}{white}
{\setlength{\tabcolsep}{0pt}
\begin{tabular}{@{}>{\cellcolor{white}}p{.12\textwidth} p{.20\textwidth} p{.06\textwidth}
               p{.18\textwidth} p{.18\textwidth} p{.16\textwidth} p{.25\textwidth}@{}}
\toprule
\rowcolor{cognition-table-green}
\cellcolor{cognition-table-green}\textbf{Attack/Defense} & \textbf{Method} & \textbf{Year} & \textbf{Category} & \textbf{Subcategory} & \textbf{Target Model} & \textbf{Dataset/Benchmark} \\
\midrule

& CHAI~\cite{burbano2025chai} & 2025 & Instruction Attack & Jailbreak Attacks & Embodied LLM & Simulation, Real World \\
& BadNAVer~\cite{bai2025badnaver} & 2025 & Instruction Attack & Jailbreak Attacks & Navigation Agent & Matterport3D \\
& Seeing-No-Evil~\cite{li2026seeing} & 2026 & Instruction Attack & Jailbreak Attacks & VLM & Custom \\
& SDoS~\cite{steinberg2026semantic} & 2026 & Instruction Attack & Jailbreak Attacks & Embodied LLM & Custom \\
& Chen et al.~\cite{chen2024multiobject} & 2024 & World Model Attack & Hallucination & VLM & Custom \\
& Tao et al.~\cite{chakraborty_heal_2025} & 2025 & World Model Attack & Hallucination & VLM & Custom \\
& HRSSM~\cite{sun2024hrssm} & 2024 & World Model Attack & Rule Violation & World Model & Custom \\
& Wen et al.~\cite{wen2025scalable} & 2025 & World Model Attack & Rule Violation & World Model & Custom \\
& TRAP~\cite{duan2026trap} & 2026 & World Model Attack & Rule Violation & World Model & Custom \\
& PhysCond-WMA~\cite{guo2026physcond} & 2026 & World Model Attack & Rule Violation & Diffusion WM & Custom \\
& CtrlAttack~\cite{xu2026ctrlattack} & 2026 & World Model Attack & Rule Violation & World Model & Custom \\
& H-CoT~\cite{kuo2025hcot} & 2025 & Reasoning Attack & CoT Hijacking & Reasoning Model & Custom \\
\multirow{-13}{*}{\parbox[c]{\linewidth}{\raggedright Cognitive\\Attack}} & Altered Thoughts~\cite{trinh2026altered} & 2026 & Reasoning Attack & CoT Hijacking & VLA & Custom \\

\midrule

& J-DAPT~\cite{marchiori2025preventing} & 2025 & Instruction Defense & Jailbreak Defenses & Embodied LLM & nuScenes, Maritime, Quadruped \\
& Hafez et al.~\cite{abuduweili2025safe} & 2025 & Instruction Defense & Jailbreak Defenses & Embodied LLM & Gazebo, Real World \\
& Ravichandran et al.~\cite{ravichandran2026safety} & 2026 & Instruction Defense & Jailbreak Defenses & LLM-Enabled Robot & Custom \\
& MASH-VLM~\cite{bae2025mash} & 2025 & World Model Defense & Hallucination & World Model & Custom \\
& SafeDreamer~\cite{huang2024safedreamer} & 2024 & World Model Defense & Rule Violation & World Model & Safety Gymnasium \\
& Drive-WM~\cite{wang2024drivewm} & 2024 & World Model Defense & Rule Violation & World Model & nuScenes \\
& VL-SAFE~\cite{Qu2025VLMSAFEVM} & 2025 & World Model Defense & Rule Violation & World Model & AD Simulation \\
& Surprise Recognition~\cite{zollicoffer2025surprise} & 2025 & World Model Defense & Rule Violation & World Model & Custom \\
& SafeDream~\cite{yan2026safedream} & 2026 & World Model Defense & Rule Violation & LLM & Custom \\
\multirow{-10}{*}{\parbox[c]{\linewidth}{\raggedright Cognitive\\Defense}} & HomeGuard~\cite{lu2026homeguard} & 2026 & World Model Defense & Contextual Risk Mitigation & VLM & Custom \\

\bottomrule
\end{tabular}
}}
\label{tab:cognition}
\end{table}


\subsection{Instruction Understanding}
\label{subsec:instruction-understanding}

Embodied agents rely on natural language instructions to bridge human intent and physical action. Unlike text-only chatbots, where misinterpretation produces merely incorrect text, failures in embodied instruction understanding can trigger unsafe physical behaviors: collisions, property damage, or human injury. This subsection surveys jailbreak attacks that bypass safety guardrails to induce harmful intent understanding, corresponding defenses, and benchmarks for evaluating instruction safety. As embodied agents extend beyond text and vision to direct auditory input, recent surveys on Large Audio Language Models~\cite{luo2026audio} indicate that LALMs will inherit and amplify these same instruction-attack surfaces, motivating proactive coverage of potential audio-borne jailbreak threats.



\noindent\textbf{Jailbreak Attacks}.\quad Jailbreak attacks manipulate language inputs to circumvent safety constraints, causing embodied agents to execute harmful physical actions that would normally be refused.
  CHAI~\cite{burbano2025chai} optimizes adversarial commands against the LVLM command layer of physical agents in the white-box setting.
  Black-box attacks at the sentence level craft semantically meaningful prompts that exploit weaknesses in safety filters.
  BadNAVer~\cite{bai2025badnaver} demonstrates that jailbreaks in embodied navigation directly trigger unsafe physical actions.
  Seeing No Evil~\cite{li2026seeing} shows that attention-guided visual perturbations can blind VLMs to safety instructions even when the safety prompt remains literally present.
  Steinberg and Gal~\cite{steinberg2026semantic} introduce semantic denial-of-service: short, safety-plausible audio injections weaponize an Embodied LLM's own safety reasoning to stall task execution.


\noindent\textbf{Jailbreak Defenses}.\quad Defenses against jailbreaks aim to detect or block adversarial instructions before they are translated into physical actions.
  J-DAPT~\cite{marchiori2025preventing} introduces multimodal domain adaptation for robotic jailbreak detection, using vision-language alignment to identify adversarial instructions before they reach the control pipeline.
  Hafez et al.~\cite{abuduweili2025safe} integrated reachability analysis with Embodied LLMs, providing formal safety guarantees by rejecting instructions whose predicted outcomes violate verified safety envelopes.
  Ravichandran et al.~\cite{ravichandran2026safety} propose a runtime safety guardrail for Embodied LLMs that addresses both LLM contextual vulnerabilities and downstream physical risks before execution.

\noindent\textbf{Benchmarks}.\quad IndustryEQA~\cite{li2024industryeqa} extends embodied QA to safety-critical industrial environments with hazard recognition and compliance verification.
  SQA3D~\cite{ma2023sqa3d} introduces situated question answering in 3D scenes requiring spatial and commonsense reasoning.
  MMRO~\cite{li2024mmro} benchmarks multimodal LLMs as cognitive engines for in-home robotics, revealing safety as a persistent weakness.
  MetaVQA~\cite{wang2025metavqa} fine-tunes VLMs with embodied scene data to improve spatial reasoning in safety-critical driving simulations.
  AGENTSAFE~\cite{liu2025agentsafe} evaluates embodied agent vulnerability to jailbreaks across adversarial scenarios with hazardous tasks.
  EmbodiedBench~\cite{yang2025embodiedbench} provides an evaluation framework with explicit safety metrics.


\subsection{World Model}
\label{subsec:world-model}

World models enable embodied agents to predict future states, reason about physical dynamics, and evaluate action consequences before execution. When these internal representations diverge from physical reality (through hallucination, sim-to-real gaps, or prediction failures), agents make decisions based on false beliefs about their environment, with potentially catastrophic physical consequences.
This subsection surveys threats to world model safety organized by two failure modes: \textbf{Hallucination in Scene Understanding} addresses VLM and world model hallucination that generates nonexistent objects, spatial relations, or actions; and \textbf{Rule Violation} covers predictive model failures and emergent misalignment that cause agents to violate physical laws, domain-specific rules, or safety constraints.


\noindent\textbf{Hallucination in Scene Understanding}.\quad VLM hallucination (generating descriptions of objects, spatial relations, or actions that do not exist in the physical scene) poses acute risks when these models serve as the perceptual backbone of embodied agents.
  Chen et al.~\cite{chen2024multiobject} showed that multi-object hallucination in VLMs remains pervasive in embodied scene understanding,
  and Tao et al.~\cite{chakraborty_heal_2025} demonstrated that hallucination is especially severe in visual-text tasks for embodied agents.
  MASH-VLM~\cite{bae2025mash} disentangles spatial and temporal tokens via DST-attention to reduce action-scene misattribution in world models.
  Beyond VLM hallucination, world models used for internal simulation exhibit distinct pathologies.
Baraldi et al.~\cite{baraldi2025safety} identified scene-generation pathology criteria spanning temporal consistency, physical conformity, and condition consistency, confirming a systematic safety gap in current world model predictions.
ContrAR~\cite{xiu2026contrar} benchmarks VLM hallucination in AR with contradictory-overlay videos, exposing an inability to resolve observed-versus-asserted reality.


\noindent\textbf{Rule Violation}.\quad 
  Beyond hallucination, world models can systematically violate physical laws, domain-specific rules, and safety constraints: failures that directly translate into unsafe embodied behavior.
  Predictive model failures occur when learned dynamics models compound errors over long horizons, producing increasingly dangerous predictions.
  Li et al.~\cite{li2025worldmodelsurvey} surveyed world model architectures across RSSM, Transformer, diffusion, and other paradigms, identifying error accumulation, distribution shift, and physical consistency as critical safety challenges; Parmar~\cite{parmar2026worldmodelrisks} maps the safety, security, and cognitive risks of world models onto MITRE ATLAS and OWASP categories.
  TRAP~\cite{duan2026trap} shows that world-model planners are vulnerable to a tail-aware ranking attack: small perturbations to the value-ranking tail of imagined trajectories steer the planner toward unsafe actions while leaving nominal accuracy untouched.
  PhysCond-WMA~\cite{guo2026physcond} perturbs the physical conditioning signals of diffusion world models in a two-stage guidance scheme, degrading downstream planner performance.
  CtrlAttack~\cite{xu2026ctrlattack} extends this to image-to-video diffusion world models with a unified attack on action-conditioned state transitions, breaking the controllability that downstream policies rely on.
  HRSSM~\cite{sun2024hrssm} learns latent dynamic robust representations to improve world model resilience to distribution shift.
  Tseng et al.~\cite{wen2025scalable} found that compounding errors in video prediction rollouts limit long-horizon reliability.
  Safety-aware world models explicitly incorporate constraints into the prediction-planning loop:
  SafeDreamer~\cite{huang2024safedreamer} integrates Lagrangian-based safety constraints into the Dreamer framework,
  VL-SAFE~\cite{Qu2025VLMSAFEVM} supervises world models using VLM-derived safety scores for autonomous driving,
  and Drive-WM~\cite{wang2024drivewm} uses multi-view diffusion for safer trajectory selection, though training planners on WM-generated data creates a cascading risk where pathologies propagate to downstream policies.
  Surprise Recognition~\cite{zollicoffer2025surprise} leverages the world model's own intrinsic surprise signal to detect out-of-distribution distractors before they translate into unsafe actions, providing a model-internal anomaly channel that complements explicit safety constraints.
  SafeDream~\cite{yan2026safedream} ports the world-model defense idiom to LLM jailbreaks: a safety world model tracks cumulative safety-alignment erosion across multi-turn conversation, enabling proactive detection before harmful content is generated.

\noindent\textbf{Contextual Risk Mitigation}.\quad
  Beyond defending against in-model hallucination and rule violation, a complementary defense thread targets \emph{contextual} safety risk, where benign instructions become hazardous due to subtle environmental states that rule-based or prompt-engineered safeguards miss.
  HomeGuard~\cite{lu2026homeguard} pairs attention-anchored perception with semantic judgment for household risk detection, with visual anchors doubling as spatial constraints for downstream planners.
  HazardArena~\cite{chen2026hazardarena} provides a paired-scenario benchmark that isolates contextual semantic risk: safe/unsafe twins share objects, layout, and action requirements but differ only in the semantic context that renders an action unsafe.
  Sermanet et al.~\cite{sermanet2025asimov} release the ASIMOV benchmark and generate robot constitutions, framing semantic safety as a first-class evaluation target for VLM-controlled robots.


\subsection{Reasoning}
\label{subsec:reasoning}

Reasoning addresses vulnerabilities in the processes embodied agents use for multi-step problem solving.

\noindent\textbf{Chain-of-Thought Hijacking}.\quad Chain-of-thought (CoT) reasoning enables transparent multi-step deliberation but exposes intermediate reasoning steps to adversarial manipulation.
  H-CoT~\cite{kuo2025hcot} demonstrates that inserting adversarial steps into chain-of-thought traces can hijack reasoning models toward harmful conclusions.
  Altered Thoughts~\cite{trinh2026altered} probes entity-substitution vulnerabilities in chain-of-thought VLA manipulation policies, where corrupted reasoning steps degrade task execution.

\section{Planning}
\label{sec:planning}

Planning forms the third layer, encompassing perception and cognition while adding the generation of action sequences, expanding the agent's capability from understanding to decision-making. This expanded capability extends the attack surface from passive interpretation to active goal pursuit: adversaries can now corrupt task decomposition, hijack trajectory optimization, and manipulate multi-agent coordination strategies. Modern embodied planners increasingly leverage LLMs for high-level task decomposition while relying on traditional methods for low-level motion generation, creating a heterogeneous attack surface spanning both learned and classical components. Because planning spans multiple abstraction levels (from symbolic task decomposition through trajectory optimization to multi-agent coordination), each level admits qualitatively distinct attack vectors. This section organizes planning into three subsections: \textbf{Task Planning} (Section~\ref{subsec:Task_Planning}) addresses vulnerabilities in LLM-based task decomposition, chain-of-thought reasoning, and goal specification (this includes jailbreak attacks that manipulate planners into generating harmful action sequences, as well as goal hijacking and reward hacking); \textbf{Trajectory Planning} (Section~\ref{subsec:Motion_Planning}) covers threats to trajectory prediction and path planning, including adversarial perturbation of collision avoidance systems; and \textbf{Multi-Agent Planning} (Section~\ref{subsec:Multi_Agent_Planning}) discusses planning-time coordination challenges including distributed task allocation, consensus failures, subgoal manipulation, and goal conflicts among cooperative agents. Note that multi-agent planning focuses on the \textbf{planning phase} (who does what), while execution-time collaboration is covered in Section~\ref{sec:action}.


\subsection{Task Planning}
\label{subsec:Task_Planning}

Task planning translates high-level objectives into actionable subgoal sequences, increasingly through LLM-based decomposition, chain-of-thought reasoning, and tool use. These capabilities enable flexible, generalizable planning but also introduce novel attack surfaces: adversaries can manipulate task specifications to induce unsafe decompositions, hijack reasoning chains to subvert intended goals, or exploit jailbreak vulnerabilities to elicit harmful plans. This subsection examines adversarial attacks on classical optimization-based planners and modern LLM-based task decomposition, jailbreak attacks that manipulate planners into generating harmful action sequences, backdoor attacks that implant hidden triggers, jailbreak defenses that prevent malicious instruction injection, and emerging risks including unforced constraints violation.

\begin{table}[!tp]
\center
\renewcommand{\arraystretch}{1.15}
\caption{A summary of \textbf{attacks and defenses} for \textbf{embodied planning}.}
\resizebox{1\textwidth}{!}{

\rowcolors{2}{tablegray}{white}
{\setlength{\tabcolsep}{0pt}
\begin{tabular}{@{}>{\cellcolor{white}}p{.15\textwidth} p{.20\textwidth} p{.06\textwidth}
               p{.16\textwidth} p{.21\textwidth} p{.21\textwidth} p{.25\textwidth}@{}}
\toprule
\rowcolor{planning-table-orange}
\cellcolor{planning-table-orange}\textbf{Attack/Defense} & \textbf{Method} & \textbf{Year} & \textbf{Category} & \textbf{Subcategory} & \textbf{Target Model} & \textbf{Environment} \\
\midrule

& Zhang et al.~\cite{zhang2022adversarial} & 2022 & White-box & Optim.-Based Attack & Trajectory Planner & Apolloscape, NGSIM, nuScenes \\
& AdvDO~\cite{cao2022advdo} & 2022 & White-box & Optim.-Based Attack & Trajectory Planner & nuScenes \\
& KING~\cite{hanselmann2022king} & 2022 & White-box & Optim.-Based Attack & Trajectory Planner & CARLA \\
& ADvLM~\cite{zhang2024visual} & 2024 & White-box & Optim.-Based Attack & Trajectory Planner & nuScenes, DriveLM \\
& Adv-GAN~\cite{fan2024adversarial} & 2024 & White-box & Model-Based Attack & Trajectory Planner & Apolloscape, NGSIM, nuScenes \\

& Islam et al.~\cite{islam2024malicious} & 2024 & Black-box & Optim.-Based Attack & Task Planner & EnvLarge-10 \\
& AdvSim~\cite{wang2021advsim} & 2021 & Black-box & Optim.-Based Attack & Trajectory Planner & UrbanScenarios \\
& STRIVE~\cite{rempe2022generating} & 2022 & Black-box & Optim.-Based Attack & Trajectory Planner & nuScenes \\
& Zheng et al.~\cite{zheng2023robustness} & 2023 & Black-box & Optim.-Based Attack & Trajectory Planner & nuScenes, Argoverse \\
& UTCIA~\cite{bai2025universal} & 2025 & Black-box & Optim.-Based Attack & Trajectory Planner & Trajectory Datasets \\
& Avatar~\cite{liu2025avatar} & 2025 & Black-box & Optim.-Based Attack & Trajectory Planner & Waymax \\
& LC~\cite{ding2020learning} & 2020 & Black-box & Model-Based Attack & Trajectory Planner & CARLA \\
& NADE~\cite{feng2021intelligent} & 2021 & Black-box & Model-Based Attack & Trajectory Planner & CARLA \\
& AdvDiffuser~\cite{xie2024advdiffuser} & 2024 & Black-box & Model-Based Attack & Trajectory Planner & nuScenes \\
& Szvoren et al.~\cite{szvoren2025adversarial} & 2025 & Black-box & Physical Attack & Trajectory Planner & Gazebo, Unitree Go1 \\
\multirow{-16}{*}{\parbox[c]{\linewidth}{\raggedright Adversarial\\Attack}} & AFM~\cite{zeng2026afm} & 2026 & White-box & Optim.-Based Attack & E2E AD Model & CARLA, Bench2Drive \\

\midrule

& EIRAD~\cite{liu2024exploring} & 2024 & White-box & Word-Level & Task Planner & AI2-THOR \\
& POEX~\cite{lu2024poex} & 2024 & White-box & Word-Level & Task Planner & CoppeliaSim, RLBench, Real world \\

& RoboPAIR~\cite{robey2024jailbreaking} & 2024 & Black-box & Sentence-Level & Task Planner & nuScenes, Real world \\
& BADROBOT~\cite{zhang2024badrobot} & 2024 & Black-box & Sentence-Level & Task Planner & RLBench, Real world \\
& Wen et al.~\cite{wen2024secure} & 2024 & Black-box & Sentence-Level & Trajectory Planner & Touchdown, Map2seq \\
& Zhang et al.~\cite{zhang2024study} & 2024 & Black-box & Sentence-Level & Trajectory Planner & EyeSim VR \\
\multirow{-7}{*}{\parbox[c]{\linewidth}{\raggedright Jailbreak\\Attack}} & PINA~\cite{liu2026pina} & 2026 & Black-box & Sentence-Level & Trajectory Planner & Indoor/Outdoor Navigation \\

\midrule

& CBA~\cite{liu2024compromising} & 2024 & Data Poisoning & Multi-Modal Triggers & Task Planner & ProgPrompt, VoxPoser, VisProg, Real \\
& BALD~\cite{jiao2024can} & 2024 & Training Manipulation & Multi-Modal Triggers & Task Planner & HighwayEnv, CARLA, nuScenes \\
\multirow{-3}{*}{\parbox[c]{\linewidth}{\raggedright Backdoor\\Attack}} & Robo-Troj~\cite{nahian2025robotroj} & 2025 & Training Manipulation & Word-Level Triggers & Task Planner & VirtualHome, AI2-THOR \\

\midrule

& Thumm et al.~\cite{thumm2023reducing} & 2023 & Robust Training & Safety Constraints & Trajectory Planner & OpenAI safety gym \\
& AR-ICRL~\cite{xu2024robust} & 2024 & Robust Training & Safety Constraints & Trajectory Planner & Blocked Half-Cheetah/Ant/Walker \\
& Yurtsever et al.~\cite{yurtsever2019risky} & 2019 & Robust Inference & Output Moderation & Trajectory Planner & NuDrive \\
& SMPC~\cite{brudigam2021stochastic} & 2021 & Robust Inference & Output Moderation & Trajectory Planner & Matlab \\
\multirow{-5}{*}{\parbox[c]{\linewidth}{\raggedright Adversarial\\Defense}} & CSP-GAN-LSTM~\cite{meng2023vehicle} & 2023 & Robust Inference & Output Moderation & Trajectory Planner & NGSIM, highD \\

\midrule

& SafeEmbodAI~\cite{zhang2024safeembodai} & 2024 & Robust Inference & Safe Prompt & Trajectory Planner & EyeSim VR \\
& NPE~\cite{wen2024secure} & 2024 & Robust Inference & Safe Prompt & Trajectory Planner & Touchdown, Map2seq \\
& SafePlan~\cite{obi2025safeplan} & 2025 & Robust Inference & Safe Prompt & Task Planner & AI2-THOR, Synthetic \\
& J-DAPT~\cite{marchiori2025preventing} & 2025 & Robust Inference & Jailbreak Detection & Trajectory Planner & nuScenes, Maritime, Quadruped \\
& RoboSafe~\cite{wang2025robosafe} & 2025 & Robust Inference & Runtime Safety & Task/Traj Planner & AI2-THOR, MetaWorld \\
& CEE~\cite{yang2025cee} & 2025 & Robust Inference & Representation Eng. & Task Planner & EI Safety Benchmarks \\
\multirow{-7}{*}{\parbox[c]{\linewidth}{\raggedright Jailbreak\\Defense}} & Zhang et al.~\cite{zhang2025enhancing} & 2025 & Robust Inference & Runtime Safety & Trajectory Planner & EyeSim, Real world \\

\bottomrule
\end{tabular}
}}
\label{tab:planning}
\end{table}

\begin{table}[!tp]
\center
\renewcommand{\arraystretch}{1.15}
\caption{A summary of \textbf{emerging risks} for \textbf{embodied planning}.}
\resizebox{1\textwidth}{!}{

\rowcolors{2}{tablegray}{white}
{\setlength{\tabcolsep}{0pt}
\begin{tabular}{@{}>{\cellcolor{white}}p{.15\textwidth} p{.20\textwidth} p{.06\textwidth}
               p{.16\textwidth} p{.21\textwidth} p{.21\textwidth} p{.25\textwidth}@{}}
\toprule
\rowcolor{planning-table-orange}
\cellcolor{planning-table-orange}\textbf{Risk} & \textbf{Method} & \textbf{Year} & \textbf{Category} & \textbf{Subcategory} & \textbf{Target Model} & \textbf{Environment} \\
\midrule

& Strobel and Ferrer~\cite{strobel_blockchain_2020} & 2020 & Multi-Agent & Byzantine Faults & Multi-Agent Planner & Physical Robots \\
& Blumenkamp et al.~\cite{blumenkamp_emergence_2021} & 2021 & Multi-Agent & Byzantine Faults & Multi-Agent Planner & Coverage, Path Planning \\
& He et al.~\cite{he_redteaming_2025} & 2025 & Multi-Agent & Byzantine Faults & Multi-Agent Planner & Multi-Agent Frameworks \\
& Zhou et al.~\cite{zhou2026ham3} & 2026 & Multi-Agent & Byzantine Faults & Multi-Agent Planner & GQA \\
& Choudhury et al.~\cite{choudhury_dynamic_2022} & 2022 & Multi-Agent & Goal Conflicts & Multi-Agent Planner & Task Allocation \\
& Bahrami and Jafarnejadsani~\cite{gao_multirobot_2025} & 2025 & Multi-Agent & Goal Conflicts & Multi-Agent Planner & Relative Localization \\
& Li et al.~\cite{li_resilient_2020} & 2020 & Multi-Agent & Potential Defenses & Multi-Agent Planner & Multi-Robot Networks \\
& Strobel et al.~\cite{strobel_swarms_2023} & 2023 & Multi-Agent & Potential Defenses & Multi-Agent Planner & 24 Physical Robots \\
& Lee and Panagou~\cite{lee_distributed_2025} & 2025 & Multi-Agent & Potential Defenses & Multi-Agent Planner & Distributed Control \\
\multirow{-10}{*}{\parbox[c]{\linewidth}{\raggedright Emerging\\Risks}} & Gandhi et al.~\cite{gandhi_roborebound_2025} & 2025 & Multi-Agent & Potential Defenses & Multi-Agent Planner & Physical Robots \\

\bottomrule
\end{tabular}
}}
\label{tab:planning_emerging}
\end{table}



\subsubsection{Adversarial Attacks}
\label{subsubsec:Task_Adversarial_Attacks}

Adversarial attacks on task planners primarily target black-box threat models where attackers perturb inputs or environmental states without model access.
Islam et al.~\cite{islam2024malicious} demonstrated that small visual perturbations can mislead CLIP-based vision--language navigation systems into attacker-defined paths, highlighting vulnerabilities in vision-grounded task planners.
Vemprala and Kapoor~\cite{vemprala_adversarial_planners_2021} showed that adversarial state configurations can degrade eigenstructure in classical optimization-based planners, forcing failure or excessive computation.
AFM~\cite{zeng2026afm} generates adversarial perturbations through a one-step flow-matching velocity field, transferring across vision-language-action and modular end-to-end driving stacks to induce hazardous maneuvers.


\subsubsection{Jailbreak Attacks}
\label{subsubsec:Jailbreak_Task_Planners}

Jailbreak attacks manipulate LLM-based planners by crafting inputs that bypass safety guardrails to elicit harmful task decompositions.

\noindent\textbf{White-box Attacks}.\quad
White-box attacks leverage gradient-based optimization to craft adversarial suffixes or token-level perturbations.
EIRAD~\cite{liu2024exploring} adapts gradient-based suffix optimization, appending adversarial tokens to benign inputs so that untargeted variants divert the agent from the intended task while targeted variants steer outputs toward harmful goals.
POEX~\cite{lu2024poex} improves suffix quality through a mutator--selector--evaluator loop that balances jailbreak success with action executability, validating attacks on real robots and introducing Harmful-RLBench, a benign--harmful task suite for sim-to-real safety evaluation.

\noindent\textbf{Black-box Attacks}.\quad
Black-box attacks manipulate instructions without model access, operating through prompt engineering and semantic manipulation.
RoboPAIR~\cite{robey2024jailbreaking} automates jailbreak generation by extending PAIR with robot-specific prompts and a syntax checker that enforces API-compliant action formats.
BADROBOT~\cite{zhang2024badrobot} identifies three embodiment-specific attack surfaces: contextual jailbreak, safety misalignment, and conceptual deception, showing that Embodied LLMs can be coerced into unsafe actions using in-the-wild prompts across both simulation and physical platforms.



\subsubsection{Backdoor Attacks}
\label{subsubsec:Backdoor_Task_Planners}

Backdoor attacks implant hidden triggers into task planners during pre-training or fine-tuning while preserving clean-task performance, enabling malicious behaviors to activate only when the trigger appears at deployment.
CBA~\cite{liu2024compromising} poisons in-context demonstrations using combined textual and visual triggers to activate harmful behaviors at deployment.
BALD~\cite{jiao2024can} taxonomizes backdoor pathways in LLM-based planners, covering word-level triggers, scenario manipulation, and RAG-based knowledge injection.
Robo-Troj~\cite{nahian2025robotroj} demonstrates backdoor attacks via poisonous fine-tuning of soft-prompts to inject malicious plans when trigger words appear in task descriptions.


\subsubsection{Jailbreak Defenses}
\label{subsubsec:Task_Planning_Defenses}

Jailbreak defenses for task planning focus on preventing malicious instruction injection and ensuring that LLM-based planners adhere to safety constraints during deployment. Current strategies employ safe prompting, jailbreak detection, runtime validation, and representation engineering techniques.

SafeEmbodAI~\cite{zhang2024safeembodai} integrates safe prompting, state management, and safety validation modules to verify and sanitize actions before execution, preventing unsafe navigation behaviors.
NPE~\cite{wen2024secure} employs structured templates such as Chain-of-Thought and Plan-and-Solve to improve planner robustness against text-based manipulations.
J-DAPT~\cite{marchiori2025preventing} integrates textual and visual embeddings via attention-based fusion and adapts general jailbreak datasets to robotics-specific domains for multimodal detection.
RoboSafe~\cite{wang2025robosafe} combines backward reflective reasoning over recent trajectories with forward predictive reasoning from safety memory to generate executable predicate-based safety logic.
Concept Enhancement Engineering (CEE)~\cite{yang2025cee} steers internal representations toward safe concepts to defend against jailbreak attacks in embodied AI systems.
SafePlan~\cite{obi2025safeplan} interposes formal logic verification at multiple points in the CoT pipeline to filter unsafe robotic task plans.
A unified framework for security and safety in Embodied LLMs~\cite{zhang2025enhancing} combines interpretable prompting, state-aware planning, and real-time validation to jointly address safety and prompt-injection security in mobile Embodied LLMs.


\subsection{Trajectory Planning}
\label{subsec:Motion_Planning}

Trajectory planning generates continuous trajectories that satisfy kinematic, dynamic, and safety constraints. Modern approaches combine learned trajectory prediction with classical path planning and collision avoidance, creating hybrid systems that inherit vulnerabilities from both paradigms. Attacks on trajectory planners can induce collisions, amplify prediction errors, or degrade trajectory quality through adversarial perturbations to perception inputs, prediction models, or planning algorithms. This subsection examines adversarial attacks on trajectory prediction and path planning, jailbreak attacks on LLM-based navigation planners, and adversarial defenses through robust training and inference.


\subsubsection{Adversarial Attacks}
\label{subsubsec:Motion_Adversarial_Attacks}

Adversarial attacks on trajectory planners exploit weaknesses in trajectory prediction and generation models, driving agents toward unsafe maneuvers or systematically amplifying prediction errors. These attacks fall into two main classes: white-box attacks, which use gradient-based methods to craft perturbations on trajectories or map context with full model access, and black-box attacks, which leverage query-based optimization or generative models to synthesize realistic adversarial scenarios without model access.

\noindent\textbf{White-box Attacks}.\quad
White-box attacks leverage model gradients to craft precise adversarial perturbations on trajectories or context maps.
Zhang et al.~\cite{zhang2022adversarial} perturbed nominal vehicle trajectories to maximize prediction error in trajectory forecasting models.
AdvDO~\cite{cao2022advdo} uses a differentiable dynamics model to construct plausible adversarial trajectories that mislead downstream planners.
KING~\cite{hanselmann2022king} employs a differentiable kinematic model to efficiently search for critical but feasible scenes.
Adv-GAN~\cite{fan2024adversarial} employs an LSTM-based generator to produce adversarial trajectory perturbations and refines them using model predictive control under realism and safety constraints.
ADvLM~\cite{zhang2024visual} addresses textual instruction variability and time-series visual scenarios in VLM-based autonomous driving through Semantic-Invariant Induction for diverse prompt libraries and Scenario-Associated Enhancement for frame-perspective optimization.

\noindent\textbf{Black-box Attacks}.\quad
Black-box attacks operate without model access, relying on query-based optimization, surrogate models, or RL to generate failure-inducing scenarios.
AdvSim~\cite{wang2021advsim} provides a general black-box adversarial scenario search framework for end-to-end planners.
STRIVE~\cite{rempe2022generating} perturbs real-world scenes in the latent space of a VAE-based traffic motion model to generate challenging scenarios for stress-testing.
Zheng et al.~\cite{zheng2023robustness} introduced adversarial corruption of context maps required by trajectory predictors.
UTCIA~\cite{bai2025universal} generates universal black-box adversarial perturbations for trajectory representation learning.
Avatar~\cite{liu2025avatar} uses RL to optimize adversarial trajectories without model access.
LC~\cite{ding2020learning} trains RL agents to act as adversaries, intentionally inducing collisions and exposing weaknesses in planners.
AdvDiffuser~\cite{xie2024advdiffuser} leverages diffusion guidance to synthesize realistic yet failure-inducing trajectories.
NADE~\cite{feng2021intelligent} generates naturalistic adversarial driving scenarios to stress-test end-to-end planners.

  Physical adversarial attacks on robot trajectory planners~\cite{wu2024characterizing} characterize how environmental manipulation can cause planner failures in deployments.
  JackZebra~\cite{sun2026beyond} demonstrates long-horizon goal hijacking through adversarial patches on an attacker vehicle, steering a victim AV to an attacker-chosen destination.


\subsubsection{Jailbreak Attacks}
\label{subsubsec:Jailbreak_Navigation_Planners}

Jailbreak attacks targeting LLM-based navigation systems manipulate natural language instructions to bypass safety constraints and elicit unsafe navigation behaviors. Unlike task planning jailbreaks that corrupt high-level goal decomposition, navigation jailbreaks directly compromise low-level trajectory generation.

Zhang et al.~\cite{zhang2024study} modeled Obvious Malicious Injection (OMI) and Goal Hijacking Injection (GHI) against LLM-integrated mobile robots.
Wen et al.~\cite{wen2024secure} demonstrated that insertion and swap attacks significantly degrade the performance of GPT-3, GPT-4, and LLaMA-based navigation planners on Touchdown and Map2Seq, with errors concentrated at intersections and other high-ambiguity locations.
PINA~\cite{liu2026pina} extends prompt injection to misguide physical navigation, leading to unsafe routes and mission failure.


\subsubsection{Adversarial Defenses}
\label{subsubsec:Motion_Planning_Defenses}

Adversarial defenses for trajectory planning operate through robust training, which incorporates safety constraints during learning, and robust inference, which applies output moderation at deployment.

\noindent\textbf{Robust Training}.\quad
Robust training integrates safety constraints into the learning process.
Thumm et al.~\cite{thumm2023reducing} proposed proactive replacement and projection methods that modify agent actions during RL training to reduce failsafe interventions, yielding policies with fewer safety violations.
AR-ICRL~\cite{xu2024robust} extends inverse RL to infer safety constraints from expert demonstrations that remain valid even under model misspecification.

\noindent\textbf{Robust Inference}.\quad
Robust inference defends planners at deployment through output moderation.
SMPC~\cite{brudigam2021stochastic} uses stochastic model predictive control with backup trajectories computed via reachable sets, overwriting planned outputs when safety constraints are violated.
Yurtsever et al.~\cite{yurtsever2019risky} identified hazardous behaviors at runtime, enabling planners to filter or down-weight high-risk maneuvers.
CSP-GAN-LSTM~\cite{meng2023vehicle} combines convolutional pooling with attention-based trajectory prediction to compute collision risk via time-to-collision metrics during inference.


\subsection{Multi-Agent Planning}
\label{subsec:Multi_Agent_Planning}

  Multi-agent planning extends single-agent task and trajectory planning to teams of embodied agents that must jointly decompose tasks, allocate subtasks, and synthesize coordinated plans. This distributed setting introduces unique attack surfaces: an adversary can compromise a single agent's planner to inject malicious subtasks that cascade through the team, manipulate inter-agent communication to corrupt plan consensus, or exploit Byzantine faults to subvert collective decision-making.


  \subsubsection{Byzantine Faults}
\label{subsubsec:Byzantine_Faults}

Byzantine faults arise when agents exhibit arbitrary or malicious behavior during distributed planning, corrupting consensus formation and task allocation.
Strobel and Ferrer~\cite{strobel_blockchain_2020} demonstrated that classical consensus algorithms break down under Byzantine attacks in swarm robotics.
Blumenkamp and Prorok~\cite{blumenkamp_emergence_2021} showed that self-interested agents in multi-robot planning tasks learn manipulative communication strategies through a differentiable shared channel, suggesting that adversarial behavior may emerge naturally from competitive pressure.
He et al.~\cite{he_redteaming_2025} introduced the Agent-in-the-Middle (AiTM) attack that intercepts and manipulates messages between LLM-based agents during cooperative planning.
Zhou et al.~\cite{zhou2026ham3} mount hierarchical attacks on multi-modal multi-agent reasoning, jointly biasing message content, interaction topology, and the cognitive pipeline to corrupt collective decisions.
Schroeder de Witt~\cite{schroeder_open_2025} taxonomizes multi-agent security threats including cascading failures, monoculture collapse, and conformity bias that drives false consensus on unsafe plans.


  \subsubsection{Goal Conflicts}
\label{subsubsec:Goal_Conflicts}

  Adversarial or self-interested agents exploit cooperative planning protocols to advance conflicting objectives.
Choudhury et al.~\cite{choudhury_dynamic_2022} formulated robust task allocation strategies that maintain plan quality under adversarial cost perturbation.
KA et al.~\cite{ka2024systematic} identified security-relevant gaps in multi-robot task allocation including lack of authentication and absence of Byzantine robustness guarantees.
Bahrami and Jafarnejadsani~\cite{gao_multirobot_2025} examined how adversarial perception attacks propagate through multi-robot relative localization to corrupt downstream coordination and planning.
Zhou and Tokekar~\cite{zhou_multirobot_2021} reviewed algorithmic trends for robust multi-robot coordination under adversarial agents, while Sookha and Benevenuto~\cite{sookha_adversarial_2024} provided a taxonomy of adversarial attacks on multi-agent reinforcement learning that can corrupt learned planning policies.


  \subsubsection{Potential Defenses}
\label{subsubsec:Multi_Agent_Planning_Defenses}

  Resilient algorithms ensure that cooperative planning converges correctly despite misbehaving agents.
Li et al.~\cite{li_resilient_2020} proposed a centerpoint-based aggregation rule that guarantees convergence to the true target state even when adversarial robots inject arbitrary state estimates.
Strobel et al.~\cite{strobel_swarms_2023} deployed smart contracts that regulate a crypto-token economy among physical robots, causing Byzantine robots to exhaust their tokens and be neutralized.
Lee and Panagou~\cite{lee_distributed_2025} designed a CBF-based distributed controller that guarantees resilient consensus and collision avoidance using only locally available information.
Gandhi et al.~\cite{gandhi_roborebound_2025} presented RoboRebound, extending Byzantine fault tolerance to physical multi-robot systems where adversarial agents can block paths or cause collisions.


\subsection{Benchmarks}
\label{subsec:Planning_Benchmarks}

Simulation platforms, scenario-generation tools, and benchmarks constitute the foundational infrastructure for developing and evaluating embodied planners. These components differ in fidelity, scalability, and safety focus, yet collectively enable systematic stress-testing of algorithms. Simulation environments provide arenas for agent interaction, scenario design frameworks construct complex and safety-critical situations, and benchmarks integrate both into standardized evaluation pipelines. Together, they enable systematic stress-testing of planning algorithms under adversarial, rare, and out-of-distribution conditions.

\noindent\textbf{Simulation Platforms}.\quad
Simulation platforms for embodied planning vary in realism, efficiency, and task coverage. For autonomous driving, SUMMIT~\cite{cai2020summit} provides high-fidelity 3D towns with diverse agents and configurable weather, while HIGH-ENV~\cite{Leurent_An_Environment_for_2018} offers a lightweight 2D setup for prototyping. CARLA~\cite{dosovitskiy2017carla} extends fidelity through configurable vehicles, pedestrians, and scenes, and MetaDrive~\cite{li2022metadrive} uses procedural generation to produce large distributions of driving layouts. NAVSIM~\cite{dauner2024navsim} complements these with dataset-replay environments for cost-effective evaluation.
In robotics, platforms emphasize physics accuracy and multi-task support. Gazebo~\cite{koenig2004design} integrates ROS and multiple physics engines for navigation and multi-robot coordination. PyBullet~\cite{coumans2021} offers a Python API and built-in robot models adopted in RL. MuJoCo~\cite{todorov2012mujoco} provides high-precision contact dynamics for locomotion and manipulation. Habitat~\cite{puig2023habitat} scales visual navigation in realistic indoor environments, iGibson~\cite{li2022igibson} supports physically grounded manipulation tasks, and NVIDIA Isaac Sim~\cite{NVIDIA_Isaac_Sim} delivers photorealistic rendering with GPU-accelerated physics.

\noindent\textbf{Scenario Design Tools}.\quad
Scenario design tools build on simulators to specify safety-critical interactions in a repeatable way. CARLA Scenario Runner~\cite{dosovitskiy2017carla} provides a Python API and OpenSCENARIO support for multi-agent coordination. SCENIC~\cite{fremont2023scenic} introduces a probabilistic programming language for expressing spatial and temporal relationships, enabling concise specification of rare and complex events. SafeBench~\cite{xu2022safebench} integrates eight categories of critical driving scenarios and multiple generation algorithms for systematic safety evaluation.
SUMO NETEDIT~\cite{lopez2018microscopic} offers graphical editing of road networks for traffic-scale simulations, and CommonRoad~\cite{wang2021commonroad} supplies XML-based scenario definitions and a Python API for standardized motion-planning research. Together, these tools span language-based specification, graphical editing, and benchmark-oriented safety testing, enabling comprehensive evaluation of embodied planners.

Benchmarks build on simulation and scenario design to create standardized, repeatable pipelines for evaluating embodied planners under adversarial, rare, and out-of-distribution conditions. Bench2Drive~\cite{jia2024bench2drive} offers a closed-loop driving suite with 220 routes spanning diverse weather, traffic, and map settings, isolating core planning skills such as lane keeping, merging, overtaking, and emergency handling. M3Bench~\cite{zhang2025m} targets mobile manipulation with 30,000 pick-and-place tasks across 119 household scenes, providing expert demonstrations and tests of generalization to novel objects and layouts. THOR-EAE~\cite{wang2023generating} assesses both action selection and natural language explanation with 840,000 samples in AI2-THOR. EAI~\cite{li2024embodied} standardizes evaluation for LLM-driven agents, unifying protocols across navigation and interaction, decomposing execution into subgoals, and benchmarking eighteen state-of-the-art models with detailed error analysis.

Safety-focused benchmarks have proliferated to address planning-specific hazards. AgentSafe~\cite{liu2025agentsafe} measures multimodal reasoning in long-horizon navigation with adversarial simulation scenarios and risk-aware task suites inspired by Asimov's Three Laws. HASARD~\cite{tomilin2025hasard} evaluates interactive household manipulation with affordance-level annotations. Safe-BeAl~\cite{huang2025framework} focuses on hazardous and adversarial settings to assess an agent's risk awareness and robustness. SafeAgentBench~\cite{yin2024safeagentbench} evaluates embodied agent safety through executable tasks spanning explicit and implicit hazards. AGENTSAFE~\cite{mao2025agentsafe} benchmarks safety of embodied agents on hazardous instructions with multi-stage evaluation across perception, planning, and execution. SafeMindBench~\cite{chen2025safemind} benchmarks safety risks in embodied LLM agents.
DESPITE~\cite{zhang2026despite} contributes a PDDL benchmark that separates planning competence from safety competence, finding LLM planners produce dangerous plans even when nominal task accuracy is high.
For domain-specific safety, Nakao and Takemoto~\cite{nakao2026safety} evaluate LLMs on a medical-ethics benchmark for robotic health-attendant control.
RoboJailBench~\cite{yeke2026robojailbench} provides a jailbreak attack/defense benchmark for embodied VLMs, pairing an ISO-derived security taxonomy with adversarial--benign intent contrast.


\section{Action and Interaction}
\label{sec:action}

Action forms the fourth layer, encompassing perception, cognition, and planning while adding physical execution, expanding the agent's capability from decision-making to real-world interaction. This expansion carries the highest stakes and broadens the attack surface to the physical domain: adversaries can now corrupt control policies to cause collisions, exploit human-agent interaction to endanger people, and poison multi-agent coordination to induce swarm-level failures. In end-to-end models like VLA, this layer represents the full system from visual input to action output. This section organizes action and interaction into three subsections: \textbf{Robot Control} (Section~\ref{subsec:action-control}) addresses robustness of low-level control policies, including RL-based controllers, diffusion policies, and VLA models; \textbf{Human-Agent Interaction} (Section~\ref{subsec:hai}) examines safety in human-agent physical interaction, including handover safety and trust manipulation; and \textbf{Multi-Agent Collaboration} (Section~\ref{subsec:mac}) covers execution-time coordination among multiple agents, focusing on infection attacks that propagate adversarial behaviors across agent populations and multi-agent collusion where autonomous agents deliberately coordinate malicious activities.


\subsection{Robot Control}
\label{subsec:action-control}

After perceiving the environment, understanding the situation, and planning a trajectory, an embodied agent must reliably execute actions in the physical world. Safe control is therefore essential to trustworthy embodied AI, ensuring robustness under uncertainty and resilience to adversarial influence. Recent surveys~\cite{mienye2026deep} chart the rapid convergence of deep reinforcement learning with foundation models, motivating the safety focus on increasingly capable yet less interpretable control policies. Existing work on safe control falls into four categories: \textbf{Adversarial Attacks}, \textbf{Adversarial Defenses}, \textbf{Backdoor Attacks}, and \textbf{Backdoor Defenses}. Adversarial attacks and defenses address inference-time perturbations to states, actions, or environments, while backdoor attacks and defenses focus on hidden triggers that remain dormant during normal operation but activate harmful behaviors when invoked.

\begin{table}[!tp]
\center
\renewcommand{\arraystretch}{1.15}
\caption{A summary of \textbf{adversarial attacks} for \textbf{robot control}. For clarity, we append suffixes to the target model (MLP, DP, DT, VLA), where \textbf{S}, \textbf{A}, \textbf{E}, \textbf{V}, and \textbf{L} denote attack surfaces on \emph{State}, \emph{Action}, \emph{Environment}, \emph{Vision}, and \emph{Language}.}

\resizebox{1\textwidth}{!}{

\rowcolors{2}{tablegray}{white}
{\setlength{\tabcolsep}{0pt}
\begin{tabular}{@{}>{\cellcolor{white}}p{.15\textwidth} p{.21\textwidth} p{.06\textwidth}
               p{.12\textwidth} p{.20\textwidth} p{.16\textwidth} p{.25\textwidth}@{}}
\toprule
\rowcolor{interaction-table-pink}
\cellcolor{interaction-table-pink}\textbf{Attack} & \textbf{Method} & \textbf{Year} & \textbf{Category} & \textbf{Subcategory} & \textbf{Target Model} & \textbf{Environment} \\
\midrule

& CPA/AA\cite{Sun_StealthyEfficientAdversarial_2020} & 2020 & White-Box & Optim.-Based Attack & MLP-S & Atari, MuJoCo \\
& RS/MAD\cite{Zhang_RobustDeepReinforcement_2020} & 2020 & White-Box & Optim.-Based Attack & MLP-S & Atari, MuJoCo \\
& Weng et al.\cite{Weng_EvaluatingRobustnessDeep_2020} & 2020 & White-Box & Optim.-Based Attack & MLP-S & MuJoCo \\
& MAS/LAS\cite{Lee_SpatiotemporallyConstrainedAction_2020} & 2020 & White-Box & Optim.-Based Attack & MLP-A & Atari, Gym \\
& DP-Attacker\cite{Chen_DiffusionPolicyAttacker_2024} & 2024 & White-Box & Optim.-Based Attack & DP-S & Robosuite \\
& Kalra et al.\cite{Kalra_HowVulnerableMy_2025} & 2025 & White-Box & Optim.-Based Attack & DP-S/DT-S & RoboMimic \\
& UADA/UPA/TMA\cite{Wang_ExploringAdversarialVulnerabilities_2025} & 2025 & White-Box & Optim.-Based Attack & VLA-V & BridgeData, LIBERO, UR10e \\
& PVEP\cite{Cheng_ManipulationFacingThreats_2024} & 2024 & White-Box & Optim.-Based Attack & VLA-V & VIMA, SIMPLER \\
& UPA-RFAS\cite{lu2025robots} & 2025 & White-Box & Optim.-Based Attack & VLA-V & BridgeData, LIBERO \\
& FreezeVLA\cite{wang2025freezevla} & 2025 & White-Box & Optim.-Based Attack & VLA-V & LIBERO \\
& EDPA\cite{xu2025model} & 2025 & White-Box & Optim.-Based Attack & VLA-V & LIBERO \\
& Tex3D~\cite{chen2026tex3d} & 2026 & White-Box & Optim.-Based Attack & VLA-V & LIBERO \\
& Zhao et al.\cite{Zhao_RethinkingIntermediateFeatures_2024} & 2024 & White-Box & Optim.-Based Attack & VLA-L & VIMA \\
& Jones et al.\cite{jones2025adversarial} & 2025 & White-Box & Optim.-Based Attack & VLA-L & LIBERO, HYDRA, SIMPLER \\
& ADVLA\cite{zhang2025attention} & 2025 & White-Box & Optim.-Based Attack & VLA-L & LIBERO \\
& VLA-Fool\cite{yan2025alignment} & 2025 & White-Box & Optim.-Based Attack & VLA-V/L & LIBERO \\
& UniAda~\cite{zhang2024uniada} & 2024 & White-Box & Optim.-Based Attack & VLM-V & nuScenes, CARLA \\
& PA-AD\cite{sun2022strongest} & 2022 & White-Box & Adversarial Policy & MLP-S & Atari, MuJoCo \\
& ANNIE-Attack\cite{huang2025annie} & 2025 & White-Box & Adversarial Policy & VLA-V & ANNIEBench \\
& SA-RL\cite{Zhang_RobustReinforcementLearning_2021} & 2021 & Black-Box & Adversarial Policy & MLP-S & MuJoCo \\
& RAT\cite{Bai_RATAdversarialAttacks_2025} & 2025 & Black-Box & Adversarial Policy & MLP-S & MuJoCo, Meta-World \\
& AP-MARL\cite{Gleave_AdversarialPoliciesAttacking_2020} & 2020 & Black-Box & Adversarial Policy & MA-MLP-S & MuJoCo \\
& IMAP\cite{Zheng_EvaluatingRobustnessReinforcement_2024} & 2024 & Black-Box & Adversarial Policy & (MA-)MLP-S & MuJoCo \\
& SUB-PLAY\cite{Ma_SUBPLAYAdversarialPolicies_2024} & 2024 & Black-Box & Adversarial Policy & MA-MLP-S & MPE \\
& LIBERO-Plus~\cite{fei2025libero} & 2025 & Black-Box & Adversarial Policy & VLA-V & LIBERO-Plus \\
& DAERT~\cite{tong2026daert} & 2026 & Black-Box & Adversarial Policy & VLA-L & CALVIN, RLBench \\
& ERT\cite{karnik_ert_2024} & 2024 & Black-Box & Adversarial Policy & VLA-L/DP-L & CALVIN, RLBench \\
& RedVLA~\cite{zhang2026redvla} & 2026 & Black-Box & Adversarial Policy & VLA-V/L & Real Robot \\
& ADVEDM~\cite{wang2025advedm} & 2025 & Black-Box & Optim.-Based Attack & VLM-V & VIMA, nuScenes \\
\multirow{-30}{*}{\parbox[c]{\linewidth}{\raggedright Adversarial\\Attack}} & JailWAM~\cite{liu2026jailwam} & 2026 & Black-Box & Optim.-Based Attack & WAM-L & LIBERO \\

\bottomrule
\end{tabular}
}}
\label{tab:interaction_adversarial_attack}
\end{table}

\subsubsection{Adversarial Attacks}

Adversarial attacks on embodied agents exploit weaknesses in RL control policies to induce unsafe or unintended behaviors. These attacks perturb inputs such as states, actions, or observations and fall into two main categories: white-box attacks, which compute or train perturbations using full access to model parameters, and black-box attacks, which rely solely on interactive queries. Both classes target multiple vulnerability surfaces (state [S], action [A], environment [E], vision [V], and language [L]) and are evaluated across diverse platforms such as MuJoCo, Gym, RoboMimic, and LIBERO (Table~\ref{tab:interaction_adversarial_attack}).

\noindent\textbf{White-box Attacks.}\quad
White-box attacks compute perturbations using gradients from the victim model.
For MLP-based agents, CPA and AA~\cite{Sun_StealthyEfficientAdversarial_2020} use trajectory sampling and adversarial policies to produce stealthy attacks, while RS and MAD~\cite{Zhang_RobustDeepReinforcement_2020} optimize perturbations by smoothing value functions or maximizing action divergence. For continuous control, Weng et al.~\cite{Weng_EvaluatingRobustnessDeep_2020} studied state and action perturbations using learned dynamics, and MAS and LAS~\cite{Lee_SpatiotemporallyConstrainedAction_2020} impose spatiotemporal constraints to preserve action plausibility.

Modern policy architectures introduce new attack surfaces. DP-Attacker~\cite{Chen_DiffusionPolicyAttacker_2024} perturbs camera observations in Diffusion Policies (DP) to exploit autoregressive dependencies, a vulnerability further analyzed by Kalra et al.~\cite{Kalra_HowVulnerableMy_2025} for Decision Transformer (DT) agents. For VLAs, UADA, UPA, and TMA~\cite{Wang_ExploringAdversarialVulnerabilities_2025} attack visual channels in BridgeData, LIBERO, and UR10e robots, inducing deviations through untargeted and targeted perturbations. Building on these visual-channel attacks, UPA-RFAS~\cite{lu2025robots} learns a physical patch in a shared feature space across models to achieve transferable adversarial manipulation. FreezeVLA~\cite{wang2025freezevla} generates cross-prompt adversarial images to induce action-freezing behaviors across diverse user instructions. Similarly, EDPA~\cite{xu2025model} generates adversarial patches to distort visual understanding in VLAs, leading to failed action execution. Moreover, PVEP~\cite{Cheng_ManipulationFacingThreats_2024} expands these attacks with blurs, typography prompts, and adversarial patches in VIMA and SIMPLER. Tex3D~\cite{chen2026tex3d} optimizes adversarial 3D textures via foreground-background decoupled differentiable rendering, turning physical objects into attack surfaces against VLAs. Language channels are similarly vulnerable: Zhao et al.~\cite{Zhao_RethinkingIntermediateFeatures_2024} and Jones et al.~\cite{jones2025adversarial} adapted GCG-style suffix optimization to manipulate decision outputs of VLAs. ADVLA~\cite{zhang2025attention} projects the visual features of adversarial perturbations into the textual feature space, thereby disrupting action prediction. To further characterize cross-modal vulnerabilities, VLA-Fool~\cite{yan2025alignment} unifies textual, visual, and cross-modal attacks and introduces an automatically crafted prompting framework.
JailWAM~\cite{liu2026jailwam} jailbreaks world-action models via visual-trajectory mapping and dual-path physical-simulator verification, exploiting the language interface as policies subsume planning.

Adversarial agents can also be trained to attack sequentially. PA-AD~\cite{sun2022strongest} extends adversarial policies to continuous control, and ANNIE-Attack~\cite{huang2025annie} evaluates VLA robustness within ANNIEBench.

\noindent\textbf{Black-box Attacks.}\quad
Black-box attacks operate without access to model parameters, relying instead on adversarial policies or interaction-driven perturbations. SA-RL~\cite{Zhang_RobustReinforcementLearning_2021} optimizes state perturbations in MuJoCo through sequential interaction, while RAT~\cite{Bai_RATAdversarialAttacks_2025} induces targeted failures across MuJoCo and Meta-World.

Multi-agent settings introduce additional vulnerabilities. AP-MARL~\cite{Gleave_AdversarialPoliciesAttacking_2020} trains red-team agents to manipulate cooperative or competitive partners. IMAP~\cite{Zheng_EvaluatingRobustnessReinforcement_2024} learns intrinsically motivated adversaries that seek out weakness-exposing states, and SUB-PLAY~\cite{Ma_SUBPLAYAdversarialPolicies_2024} exploits partial observability to create deceptive multi-agent interactions. ERT~\cite{karnik_ert_2024} further extends black-box threats to VLM-driven robots via instruction-grounded red-teaming, and DAERT~\cite{tong2026daert} extends the same lineage with diversity-aware paraphrasing that uncovers linguistic fragility unreachable by template-bound prompts. RedVLA~\cite{zhang2026redvla} closes the loop with physical red-teaming on real VLA-controlled robots, exposing failure modes that simulation-only protocols miss.
LIBERO-Plus~\cite{fei2025libero} introduces a comprehensive benchmark for safety evaluation of VLAs, investigating performance drops under diverse conditions ranging from lighting changes to camera pose variations.
LIBERO-X~\cite{wang_liberox_2026} proposes a hierarchical evaluation protocol from spatial perturbation to semantic reformulation, finding VLAs strong on standard benchmarks collapse under minor distributional shifts.
ADVEDM~\cite{wang2025advedm} proposes a fine-grained black-box adversarial attack framework against VLM-based policies that semantically edits only a few key objects while preserving the remaining regions, reducing conflicts with task context and inducing valid but incorrect decisions.
For end-to-end autonomous driving, UniAda~\cite{zhang2024uniada} unifies multi-objective universal adversarial attacks across perception, prediction, and planning modules.

\subsubsection{Adversarial Defenses}

\begin{table}[!tp]
\center
\renewcommand{\arraystretch}{1.15}
\caption{A summary of \textbf{adversarial defenses} for \textbf{robot control}. For clarity, we append suffixes to the target model (MLP, DP, DT, VLA), where \textbf{S}, \textbf{A}, \textbf{E}, \textbf{V}, and \textbf{L} denote attack surfaces on \emph{State}, \emph{Action}, \emph{Environment}, \emph{Vision}, and \emph{Language}.}

\resizebox{1\textwidth}{!}{

\rowcolors{2}{tablegray}{white}
{\setlength{\tabcolsep}{0pt}
\begin{tabular}{@{}>{\cellcolor{white}}p{.12\textwidth} p{.20\textwidth} p{.06\textwidth}
               p{.16\textwidth} p{.20\textwidth} p{.16\textwidth} p{.25\textwidth}@{}}
\toprule
\rowcolor{interaction-table-pink}
\cellcolor{interaction-table-pink}\textbf{Defense} & \textbf{Method} & \textbf{Year} & \textbf{Category} & \textbf{Subcategory} & \textbf{Target Model} & \textbf{Environment} \\
\midrule

& Xu et al.\cite{xu2025model} & 2025 & Robust Training & Adversarial Training & VLA-V & LIBERO \\
& ATLA\cite{Zhang_RobustReinforcementLearning_2021} & 2021 & Robust Training & Adversarial Training & MLP-S & MuJoCo \\
& Liu et al.\cite{Liu_RethinkingAdversarialPolicies_2024} & 2024 & Robust Training & Adversarial Training & MLP-S & RoboSumo \\
& S-DQN/S-PPO\cite{sun2024breaking} & 2024 & Robust Training & Adversarial Training & MLP-S & Atari, MuJoCo \\
& VALT\cite{Nakanishi_OffPolicyActorCriticAdversarial_2025} & 2025 & Robust Training & Adversarial Training & MLP-S & MuJoCo \\
& ACoE\cite{Belaire_MinimizingAdversarialCounterfactual_2025} & 2024 & Robust Training & Adversarial Training & MLP-S & Highway, Atari, MuJoCo \\
& EPOpt\cite{Rajeswaran_EPOptLearningRobust_2017} & 2017 & Robust Training & Adversarial Training & MLP-E & MuJoCo \\
& RARL\cite{Pinto_RobustAdversarialReinforcement_2017} & 2017 & Robust Training & Adversarial Training & MLP-E & MuJoCo \\
& ARPL\cite{Mandlekar_AdversariallyRobustPolicy_2017} & 2017 & Robust Training & Adversarial Training & MLP-E & MuJoCo \\
& MRPO\cite{Jiang_MonotonicRobustPolicy_2021} & 2021 & Robust Training & Adversarial Training & MLP-E & MuJoCo \\
& Stack-PG\cite{Huang_RobustReinforcementLearning_2022} & 2022 & Robust Training & Adversarial Training & MLP-E & Highway, Gym \\
& DiAMetR\cite{Ajay_DistributionallyAdaptiveMeta_2022} & 2022 & Robust Training & Adversarial Training & Meta-MLP-E & MuJoCo, Gym \\
& RAPPO\cite{lien2023revisiting} & 2023 & Robust Training & Adversarial Training & MLP-E & MuJoCo \\
& RoML\cite{Greenberg_TrainHardFight_2023} & 2023 & Robust Training & Adversarial Training & Meta-MLP-E & MuJoCo \\
& UOR-RL\cite{You_UserOrientedRobustReinforcement_2023} & 2023 & Robust Training & Adversarial Training & MLP-E & MuJoCo \\
& RNAC\cite{Zhou_NaturalActorCriticRobust_2023} & 2023 & Robust Training & Adversarial Training & MLP-E & MuJoCo, TurtleBot \\
& EWoK\cite{Gadot_BringYourOwn_2024} & 2024 & Robust Training & Adversarial Training & MLP-E & MuJoCo \\
& BAT\cite{Wang_RobustDeepReinforcement_2025} & 2025 & Robust Training & Adversarial Training & MLP-E & Overcooked \\
& PR/NR-MDP\cite{Tessler_ActionRobustReinforcement_2019} & 2019 & Robust Training & Adversarial Training & MLP-A & MuJoCo \\
& RAP\cite{Vinitsky_RobustReinforcementLearning_2020} & 2020 & Robust Training & Adversarial Training & MLP-A & MuJoCo \\
& Tan et al.\cite{Tan_RobustifyingReinforcementLearning_2020} & 2020 & Robust Training & Adversarial Training & MLP-A & Gym \\
& OA-PI\cite{Nie_ActionRobustReinforcement_2025} & 2025 & Robust Training & Adversarial Training & MLP-A & MuJoCo \\
& TBRR\cite{Huang_TradeOffRobustnessRewards_2023} & 2023 & Robust Training & Adversarial Training & MLP-S/E & MuJoCo, PyBullet, KUKA \\
& ROMANCE\cite{Yuan_RobustMultiAgentCoordination_2023} & 2023 & Robust Training & Adversarial Training & MA-MLP-S & SMAC \\
& PATROL\cite{guo2023patrol} & 2023 & Robust Training & Adversarial Training & MA-MLP-S & Atari, MuJoCo, SMAC \\
& AME\cite{sun2023certifiably} & 2023 & Robust Training & Adversarial Training & MA-MLP-S & Customized \\
& GRAD\cite{Liang_GameTheoreticRobustReinforcement_2024a} & 2024 & Robust Training & Adversarial Training & MLP-S/A & MuJoCo \\
& RIQL\cite{Yang_RobustOfflineReinforcement_2024} & 2024 & Robust Training & Adversarial Training & Offline-MLP-S & D4RL \\
& ARDT\cite{Tang_AdversariallyRobustDecision_2024} & 2024 & Robust Training & Adversarial Training & DT-S & MuJoCo \\
& SafeVLA\cite{zhang2025safevla} & 2025 & Robust Training & Adversarial Training & VLA-E & Safety-CHORES \\

& STRONG-VLA\cite{xie2026strongvla} & 2026 & Robust Training & Adversarial Training & VLA-V/L & LIBERO \\
& DDP\cite{hu2026dream} & 2026 & Robust Training & Adversarial Training & DP-V & Custom \\

& SA-MDP\cite{Zhang_RobustDeepReinforcement_2020} & 2020 & Robust Training & Robust Regularizer & MLP-S & Atari, MuJoCo \\
& RADIAL\cite{Oikarinen_RobustDeepReinforcement_2021} & 2021 & Robust Training & Robust Regularizer & MLP-S & Atari, MuJoCo \\
& WocaR\cite{Liang_EfficientAdversarialTraining_2022} & 2022 & Robust Training & Robust Regularizer & MLP-S & Atari, MuJoCo \\
& RAD\cite{Belaire_RegretBasedDefenseAdversarial_2024} & 2024 & Robust Training & Robust Regularizer & MLP-S & Atari, MuJoCo, Highway \\
& SCPO\cite{Kuang_LearningRobustPolicy_2022} & 2022 & Robust Training & Robust Regularizer & MLP-E & MuJoCo \\
& RoMFAC\cite{Zhou_RobustMeanFieldActorCritic_2023} & 2023 & Robust Training & Robust Regularizer & MA-MLP-S & MAgent \\
& TRACER\cite{Yang_UncertaintyBasedOfflineVariational_2024} & 2024 & Robust Training & Robust Regularizer & Offline-MLP-S & D4RL \\
& MIR3\cite{Li_RobustMultiAgentReinforcement_2025} & 2025 & Robust Training & Robust Regularizer & MA-MLP-S & SMAC \\

& CROP\cite{Wu_CROPCertifyingRobust_2022} & 2022 & Robust Inference & Input Moderation & MLP-S & Atari, Highway, Gym \\
& VQ-RL\cite{Luu_MitigatingAdversarialPerturbations_2024} & 2024 & Robust Inference & Input Moderation & MLP-S & Atari, MuJoCo \\
& BYOVLA\cite{hancockrun} & 2025 & Robust Inference & Input Moderation & VLA-V & BridgeData \\

& PROTECTED\cite{Liu_WorstCaseAttacksRobust_2024} & 2024 & Robust Inference & Output Moderation & MLP-A & MuJoCo \\
& VLSA\cite{hu2025vlsa} & 2025 & Robust Inference & Output Moderation & VLA-E & LIBERO, SafeLIBERO \\
\multirow{-46}{*}{\parbox[c]{\linewidth}{\raggedright Adversarial\\Defense}} & AERMANI-VLM\cite{Mishra_AERMANIVLMStructuredPrompting_2025} & 2025 & Robust Inference & Output Moderation & VLA-A & real world \\

\bottomrule
\end{tabular}
}}
\label{tab:interaction_adversarial_defense}
\end{table}

Defenses against adversarial attacks in embodied interaction fall into two categories: robust training, which incorporates adversaries or regularization during learning, and robust inference, which protects policies at deployment through input filtering or ensemble strategies. These methods address a wide range of attack surfaces and have been evaluated across MuJoCo, SMAC, and real robotic systems (Table~\ref{tab:interaction_adversarial_defense}).

\noindent\textbf{Robust Training.}\quad
Robust training defenses embed adversarial resilience directly into the learning process. At the environment level, methods such as EPOpt \cite{Rajeswaran_EPOptLearningRobust_2017} and RARL \cite{Pinto_RobustAdversarialReinforcement_2017} train policies against ensembles of perturbed environments or destabilizing opponents, while ARPL \cite{Mandlekar_AdversariallyRobustPolicy_2017} generates plausible adversarial examples during training. Successors including MRPO \cite{Jiang_MonotonicRobustPolicy_2021}, Stack-PG \cite{Huang_RobustReinforcementLearning_2022}, and RAPPO \cite{lien2023revisiting} formalize robustness through simulator sampling, Stackelberg games, or improved domain randomization; scalable variants such as RNAC \cite{Zhou_NaturalActorCriticRobust_2023}, EWoK \cite{Gadot_BringYourOwn_2024}, and BAT \cite{Wang_RobustDeepReinforcement_2025} introduce efficient uncertainty sets, kernel estimators, and boosted fine-tuning. Meta-RL approaches DiAMetR \cite{Ajay_DistributionallyAdaptiveMeta_2022} and RoML \cite{Greenberg_TrainHardFight_2023} strengthen cross-task generalization through population-based training and gradient debiasing.

Action-space defenses (MLP-A) address actuator perturbations. PR and NR-MDP \cite{Tessler_ActionRobustReinforcement_2019} and RAP \cite{Vinitsky_RobustReinforcementLearning_2020} formalize robustness to bounded or stochastic action noise, while Tan et al. \cite{Tan_RobustifyingReinforcementLearning_2020} showed that action-adversarial training improves resilience without degrading nominal performance. OA-PI \cite{Nie_ActionRobustReinforcement_2025} further extends action-robust policy optimization.

State-space defenses (MLP-S) include ATLA \cite{Zhang_RobustReinforcementLearning_2021}, which co-trains adversaries with the policy, and TBRR \cite{Huang_TradeOffRobustnessRewards_2023}, which trades reward for robustness under severe perturbations. GRAD \cite{Liang_GameTheoreticRobustReinforcement_2024a} models temporally coupled threats via zero-sum games, and Liu et al. \cite{Liu_RethinkingAdversarialPolicies_2024} proposed a flexible adversary formulation with provable convergence. Recent advances such as VALT \cite{Nakanishi_OffPolicyActorCriticAdversarial_2025} and ACoE \cite{Belaire_MinimizingAdversarialCounterfactual_2025} exploit policy evaluation symmetries or minimize adversarial counterfactual errors. Offline and multi-agent robustness are addressed by RIQL \cite{Yang_RobustOfflineReinforcement_2024}, ROMANCE \cite{Yuan_RobustMultiAgentCoordination_2023}, PATROL \cite{guo2023patrol}, AME \cite{sun2023certifiably}, and ARDT \cite{Tang_AdversariallyRobustDecision_2024}, which handle corrupted datasets, adversarial coordination, and communication failures. Specifically, for VLAs, SafeVLA~\cite{zhang2025safevla} constrains VLA policies from a min-max perspective against elicited safety risks via safe RL, and Xu et al.~\cite{xu2025model} fine-tune the visual encoder using adversarial visual samples to enhance model robustness.

Regularization enhances robustness. SA-MDP \cite{Zhang_RobustDeepReinforcement_2020}, RADIAL \cite{Oikarinen_RobustDeepReinforcement_2021}, and WocaR \cite{Liang_EfficientAdversarialTraining_2022} penalize perturbation sensitivity. RAD \cite{Belaire_RegretBasedDefenseAdversarial_2024}, TRACER \cite{Yang_UncertaintyBasedOfflineVariational_2024}, RoMFAC \cite{Zhou_RobustMeanFieldActorCritic_2023}, MIR3 \cite{Li_RobustMultiAgentReinforcement_2025}, and SCPO \cite{Kuang_LearningRobustPolicy_2022} use regret minimization, uncertainty modeling, mean-field regularization, information bottlenecks, or gradient penalties.

\noindent\textbf{Robust Inference.}\quad
At deployment time, defenses focus on maintaining robustness without retraining. Input moderation methods provide complementary protection: CROP \cite{Wu_CROPCertifyingRobust_2022} certifies robustness on a per-state basis, and VQ-RL \cite{Luu_MitigatingAdversarialPerturbations_2024} compresses observation spaces for lightweight resilience. For VLA systems, BYOVLA \cite{hancockrun} enhances robustness by detecting and minimally editing vulnerable image regions during inference. STRONG-VLA~\cite{xie2026strongvla} decouples robustness learning across modalities, training the visual and language branches independently against multimodal perturbations rather than jointly. DDP~\cite{hu2026dream} (Dream Diffusion Policy) integrates a world-model regularizer into diffusion-policy training, using imagined-future supervision to lift visual out-of-distribution robustness without adversarial perturbations. Output moderation approaches such as PROTECTED \cite{Liu_WorstCaseAttacksRobust_2024} minimize regret across policy sets to withstand adversarial conditions. VLSA~\cite{hu2025vlsa} adds a plug-and-play safety constraint layer that leverages VLM reasoning to improve VLA safety. AERMANI-VLM~\cite{Mishra_AERMANIVLMStructuredPrompting_2025} applies structured prompting to reduce VLM hallucinations in manipulation policies.
\begin{table}[!tp]
\center
\renewcommand{\arraystretch}{1.15}
\caption{A summary of \textbf{backdoor} attacks and defenses for \textbf{robot control (Part II)}, where \textbf{S}, \textbf{A}, \textbf{E}, \textbf{R}, \textbf{V}, and \textbf{L} denote attack surfaces on \emph{State}, \emph{Action}, \emph{Environment}, \emph{Reward}, \emph{Vision}, and \emph{Language}.}
\resizebox{1\textwidth}{!}{

\rowcolors{2}{tablegray}{white}
{\setlength{\tabcolsep}{0pt}
\begin{tabular}{@{}>{\cellcolor{white}}p{.12\textwidth} p{.20\textwidth} p{.06\textwidth}
               p{.18\textwidth} p{.18\textwidth} p{.16\textwidth} p{.25\textwidth}@{}}
\toprule
\rowcolor{interaction-table-pink}
\cellcolor{interaction-table-pink}\textbf{Attack/Defense} & \textbf{Method} & \textbf{Year} & \textbf{Category} & \textbf{Subcategory} & \textbf{Target Model} & \textbf{Environment} \\
\midrule

& BackdooRL~\cite{Wang_BackdooRLBackdoorAttack_2021} & 2021 & Training Manipulation & Trajectory Manipulation & MA-MLP-E & MuJoCo \\
& MARNet~\cite{Chen_MARNetBackdoorAttacks_2023} & 2023 & Training Manipulation & Trajectory Manipulation & MA-MLP-E & Predator Prey, SMAC \\
& PNAct~\cite{Guo_PNActCraftingBackdoor_2025} & 2025 & Training Manipulation & Trajectory Manipulation & Safe-MLP-S,A & Safety-Gymnasium \\
& BadVLA~\cite{zhou2025badvla} & 2025 & Training Manipulation & Model Manipulation & VLA-V & LIBERO \\
& Xie et al.~\cite{xie2026prompt} & 2026 & Training Manipulation & Model Manipulation & VLA-L & ROS2 \\
& FlowHijack~\cite{an2026flowhijack} & 2026 & Training Manipulation & Model Manipulation & VLA-V/L & LIBERO \\

& TooBadRL\cite{Li_TooBadRLTriggerOptimization_2025} & 2025 & Data Poisoning & State Trigger & MLP-S,A,R & MuJoCo \\
& Baffle\cite{gong2024baffle} & 2024 & Data Poisoning & Trajectory Trigger & Offline-MLP-S,A,R & MuJoCo \\
& Ashcraft et al.~\cite{Ashcraft_BackdoorsDRLFour_2025} & 2025 & Data Poisoning & Environment Trigger & MLP-E & Minigrid, Safety Gymnasium \\
& TrojanRobot~\cite{wang2024trojanrobot} & 2024 & Data Poisoning & Visual Trigger & VLA-V & LIBERO, UR3e \\
& DropVLA\cite{xu2025tabvla} & 2025 & Data Poisoning & Visual Trigger & VLA-V & LIBERO \\
& GoBA\cite{zhou_goal_backdoor_2025} & 2025 & Data Poisoning & Visual Trigger & VLA-V & LIBERO \\
& BEAT~\cite{zhan2026beat} & 2026 & Data Poisoning & Visual Trigger & VLA-V & ALFWorld, VirtualHome \\
& SilentDrift~\cite{xu2026silentdrift} & 2026 & Data Poisoning & Visual Trigger & VLA-V & LIBERO \\
\multirow{-15}{*}{\parbox[c]{\linewidth}{\raggedright Backdoor\\Attack}} & BackdoorVLA\cite{li2025attackvla} & 2025 & Data Poisoning & Visual/Textual Trigger & VLA-V/L & LIBERO, Franka \\

\midrule

\multirow{-1}{*}{\parbox[c]{\linewidth}{\raggedright Backdoor\\Defense}} & PolicyCleanse\cite{Guo_PolicyCleanseBackdoorDetection_2023} & 2023 & Robust Inference & Detection \& Removal & MA-MLP-E & MuJoCo \\

\bottomrule
\end{tabular}
}}
\label{tab: safe_interaction_2}
\end{table}

\subsubsection{Backdoor Attacks}

Backdoor attacks embed covert triggers during training so that a control policy behaves normally on benign inputs but executes attacker-chosen (often unsafe) behaviors when the trigger appears. In embodied agents these attacks target states, actions, rewards, environments, or visual inputs, and have been demonstrated across MuJoCo, Safety-Gymnasium, LIBERO, and other platforms (Table~\ref{tab: safe_interaction_2}). Two practical threat models dominate: training manipulation attacks, which compromise the training process itself, and data poisoning attacks, which poison training data.

\noindent\textbf{Training Manipulation Attacks.}\quad
With access to the training pipeline, attackers can implant robust triggers.
BackdooRL~\cite{Wang_BackdooRLBackdoorAttack_2021} and MARNet~\cite{Chen_MARNetBackdoorAttacks_2023} combine trigger injection with action or reward manipulation to induce targeted failures in single- and multi-agent control.
PNAct~\cite{Guo_PNActCraftingBackdoor_2025} ties triggers to unsafe actions in safe-RL settings with targeted positive-negative sampling.
BadVLA~\cite{zhou2025badvla} extends to VLAs by decoupling objectives and fine-tuning action heads to maintain nominal behavior while enabling triggered failures.
FlowHijack~\cite{an2026flowhijack} targets flow-matching VLAs like $\pi_0$ via tau-conditioned vector-field injection with a dynamics-mimicry regularizer, producing kinematically indistinguishable triggered actions where prior autoregressive backdoors do not transfer.

\noindent\textbf{Data Poisoning Attacks.}\quad
Without access to the training code, attackers rely on dataset poisoning. TooBadRL~\cite{Li_TooBadRLTriggerOptimization_2025} jointly optimizes trigger placement, timing, and magnitude for state-based triggers. In offline RL, Baffle~\cite{gong2024baffle} biases policies by injecting high-reward trajectories generated by weak agents.
TrojanRobot~\cite{wang2024trojanrobot} embeds a backdoor-finetuned VLM as a malicious perception module within modular robotic policies, demonstrating physical-world backdoor attacks through permutation, stagnation, and intentional trigger strategies.
For VLAs, DropVLA~\cite{xu2025tabvla} embeds visual trigger to induce the \texttt{open\_gripper} action when the trigger appears. Additionally, GoBA~\cite{zhou_goal_backdoor_2025} and AttackVLA~\cite{li2025attackvla} implant a trigger to activate a predefined long-horizon action sequence while preserving normal performance on clean inputs. SilentDrift~\cite{xu2026silentdrift} exploits the intra-chunk visual open-loop of action-chunked VLAs, hiding a trigger that evades frame-level inspection at low poisoning rates.
BEAT~\cite{zhan2026beat} backdoors MLLM-driven embodied agents using environmental objects as triggers, enabling multi-step malicious policy execution through contrastive trigger learning.
Moving up the stack, Xie and Wei-Kocsis~\cite{xie2026prompt} demonstrate that LoRA-poisoned LLMs can implant structured-JSON backdoors that propagate from natural-language prompts into ROS2 robotic control commands.

\subsubsection{Backdoor Defenses}

Backdoor defenses in embodied interaction aim to detect or neutralize hidden triggers implanted during training. Current strategies focus primarily on robust inference, which identifies and removes malicious influences at deployment (Table~\ref{tab: safe_interaction_2}).

In competitive MARL, PolicyCleanse~\cite{Guo_PolicyCleanseBackdoorDetection_2023} detects adversarial triggers through reward degradation signals and restores robustness via machine unlearning. This remains the only defense evaluated beyond Atari-only settings, highlighting a significant gap: as backdoor attacks increasingly target VLA and multi-agent embodied systems, defenses have yet to follow.


\subsection{Human-Agent Interaction}
\label{subsec:hai}

The presence of humans in a shared workspace fundamentally changes the safety requirements for embodied agents~\cite{zou_humanagent_2025}: the robot must not only complete its task but also continuously guarantee no physical or psychological harm to its human co-worker. Unlike the adversarial and backdoor threats discussed in Section~\ref{subsec:action-control}, which attack the policy itself, HRI safety addresses the broader challenge of ensuring that a competent policy still operates safely around people. We organize recent work into two categories: \textbf{Handover Safety} ensures safe object transfer between humans and robots; and \textbf{Trust Manipulation} addresses adversarial exploitation of the human-agent trust relationship.


\noindent\textbf{Handover Safety.}\quad
Object handover, transferring items between human and robot, requires coordinated grasp planning, force regulation, and intent detection to prevent drops, collisions, or discomfort.
For robot-to-human (R2H) handover, Yang et al.~\cite{handover_mobile_2024} developed a mobile cooperation system ensuring collision-free transfer trajectories, Makenova et al.~\cite{trust_handover_biomimetic_2025} demonstrated that trust significantly impacts movement dynamics and grip forces during R2H transfer, and compliant blind handover~\cite{blind_handover_2024} addresses scenarios where operators lack visual contact with the robot.
For human-to-robot (H2R) handover, Ding et al.~\cite{intention_fuzzy_wearable_2023} used wearable IMU sensors with fuzzy-rule-based inference to detect handover intentions, while Rosenberger et al.~\cite{handover_haptic_cues_2021} employed haptic cues as a communication channel during physical transfer.
Belmonte et al.~\cite{handover_adaptive_transport_2023} showed that adaptive transport methods significantly affect perceived safety, and Yang et al.~\cite{handover_survey_2024} provided a comprehensive review advocating simulation-based training with safety constraints.


\noindent\textbf{Trust Manipulation.}\quad
Miscalibrated trust, both over-trust and under-trust, leads to safety-critical failures in human-agent interaction~\cite{trust_review_hri_2021}. Over-trust causes operators to accept unsafe robot suggestions without scrutiny, while under-trust leads to disuse of capable systems in time-critical situations.
Lasota et al.~\cite{perceived_safety_survey_2021} demonstrated that robots meeting all engineering safety criteria can still induce anxiety if their motions appear unpredictable, and Rubagotti et al.~\cite{perceived_safety_taxonomy_2023} proposed a taxonomy of factors influencing perceived safety.
Beyond passive miscalibration, the human-agent interaction interface itself becomes a bidirectional attack surface. PsySafe~\cite{zhang2024psysafe} demonstrates that embedding dark personality traits into multi-agent system prompts induces collectively harmful behaviors that propagate through inter-agent dialogue rounds, revealing two distinct interaction-layer threats: \emph{trust exploitation} (Agent$\to$Human), where a personality-corrupted agent leverages conversational rapport to deliver psychologically manipulative responses, and \emph{interface poisoning} (Human$\to$Agent), where an adversarial user exploits the natural-language channel to embed persistent harmful tendencies that spread beyond the directly targeted agent.


\subsection{Multi-Agent Collaboration}
\label{subsec:mac}

When multiple embodied agents operate in a shared physical space, emergent safety risks arise. Unlike RL policy robustness (discussed in Section~\ref{subsec:action-control}) and planning-time coordination threats (discussed in Section~\ref{subsec:Multi_Agent_Planning}), this subsection focuses on execution-time threats that compromise collaboration through \textbf{Infection Attacks} that propagate adversarial behaviors across agent populations, and \textbf{Collusion Attacks} where autonomous agents deliberately coordinate malicious activities.


\noindent\textbf{Infection Attacks.}\quad
Infection attacks exploit inter-agent communication and memory-sharing channels to propagate adversarial behaviors from a single compromised agent to the broader population.
Agent Smith~\cite{gu2024agentsmith} shows that a single compromised agent can infect multimodal LLM agents exponentially fast, with adversarial content spreading through inter-agent communication without attacker intervention.


\noindent\textbf{Collusion Attacks.}\quad
Beyond passive infection, autonomous agents can deliberately coordinate malicious activities, a threat that intensifies as multi-agent systems gain tool-use capabilities and the ability to communicate freely.
Ren et al.~\cite{ren2025collusion} demonstrated that decentralized groups of AI agents outperform centralized ones at executing coordinated harmful actions such as misinformation campaigns and fraud, and can dynamically adjust tactics to evade detection even under active countermeasures.
The distributional AGI safety framework~\cite{tomasev_distributional_agi_2025} formalizes this concern through the ``patchwork AGI'' hypothesis: general intelligence capabilities may first arise through coordinated groups of specialized sub-AGI agents, making collusion risks relevant even before any individual system reaches superintelligent capability.


\section{Agentic System}
\label{sec:agentic}

The agentic system layer wraps the entire cognitive pipeline (perception $\rightarrow$ cognition $\rightarrow$ planning $\rightarrow$ action) with capabilities that define modern AI agents (tool use and skill use, memory, and self-evolution~\cite{yang2026toward,jiang_sok_skills_2026}), expanding the agent's capability from task execution to open-ended autonomy. This final expansion creates the broadest attack surface: adversaries can inject malicious tools or poisoned skills into the agent's action space, poison agent memory to cause persistent unsafe behavior, hijack self-evolution to erode alignment guarantees, and trigger cascading failures that propagate through all inner layers~\cite{wu2026isc}. Broader surveys on large-model and agent safety~\cite{ma2025safety,shahriar_agentic_security_2025,lazer_agentic_cybersecurity_2026,kim2026agentic,zhou_externalization_2026,li_harness_2026} provide complementary coverage of these threats. While Sections~\ref{sec:perception}--\ref{sec:action} address layer-specific vulnerabilities within the sense-think-act loop, this section examines emerging threats unique to agentic systems in embodied AI. This section organizes agentic system concerns into four subsections: \textbf{Tool and Skill Use}~(\ref{subsec:tool-use}) covers tool creation, tool manipulation, skill injection, skill stealing, and skill-auditing defenses; \textbf{Memory}~(\ref{subsec:memory}) examines memory poisoning, memory leakage, and memory defenses; \textbf{Self-Evolving}~(\ref{subsec:self-evolving}) covers misalignment and capability expansion risks in agents that autonomously modify themselves, together with embodied alignment defenses; and \textbf{Cascading Risks}~(\ref{subsec:cascading}) examines cross-layer attack propagation, supply-chain compromise, and infrastructure failures.

\begin{table}[!tp]
\center
\renewcommand{\arraystretch}{1.15}
\caption{A summary of \textbf{agentic attacks} for \textbf{agentic systems}.}
\resizebox{1\textwidth}{!}{

\rowcolors{2}{tablegray}{white}
{\setlength{\tabcolsep}{0pt}
\begin{tabular}{@{}>{\cellcolor{white}}p{.15\textwidth} p{.20\textwidth} p{.06\textwidth}
               p{.13\textwidth} p{.22\textwidth} p{.16\textwidth} p{.22\textwidth}@{}}
\toprule
\rowcolor{agentic-table-purple}
\cellcolor{agentic-table-purple}\textbf{Attack} & \textbf{Method} & \textbf{Year} & \textbf{Category} & \textbf{Subcategory} & \textbf{Target} & \textbf{Benchmark/Evaluation} \\
\midrule

& RoboCodeX~\cite{yu_robocodex_2024} & 2024 & Tool Use & Tool Creation Risks & VLA & RLBench, CALVIN \\
& ToolHijacker~\cite{wang_toolhijacker_2025} & 2025 & Tool Use & Tool Manipulation Attacks & LLM Agent & Custom \\
& STAC~\cite{li2025stac} & 2025 & Tool Use & Tool Manipulation Attacks & LLM Agent & Custom \\
& BackdoorAgent~\cite{feng2026backdooragent} & 2026 & Tool Use & Tool Manipulation Attacks & LLM Agent & AgentBoard \\
& MCP-FHA~\cite{belkhiter2026fha} & 2026 & Tool Use & Tool Manipulation Attacks & MCP Agent & Custom \\
& MalTool~\cite{hu2026maltool} & 2026 & Tool Use & Tool Manipulation Attacks & LLM Agent & Custom \\
& SkillJect~\cite{jia_skillject_2026} & 2026 & Tool Use & Skill Injection and Poisoning Attacks & Agent Skill & Custom \\
& Liu et al.~\cite{liu_agent_skills_2026} & 2026 & Tool Use & Skill Injection and Poisoning Attacks & Agent Skill & Custom \\
& Qu et al.~\cite{qu_skill_supply_chain_2026} & 2026 & Tool Use & Skill Injection and Poisoning Attacks & Agent Skill & Custom \\
& Wang et al.~\cite{wang_skill_stealing_2026} & 2026 & Tool Use & Skill Injection and Poisoning Attacks & LLM Agent & Custom \\

& AgentPoison~\cite{chen2024agentpoison} & 2024 & Memory & Memory Poisoning & RAG Agent & Autonomous Driving, Healthcare \\
& OEP~\cite{wang2026oep} & 2026 & Memory & Memory Poisoning & LLM Agent & Custom \\
& Pulipaka et al.~\cite{pulipaka2026sleepermem} & 2026 & Memory & Memory Poisoning & LLM Agent & Custom \\
& MEXTRA~\cite{wang_mextra_2025} & 2025 & Memory & Memory Leakage & LLM Agent & Custom \\
& MemoAnalyzer~\cite{zhang_memoanalyzer_2024} & 2024 & Memory & Memory Leakage & LLM Agent & Custom \\
& Zheng et al.~\cite{zheng_justask_2026} & 2026 & Memory & Memory Leakage & LLM Agent & Custom \\
& MAMA~\cite{liu_mama_2025} & 2025 & Memory & Memory Leakage & Multi-Agent & Custom \\
& ImmersedPrivacy~\cite{wang2026immersedprivacy} & 2026 & Memory & Memory Leakage & VLM Agent & ImmersedPrivacy \\

& Shao et al.~\cite{shao_misevolving_2025} & 2025 & Self-Evolving & Misalignment & LLM Agent & Custom \\
& Self-Improving EFM~\cite{selfimproving_efm_2025} & 2025 & Self-Evolving & Capability Expansion & Embodied Agent & Custom \\

& RAVEN~\cite{yeke_raven_2025} & 2025 & Cascading & Cross-Layer Propagation & Multi-Agent & Custom \\
& SAPIA~\cite{geng2026white} & 2026 & Cascading & Cross-Layer Propagation & Embodied Agent & Custom \\
& Zhou et al.~\cite{zhou_goal_backdoor_2025} & 2025 & Cascading & Supply Chain Attacks & VLA & Custom \\
\multirow{-24}{*}{\parbox[c]{\linewidth}{\raggedright Agentic\\Attack}} & TrojanRobot~\cite{wang2024trojanrobot} & 2024 & Cascading & Supply Chain Attacks & VLM Agent & Custom \\

\bottomrule
\end{tabular}
}}
\label{tab:agentic_attack}
\end{table}

\begin{table}[!tp]
\center
\renewcommand{\arraystretch}{1.15}
\caption{A summary of \textbf{agentic defenses} for \textbf{agentic systems}.}
\resizebox{1\textwidth}{!}{

\rowcolors{2}{tablegray}{white}
{\setlength{\tabcolsep}{0pt}
\begin{tabular}{@{}>{\cellcolor{white}}p{.12\textwidth} p{.20\textwidth} p{.06\textwidth}
               p{.18\textwidth} p{.18\textwidth} p{.16\textwidth} p{.25\textwidth}@{}}
\toprule
\rowcolor{agentic-table-purple}
\cellcolor{agentic-table-purple}\textbf{Defense} & \textbf{Method} & \textbf{Year} & \textbf{Category} & \textbf{Subcategory} & \textbf{Target} & \textbf{Benchmark/Evaluation} \\
\midrule

& Safety Chip~\cite{brown_safetychip_2024} & 2024 & Tool Use & Tool Use Defenses & LLM Agent & Custom \\
& SELP~\cite{ahn_selp_2024} & 2024 & Tool Use & Tool Use Defenses & LLM Agent & Custom \\
& AgentSpec~\cite{chen_agentspec_2026} & 2026 & Tool Use & Tool Use Defenses & LLM Agent & Code, Autonomous Driving, Embodied \\
& Chang et al.~\cite{chang2026trustboundary} & 2026 & Tool Use & Tool Use Defenses & VLM Agent & Custom \\
& Lv et al.~\cite{lv_skill_audit_2026} & 2026 & Tool Use & Skill Defenses & Agent Skills & Custom \\
& RouteGuard~\cite{xiao_routeguard_2026} & 2026 & Tool Use & Skill Defenses & LLM Agent & Custom \\

& MemOS~\cite{hu_memos_2025} & 2025 & Memory & Memory Defenses & LLM Agent & Custom \\
& SafeHarbor~\cite{liu2026safeharbor} & 2026 & Memory & Memory Defenses & LLM Agent & Custom \\
& PRISM~\cite{tapwal2026prism} & 2026 & Memory & Memory Defenses & Multi-Agent & Custom \\

& Moral Anchor~\cite{moralanchor_2025} & 2025 & Self-Evolving & Embodied Alignment & LLM Agent & Custom \\
& Q-DIG~\cite{srikanth2026red} & 2026 & Self-Evolving & Embodied Alignment & LLM Agent & Custom \\
& Nay~\cite{nay_law_alignment_2025} & 2025 & Self-Evolving & Embodied Alignment & LLM Agent & Custom \\
& C3AI~\cite{wu_c3ai_2025} & 2025 & Self-Evolving & Embodied Alignment & LLM Agent & Custom \\
& ERT~\cite{karnik_ert_2024} & 2024 & Self-Evolving & Embodied Alignment & Robot & Custom \\
\multirow{-15}{*}{\parbox[c]{\linewidth}{\raggedright Agentic\\Defense}} & HEAL~\cite{chakraborty_heal_2025} & 2025 & Self-Evolving & Embodied Alignment & Embodied Agent & Custom \\

\bottomrule
\end{tabular}
}}
\label{tab:agentic_defense}
\end{table}


\subsection{Tool and Skill Use}\label{subsec:tool-use}

Agentic embodied systems extend their action space through two complementary mechanisms: \emph{tools} (typed APIs, code interpreters, and external services that the agent invokes at runtime) and \emph{skills} (packaged, versioned behavioral routines distributed through community skill ecosystems~\cite{jiang_sok_skills_2026}). Both translate model decisions into physical or digital effects and both expose attack surfaces where a misdirected call can cause concrete harm. The OWASP Top~10 for Agentic Applications~\cite{owasp_agentic_2026} flags tool misuse and delegated trust as critical agentic vulnerability classes, while emerging skill marketplaces add a parallel supply-chain surface that is largely untouched by traditional code-review pipelines~\cite{li_secure_skills_2026}.

\noindent\textbf{Tool Creation Risks.}\quad Agents that generate or ingest tools introduce vulnerabilities that translate directly into physical harm when the tools control actuators.
In the code-as-action paradigm, RoboCodeX~\cite{yu_robocodex_2024} synthesizes control code via LLMs, inheriting all vulnerabilities of the underlying model.

\noindent\textbf{Tool Manipulation Attacks.}\quad Adversaries can trick agents into selecting or sequencing tools in harmful ways.
ToolHijacker~\cite{wang_toolhijacker_2025} injects malicious tool documents that compel agents to select attacker-controlled tools.
STAC~\cite{li2025stac} composes individually benign tool calls into dangerous multi-turn sequences.
BackdoorAgent~\cite{feng2026backdooragent} embeds persistent triggers across planning, memory, and tool-use stages.
At the protocol layer, function-hijacking attacks against the Model Context Protocol~\cite{belkhiter2026fha} subvert function-calling agents by impersonating or shadowing legitimate tool endpoints.
MalTool~\cite{hu2026maltool} synthesizes malicious tool code via coding-LLMs under a CIA-triad taxonomy, embedding attack behavior in tool implementations distributed through agent tool platforms.
An in-the-wild study~\cite{khodayari2026indirect} documents indirect prompt injections embedded in webpages and documents that agents passively ingest.

\noindent\textbf{Tool Use Defenses.}\quad Defenses against tool misuse span runtime enforcement and code-level safety.
Safety Chip~\cite{brown_safetychip_2024} intercepts and validates generated code before execution.
SELP~\cite{ahn_selp_2024} filters LLM-generated plans through safety verification before physical execution.
AgentSpec~\cite{chen_agentspec_2026} specifies and enforces runtime constraints on LLM agents via a lightweight DSL, validated on embodied and autonomous driving tasks.
RoboSafe~\cite{wang2025robosafe} prevents implicit risks in VLM-driven agents via executable safety logic.
Chang et al.~\cite{chang2026trustboundary} address \emph{trust-boundary confusion}, where a VLM agent treats text injected into an image as a trusted instruction. Their defense renegotiates the per-modality trust contract before tool invocation.

\noindent\textbf{Skill Injection and Poisoning Attacks.}\quad Because skill artifacts encode persistent, reusable embodied routines (potentially driving actuator commands across sessions), a single poisoned skill amplifies into wide deployment of unsafe physical behavior before detection.
Distinct from prompt-time tool manipulation, skill-based attacks compromise the persistent skill artifacts an agent loads, executes, or composes.
SkillJect~\cite{jia_skillject_2026} automates closed-loop skill prompt injection, concealing malicious payloads in auxiliary scripts that evade manual review.
At the supply-chain layer, an empirical analysis of published skills finds many contain security vulnerabilities spanning prompt injection, credential exfiltration, and privilege escalation~\cite{liu_agent_skills_2026}. Coding-agent ecosystems are similarly compromised by supply-chain skill poisoning that propagates downstream~\cite{qu_skill_supply_chain_2026}.
Adversaries can also exfiltrate the skill artifacts themselves: black-box skill stealing reconstructs proprietary skills from query access alone~\cite{wang_skill_stealing_2026}.
As a unifying view, the DTap red-teaming platform~\cite{chen_dtap_2026} treats skill as a first-class injection vector alongside prompt, tool, and environment attacks, exposing systematic vulnerabilities across production domains.
SkillSafetyBench~\cite{jin2026skillsafety} measures how malicious skill materials and local artifacts steer agents toward unsafe actions even on benign requests.
AgentTrap~\cite{zhuang2026agenttrap} dynamically tests whether agents resist malicious runtime behavior disguised as routine workflow inside installed third-party skills.

\noindent\textbf{Skill Defenses.}\quad Defending the skill layer requires both per-skill auditing and runtime privilege control.
Li et al.~\cite{li_secure_skills_2026} formalize a skill threat taxonomy and propose architectural mitigations, while structured security auditing~\cite{lv_skill_audit_2026} hardens deployed skills against runtime drift.
RouteGuard~\cite{xiao_routeguard_2026} flags skill poisoning by monitoring the agent's internal routing signals.


\subsection{Memory}\label{subsec:memory}

Agent memory (episodic logs, RAG corpora, and conversation histories) enables cross-session experience accumulation but creates a durable attack surface for both integrity and confidentiality violations~\cite{hu_memory_agents_2025,zhang_memory_mechanism_2025,luo_storage_experience_2026,wu2025memory}. The OWASP Top~10 for Agentic Applications classifies memory poisoning~(ASI06) as a critical agentic risk~\cite{owasp_agentic_2026}.

\noindent\textbf{Memory Poisoning.}\quad Agents that store and retrieve past experiences are vulnerable to attacks implanting malicious records that persist across sessions.
AgentPoison~\cite{chen2024agentpoison} backdoors RAG-based agents by poisoning memory, validated on autonomous driving and healthcare agents.
OEP~\cite{wang2026oep} poisons self-evolving agents with locally correct but non-transferable experiences, biasing experience reflection into over-generalized rules that cause downstream failures.
Pulipaka et al.~\cite{pulipaka2026sleepermem} demonstrate sleeper memory poisoning, where an assistant stores a fabricated user memory from manipulated context that lies dormant before resurfacing to steer later sessions.
Al-Tawaha et al.~\cite{tawaha2026temporal} measure temporal memory contamination, showing agent safety degrades as memory accumulates across unrelated tasks.
For embodied agents with personalized memory~\cite{zhang2026smartagent}, persistent poisoning can cause repeated unsafe physical behaviors.
In multi-agent systems, memory poisoning creates cascading failures through semantic opacity and temporal compounding~\cite{adversa_cascading_2025}.

\noindent\textbf{Memory Leakage.}\quad Embodied agents that log interactions, sensor readings, and user preferences create confidentiality risks when adversaries extract private data from memory stores.
MEXTRA~\cite{wang_mextra_2025} introduces black-box memory extraction attacks that recover private data without model access.
MemoAnalyzer~\cite{zhang_memoanalyzer_2024} analyzes privacy leakage from persistent conversation memory in LLM agents.
MAMA~\cite{liu_mama_2025} shows that multi-agent topologies amplify leakage risks through inter-agent communication channels.
System prompt extraction techniques~\cite{zheng_justask_2026} suggest that agents can be induced to reveal privileged context through natural-language queries, a threat model that extends to memory stores.
For embodied VLMs, ImmersedPrivacy~\cite{wang2026immersedprivacy} measures how poorly current models recognize private content in the surrounding physical scene, an upstream sensor-side leakage source that bypasses memory defenses.

\noindent\textbf{Memory Defenses.}\quad Defenses against memory attacks focus on provenance tracking and architectural safeguards.
MemOS~\cite{hu_memos_2025} proposes a memory operating system with provenance tagging, lifecycle tracking, and permission enforcement.
A-MEM~\cite{xu2025amem} uses interconnected knowledge networks where poisoning can be detected through link consistency checks.
SafeHarbor~\cite{liu2026safeharbor} adds a hierarchical memory-augmented guardrail that intercepts unsafe agent behavior at runtime by retrieving safety-relevant precedents from a structured memory bank.
PRISM~\cite{tapwal2026prism} detects secret leakage at generation time, fusing many runtime signals into a calibrated risk score that intervenes before a credential is fully reconstructed.
SPINE~\cite{fan2026spine} argues that embodied AI requires explicit privacy-utility trade-offs at the memory layer.
The OWASP mitigation framework recommends memory segmentation, provenance tracking, and automatic expiry of suspicious entries~\cite{owasp_agentic_2026}.


\subsection{Self-Evolving}\label{subsec:self-evolving}

Self-evolving agents (systems that autonomously modify their own models, memory, tools, or workflows) introduce risks absent from static systems~\cite{zheng_lifelong_2025,fang_selfevolving_2025,gao2025selfevolving,xiang2025selfevolving}. Shao et al.~\cite{shao_misevolving_2025} demonstrated misevolution degrades safety via four pathways: parametric drift, memory accumulation, tool corruption, and workflow degradation.

\noindent\textbf{Misalignment.}\quad Two concrete pathways from parametric drift and memory accumulation erode alignment: self-training degrades safety refusal rates, while accumulated experiences encode unsafe patterns that override original constraints.
The Moral Anchor System~\cite{moralanchor_2025} detects and mitigates value drift via Bayesian monitoring with adaptive governance.
The ``safe-by-coevolution'' paradigm~\cite{sun2025texttt} argues that safety must coevolve alongside capability growth.
Agent-SafetyBench~\cite{zhang_agentsafetybench_2024} benchmarks agent safety across multiple risk dimensions, finding that no agent passes 60\% of safety evaluations.

\noindent\textbf{Capability Expansion.}\quad Self-evolving agents may acquire capabilities beyond their original design scope.
Self-Improving Embodied Foundation Models~\cite{selfimproving_efm_2025} demonstrate robots autonomously acquiring manipulation skills beyond their training distribution, showing how capability growth can be rapid and unsupervised.
A survey on safe continual RL~\cite{safecontinualrl_2026} identifies the tension between adaptation and constraint preservation in lifelong embodied learning.


\noindent\textbf{Embodied Alignment.}\quad Aligning embodied agents requires bridging abstract human values and concrete physical actions. Agentic RL~\cite{zhang_agentic_rl_2025} provides a landscape of reinforcement learning approaches for LLM-based agents, where reward design and policy optimization directly shape alignment outcomes.
VLSA~\cite{hu2025vlsa} adds a plug-and-play safety layer using control barrier functions for VLA models.
For red teaming, ERT~\cite{karnik_ert_2024} audits robotic foundation models by generating adversarial instructions refined through robot execution feedback.
Q-DIG~\cite{srikanth2026red} discovers failure modes through quality-diversity prompt generation.
For constitutional alignment, Nay~\cite{nay_law_alignment_2025} grounds agent alignment in legal principles, extended by C3AI~\cite{wu_c3ai_2025} with graph-based principle selection.
HEAL~\cite{chakraborty_heal_2025} targets hallucination in embodied agents as a safety-critical failure mode.
The FLI AI Safety Index~\cite{fli_safety_index_2025} reports that no major AI company has achieved satisfactory existential safety planning.


\subsection{Cascading Risks}\label{subsec:cascading}

All capability layers operate within a closed physical loop, so compromising any single layer can propagate to unsafe physical actions. This subsection examines cross-layer penetration, supply-chain compromise, infrastructure failures, and mitigation strategies.

\noindent\textbf{Cross-Layer Attack Propagation.}\quad Adversaries can exploit one pipeline layer to trigger unsafe behavior downstream.
In autonomous driving, Wu et al.~\cite{han_adversarial_driving_2021} showed camera perturbations propagating directly to steering outputs.
Liu et al.~\cite{liu_dynamic_ad_2023} demonstrated dynamic adversarial patches manipulating downstream decisions, and Cheng et al.~\cite{cheng_attacking_ad_2025} evaluated multi-sensor pipeline attacks producing system-level failures.
Surveys further systematize sensor-to-control propagation across 2D perception~\cite{chen_revisiting_adv_2025}, 3D perception~\cite{mahima2024toward}, and full sensor pipelines~\cite{liu_sok_sensor_2025}.
In embodied navigation, Liu et al.~\cite{liu_physical_nav_2024} demonstrated physically-realizable patches cascading to navigation failure, and Jia et al.~\cite{jia_spatiotemporal_2020} exploited temporal vulnerabilities across perception-action loops.
Language model integration creates a new cross-layer surface: Liu et al.~\cite{liu2024exploring} showed LLM decision-layer attacks cascading through the full pipeline.
In multi-robot systems, Yeke et al.~\cite{yeke_raven_2025} automated discovery of semantic attacks causing coordination failures, and Bahrami and Jafarnejadsani~\cite{gao_multirobot_2025} showed misclassifications propagating to fleet-level breakdowns.
SAPIA~\cite{geng2026white} demonstrates prompt injection propagating to embodied actions.
For VLAs, Wang et al.~\cite{Wang_ExploringAdversarialVulnerabilities_2025} showed perception attacks propagating through language understanding to control, Li et al.~\cite{li_adv_vla_2025} obtained complete control authority via the language interface, and Wang et al.~\cite{wang2025advedm} demonstrated that fine-grained semantic edits to a few key objects induce valid but incorrect VLA-policy decisions.

\noindent\textbf{Supply Chain Attacks.}\quad Pre-trained models and third-party plugins create trust boundaries that adversaries can compromise before deployment.
Wang et al.~\cite{wang2024trojanrobot} embedded backdoor-finetuned VLMs as malicious perception modules within modular robotic policies.
Zhou et al.~\cite{zhou_goal_backdoor_2025} manipulated VLAs through physical object triggers achieving cross-layer penetration to goal-oriented actions.
Beyond model-level poisoning, community skill marketplaces present a parallel supply-chain surface that is treated as a first-class subject in Section~\ref{subsec:tool-use}.

\noindent\textbf{Infrastructure Failures.}\quad Embodied agents depend on cloud infrastructure for inference, storage, and coordination, creating dependencies that compromise safety when infrastructure fails.
Data poisoning at the infrastructure level propagates through the agent lifecycle~\cite{bengio_safety_2025_safeguards}, and network partitioning leads to inconsistent world models and uncoordinated actions~\cite{adversa_cascading_2025}.
Abdelfattah et al.~\cite{lopez2025securing} mapped perception-to-control propagation in vision-based autonomous systems.
Zhang et al.~\cite{zhang_robotics_cyber_2021} documented robotic vulnerabilities across hardware, middleware, and application layers.
Wang et al.~\cite{wang_sok_humanoid_2025} addressed cross-layer propagation in humanoid ecosystems.
Khalid et al.~\cite{haskard2025secure} examined safety-trust-cybersecurity intersections, and Neupane et al.~\cite{neupane_security_ai_robotics_2023} mapped AI-enabled attack surfaces in hybrid architectures.
The OWASP ASI08 standard~\cite{owasp_asi08_2026} provides industry guidance on cascading failure assessment.

\noindent\textbf{Benchmarks and Mitigation.}\quad Evaluating cascading risks requires benchmarks spanning multiple pipeline layers.
SafeAgentBench~\cite{yin2024safeagentbench} provides tasks spanning the full perception-to-action pipeline with safety hazards.
Agent-SafetyBench~\cite{zhang_agentsafetybench_2024} identifies robustness and risk-awareness as fundamental gaps across agent safety evaluations.
HarnessAudit~\cite{liu2026harnessaudit} audits full agent execution trajectories for boundary compliance, fidelity, and stability, catching mid-trajectory violations that output-level evaluation misses.
For mitigation, the International AI Safety Report~\cite{bengio_safety_2025_safeguards} recommends layered safeguards across training, deployment, and post-deployment stages.
Mechanical fail-safes (physical stops, force limits, and emergency brakes) provide protection independent of AI control.
Pre-certified boundaries~\cite{amodei_adolescence_2025} constrain agents to safe envelopes, with crossings requiring human approval.

\section{Open Challenges} \label{sec:challenges}
Despite rapid progress, safety in embodied AI remains at a formative stage. Current systems are fragile, narrow in capability, and far from understanding safety in any autonomous sense. Their deployment in human-centered environments exposes failure modes that span physical, cognitive, and social dimensions. Below we outline several cross-cutting open problems that must be addressed before embodied intelligence can be safely integrated into the real world.

\subsubsection*{1. Safety Evaluation Without Causing Real-World Risks}

Many of the most serious safety risks in embodied AI cannot be directly evaluated in the real world. Unlike purely digital systems, where unsafe outputs can often be studied offline, failures in embodied agents can cause property damage, injury, or loss of life. It is neither ethical nor legally permissible to systematically test scenarios in which robots may harm humans, damage infrastructure, or deliberately operate at the edge of instability. This creates a fundamental gap between the risks researchers need to characterize and the experiments they can actually run.

Existing simulators offer only a partial solution. They typically fall short in representing the richness and uncertainty of real-world physics, human behavior, and long-horizon interactions in cluttered environments. Rare but catastrophic failures are especially difficult to model, both because they are statistically infrequent and because they often arise from coupled perception--control--interaction dynamics that are poorly captured by current tools. As a result, safety guarantees derived from virtual tests often fail to transfer when systems leave the lab.

A core open challenge is to design evaluation methodologies that provide strong evidence about real-world safety without exposing humans to danger. This includes (i) high-fidelity digital twins that capture contact dynamics, sensing noise, and environmental variability; (ii) scalable stress testing and adversarial scenario generation targeting long-tail failures; and (iii) formal verification and offline analysis frameworks that can reason about unexecuted trajectories and counterfactual interactions. For safety research itself, we also need protocols that allow realistic red-teaming and physical stress tests while maintaining bounds on allowed risk.

\subsubsection*{2. Safety Generalization Across Embodiments and Tasks}

Embodied AI spans humanoids, mobile manipulators, autonomous vehicles, drones, wheeled platforms, and micro-robots. These embodiments differ dramatically in dynamics, perception stacks, interaction modes, and the magnitude of potential harm. Consequently, vulnerabilities identified on one embodiment often fail to transfer to another, and safety interventions are frequently tailored to a specific robot, task, or environment.

The field currently lacks shared abstractions for thinking about safety across embodiments. There are no widely adopted taxonomies of failure that cut across platforms, nor common stress-testing protocols analogous to standard benchmarks in perception or language. This fragmentation impedes cumulative scientific progress: results are difficult to reproduce, compare, or build upon, and it is unclear how to translate insights from one domain (e.g., warehouse robots) to another (e.g., assistive humanoids).

An important open problem is to identify cross-embodiment safety principles and interfaces. This includes modular safety layers that sit above low-level control but below high-level task specification, common representations for unsafe states and risk signals, and evaluation protocols that factor out embodiment-specific details while still accounting for different harm profiles. Achieving such generality will require sustained collaboration across robotics, control, learning, and safety engineering, and may ultimately resemble the role that system-level standards play in cybersecurity.

\subsubsection*{3. Safety Protocols for Human-Robot Interaction}

As robots move from industrial cages into homes, hospitals, and public spaces, human-robot interaction (HRI) becomes a primary axis of safety risk. Embodied agents must interpret ambiguous language, gestures, gaze, and proxemics; adapt to diverse social norms; and remain robust to human error, frustration, and strategic behavior. Failures in HRI are rarely just perception errors or control glitches; they are often failures of shared mental models, trust calibration, and social context understanding.

Today, we lack systematic tools to assess safety vulnerabilities in HRI. Robots may misinterpret human intent, overlook subtle cues of distress or danger, or over-trust misleading instructions. Adversarial or curious users can exploit these weaknesses: issuing conflicting commands, providing deceptive demonstrations, or manipulating the robot’s social compliance to bypass safeguards. Such behaviors are difficult to explore experimentally because they sit at the intersection of technical safety and human subjects research.

A major challenge in ensuring safety in HRI is the development of comprehensive and safe HRI protocols. These protocols must account for diverse user groups, such as children, the elderly, and adversaries, while considering varying emotional states, cultural norms, and social edge cases, all without exposing real participants to harm. This calls for research into simulated or mixed-reality humans, data-driven models of human behavior, and the creation of protocols for studying conversational manipulation, physical proxemics, and social pressure in controlled yet realistic environments. Ultimately, embodied safety demands protocols that allow models to jointly reason about physical risks and social context, ensuring safe and effective interactions in a wide range of scenarios.

\subsubsection*{4. Safety-aware Embodiments: From Algorithms to Hardware}
Safety in embodied AI is shaped not only by the algorithms that govern behavior but also by the physical systems on which they rely. Sensors such as cameras, LiDAR, IMUs, microphones, and tactile arrays are vulnerable to manipulation through environmental factors like lighting patterns, reflective materials, acoustic and electromagnetic interference, or mechanical disturbances. Similarly, actuators may saturate, overheat, or behave nonlinearly under load, leading to potential failures. These hardware-specific vulnerabilities cannot always be mitigated by digital defenses alone and can directly result in hazardous behavior.

Addressing hardware-level vulnerabilities presents significant challenges. The attack surface depends on factors such as manufacturing tolerances, material properties, mechanical resonances, and proprietary signal-processing methods, all of which can vary widely across devices and vendors. To systematically discover these vulnerabilities, new methodologies are required—ones that combine physical experimentation with model-based analysis. Tools are also needed to characterize worst-case perturbations under realistic constraints, ensuring that both known and unforeseen risks are addressed. We must develop principled approaches to translate insights from controlled lab environments to more complex, real-world deployments.

Beyond ensuring robustness, safety must be inherently designed into the embodiment itself. Robots built with rigid frames, high-torque actuators, and sharp edges inherently pose risks, even when their software is functioning correctly. Future research should focus on soft and compliant robotics, low-impact actuation, energy-efficient motion planning, and mechanical fail-safes (such as passive compliance, safe braking, and redundancy). These strategies aim to minimize the potential damage caused by any single failure. A comprehensive approach to embodied safety will integrate algorithmic defenses, fault-tolerant hardware, and physical attack simulations into a cohesive design philosophy.

\subsubsection*{5. Safety for Generalist Embodied Foundation Models}

The emergence of foundation models for robotics and embodied agents promises broad generalization across tasks, environments, and modalities. However, these generalist models also introduce new safety challenges. They are trained on heterogeneous data with weak or implicit supervision, may acquire capabilities that were not anticipated by designers, and can be rapidly adapted or fine-tuned by end users. In such systems, the space of possible behaviors is too large to enumerate or exhaustively test.

Designing safety mechanisms for generalist embodied models requires rethinking traditional assumptions. Hard-coded rule sets and task-specific guardrails do not scale when the model can compose novel behaviors on the fly. Instead, we need representations that encode hazard and value information, uncertainty-aware planning that detects and avoids novel risks, and alignment techniques that transfer safety constraints across tasks and embodiments. These mechanisms must remain robust under continual learning, distribution shift, and model updates, without catastrophic forgetting of previously learned safety properties.

Moreover, the attack surface expands as these models become more programmable by natural language or demonstration: malicious prompts, poisoned demonstrations, and subtle changes in training data can all induce unsafe behavior. Developing principled red-teaming methodologies, scalable oversight strategies, and defenses against training-time and deployment-time manipulation is an open and pressing problem. Addressing it will require closer integration between embodied learning, foundation model safety, and security-oriented machine learning.

\subsubsection*{6. Governance, Standards, and Shared Safety Infrastructure}

Technical progress alone will not ensure safe deployment of embodied AI. Regulation, standards, and institutional practices must co-evolve with capability. At present, there are few comprehensive legal or safety frameworks specific to humanoids, service robots, or multimodal embodied systems. Liability regimes are unclear when failures arise from complex human--robot--environment interaction chains, and there is little consensus on acceptable levels of risk in public or domestic settings.

The field needs shared safety infrastructure: standardized incident reporting, benchmarks and test suites for safety-critical scenarios, certification procedures for hardware and software stacks, and guidelines for data governance, logging, and post-incident analysis. Requirements for red-teaming, auditing, and third-party evaluation should be aligned with what is feasible in research and industry, while still providing meaningful protection for end users.

Longer-term societal impacts also need to be incorporated into our notion of safety. Widespread deployment of embodied agents will affect labor markets, social trust, fairness in access to assistance, and the psychological experience of living with quasi-autonomous machines. Addressing these issues demands interdisciplinary collaboration among roboticists, machine learning researchers, human factors experts, ethicists, and policymakers. For embodied AI, ``safety by design'' must extend beyond individual systems to include the institutions and norms that govern their development and use.

\vspace{0.5em}
The safety challenges of embodied AI span evaluation, across-embodiment generalization, human interaction, hardware, foundation models, and governance. Addressing them will require a coordinated research program that links mechanism-level defenses, adversarial testing, formal guarantees, embodied cognition, and societal oversight. Embodied agents can deliver transformative benefits, but only if safety is treated as a central scientific objective rather than an afterthought in capability development.

\section{Future Trends} \label{sec:future}

Embodied AI safety is on the verge of a major evolution as robots transition from narrow, task-specific systems to general-purpose agents operating in dynamic, human-centered environments. The next decade is expected to reshape not only the training and deployment of embodied agents but also the way safety is conceptualized, measured, and ensured. Below, we outline several trends that will guide this transformation.

\subsubsection*{1. Generalist Embodied Foundation Models}

The rise of embodied foundation models marks a shift away from specialized controllers toward unified architectures capable of grounding language, vision, and action in a single expressive representation. These models will generalize across tasks, adapt to new embodiments with minimal retraining, and enable more fluid forms of human–robot collaboration. At the same time, their broad capability surfaces will demand new safety frameworks. Future research will focus on embedding safety directly into the model's representations and planning mechanisms, allowing robots to reason about risk, uncertainty, and value alignment as intrinsic cognitive processes rather than as externally imposed constraints. As these systems become increasingly programmable via language and demonstration, maintaining robust alignment under adaptation and fine-tuning will become a defining challenge.

\subsubsection*{2. World Model and Safety Simulation}
World Models are a class of generative models that learn an internal representation of an environment, allowing systems to simulate and predict future states based on past observations. They enable agents to learn in simulated environments instead of relying solely on real-world data. These models will be crucial for simulating safety scenarios in embodied AI systems. By capturing complex dynamics such as sensor noise, human behavior, and environmental variability, World Models allow safety protocols to be tested without real-world risks. They can simulate rare or adversarial hazard scenarios, identify system vulnerabilities, and test failure boundaries at scale, including long-term interactions and extreme conditions that are difficult or unsafe to replicate physically. Moreover, World Models enable continuous validation of safety properties by replicating near-real-time conditions, offering dynamic assessments as robots adapt. By integrating training, testing, and monitoring, they proactively detect risks and predict future threats, reducing the likelihood of real-world failures. An emerging paradigm of WAMs~\cite{wang2026wams} pushes this direction further by unifying predictive state modeling with action generation in a single foundation model; their broader action surface and tighter dynamics coupling raise new safety questions that remain unexplored.

\subsubsection*{3. Safe-reasoning Embodied Agents}

Future embodied agents will increasingly learn about danger, physical causality, and risk through large-scale self-supervision. Instead of depending primarily on human-labeled safety rules, robots will acquire intuitive physical knowledge by predicting the outcomes of their interactions, identifying precursors to unsafe states, and modeling the causal structure that governs real-world hazards. This trend points toward embodied AI systems that treat safety reasoning as a core part of world modeling: agents will anticipate multi-step consequences, detect when uncertainty is rising, and adjust policies before danger materializes. As World Models become more advanced, safety will transition from reactive enforcement to proactive management, driven by internalized predictive structures.

\subsubsection*{4. Integration of Physical, Cyber, and Social Safety}

Embodied AI dissolves traditional boundaries between cyber security, physical robustness, and social interaction safety. Misleading language can trigger unsafe actions; sensor spoofing can destabilize mechanical behavior; cyber compromises can lead to real-world motion that endangers humans. Future systems will require unified safety architectures that reason jointly across these domains rather than treating them as disjoint fields. Robustness to adversarial human interaction, resilience to perception and actuation manipulation, and protection against cyber exploits will form a coherent safety stack. This integration will push research toward multi-layered monitoring mechanisms capable of tracking intent, environmental anomalies, control divergence, and communication risks within a shared threat model.

\subsubsection*{5. Safety-Centered Robotic Design}

Safety will increasingly be integrated into the physical design of robots. Advances in soft robotics, variable-impedance actuators, compliant mechanisms, and low-impact materials will minimize intrinsic physical hazards, making robots safer through their construction rather than relying solely on software control. Mechanical structures may include passive safety features, such as energy-dissipating actuators or fail-safe collapsible components, which significantly reduce the severity of unexpected impacts. These innovations will promote co-design methodologies, where both hardware and algorithms contribute to safety guarantees, ensuring that systems maintain a controlled risk profile, even under severe perception or control failures.

\subsubsection*{6. Continuous Red-Teaming and Safety Monitoring}

As embodied agents operate continuously, update models over time, and encounter new states far beyond their training distributions, static evaluation will no longer be sufficient. The field is moving toward continual red-teaming frameworks in which adversarial agents, simulated humans, or automated perturbation generators probe robot policies for emergent vulnerabilities. In parallel, real-time safety monitors will track system uncertainty, detect anomalous internal activations, and intervene when the agent’s behavior drifts outside known-safe regimes. This continuous oversight will form a persistent layer of defense that evolves alongside the agent, supporting early detection of unsafe adaptation or model degradation. Embodied AI safety research will likewise need controllable, interactive red-teaming platforms in the spirit of DTap~\cite{chen_dtap_2026}, but extended to physical action spaces, sensor pipelines, and human-in-the-loop interaction.

\subsubsection*{7. Institutional Governance and Safety Infrastructure}

As embodied AI enters homes, hospitals, public spaces, and workplaces, technical safety will be inseparable from institutional policy and societal expectations. We anticipate the emergence of shared safety infrastructure including standardized incident reporting pipelines, third-party certification frameworks, requirements for transparent logging and post-incident analysis, and legal guidelines for deployment in human-centered environments. Governance mechanisms will increasingly require obligatory red-teaming, regular safety audits, and documented risk analyses prior to deployment. At a broader level, the societal impacts of embodied AI, ranging from labor displacement to public trust, will reshape how safety is defined and regulated. Technical research will need to operate in concert with ethics, human factors, policy, and law to ensure responsible integration at scale.

\vspace{0.5em}
The trajectory of embodied AI safety is moving toward integration of model cognition, simulation, hardware design, continuous oversight, and governance. Safety will become a performance axis that shapes every layer of the embodied intelligence stack, from foundation models to construction to deployment policy. This evolution redefines how embodied agents are built, evaluated, and trusted in human environments.

\section{Conclusion} \label{sec:conclusion}

Embodied AI is undergoing a rapid transition from controlled laboratory demonstrations to deployment in open, dynamic, and inherently safety-critical real-world environments. This survey has provided the first systematic and comprehensive treatment of \textbf{safety in embodied AI}, organizing attacks, vulnerabilities, and defenses across the intertwined stages of perception, cognition, planning, and interaction. By integrating insights from over \numpapers{} works spanning traditional AI safety, robotics, foundation models, and multimodal systems, we highlight how embodied safety requires a different perspective from digital-only AI: one that treats safety not as an isolated module but as a property emerging from the entire perceive–think–act loop.

Our analysis reveals that despite rapid progress in embodied perception, reasoning, planning, and control, current systems remain fragile and far from internalizing robust notions of risk, hazard, or alignment. The open challenges identified in Section~\ref{sec:challenges} underscore the breadth of unresolved problems. Safety evaluation remains constrained by the impossibility of real harm experiments; embodiments and tasks vary so widely that cross-platform safety abstractions are still lacking; human–robot interaction exposes systems to complex social, adversarial, and unpredictable dynamics; and hardware-level vulnerabilities can bypass software safeguards entirely. At the same time, generalist embodied foundation models introduce new avenues for emergent behaviors, unpredictable generalization, and manipulation via language or demonstration. These challenges collectively demonstrate that embodied AI safety is still in its infancy and demands coordinated progress across algorithms, simulation, hardware, and governance.

Looking forward, the trends outlined in Section~\ref{sec:future} point toward a redefinition of safety in embodied intelligence. Generalist embodied models will reshape the space of possible behaviors, necessitating safety representations embedded directly within the model’s cognitive structure. High-fidelity simulation and digital twins will become indispensable for stress testing, rare-event generation, and continuous monitoring. Advances in self-supervised world modeling will allow agents to anticipate and reason about physical risk ahead of action, moving safety from reactive constraints to proactive understanding. The convergence of cyber, physical, and social safety will push researchers toward holistic threat models and unified safety stacks. Meanwhile, progress in soft and compliant robotics will embed safety into the embodiment itself, reducing the intrinsic risk of physical interaction. Finally, continuous red-teaming, automated oversight, and emerging regulatory frameworks will form the institutional backbone that supports the responsible deployment of embodied agents in human-centered environments.

Together, these challenges and trends highlight a pivotal moment for the field. Embodied agents have the potential to transform transportation, healthcare, manufacturing, and daily life, but only if their intelligence is matched by robust, principled safety design. Achieving this vision will require new scientific foundations that unify perception, causal cognition, safe planning, and human-centered interaction; engineering practices that integrate hardware, simulation, and continuous monitoring; and governance structures ensuring transparency, accountability, and public trust. We hope this survey serves as both a reference and a catalyst for future work, guiding the community toward embodied AI systems that are capable, aligned, and safe for the real world.

\clearpage

\bibliographystyle{plainnat}
\bibliography{main}






\end{document}